\documentclass[useAMS,usenatbib]{lib/mnras}

\usepackage{natbib}
\usepackage{epsfig}
\usepackage{scrtime}
\usepackage{graphicx}	
\usepackage{amsmath}	
\usepackage{amssymb}	
\usepackage{float}
\usepackage[caption = false]{subfig}
\usepackage{multirow}
\usepackage{rotating}
\usepackage{ulem}
\usepackage{booktabs}

\newcommand{\Oiii}{\mathrm{O\ III}}

\newcommand{\lsigG}{$L - \sigma$}
\newcommand{\lsigb}{$L(\mathrm{H}\beta) - \sigma$}

\newcommand{\mincir}{\raise-3.truept\hbox{\rlap{\hbox{$\sim$}}\raise4.truept\hbox{$<$}\ }}

\title[Independent Cosmological Constraints from HIIG]{Independent Cosmological Constraints from high-z HII~Galaxies  \thanks{The new data presented in this paper were obtained using MOSFIRE at the W. M. Keck Observatory, which is operated as a scientific partnership among the California Institute of Technology, the University of California and the National Aeronautics and Space Administration. The Observatory was made possible by the generous financial support of the W. M. Keck Foundation.}}
\author[Ana Luisa Gonz\'alez-Mor\'an et al.] {Ana Luisa Gonz\'alez-Mor\'an $^{1}$\thanks{Contact e-mail: analuisagm@inaoep.mx},
Ricardo Ch\'avez$^{1,2,3,4}$\thanks{Contact e-mail: r.chavez@irya.unam.mx},
Roberto Terlevich$^{1,5}$,  \newauthor
Elena Terlevich$^{1}$,
Fabio Bresolin$^{6}$,
David Fern\'andez-Arenas$^{7,1}$,
Manolis Plionis$^{8,9}$, \newauthor
Spyros Basilakos$^{8,10}$,
Jorge Melnick$^{11,12}$ and
Eduardo Telles$^{12}$
\\ \\
$^{1}$ Instituto Nacional de Astrof\'\i sica, \'Optica y Electr\'onica,Tonantzintla, C.P. 72840, Puebla, M\'exico \\
$^{2}$ Instituto de Radioastronom\'ia y Astrof\'isica, UNAM, Campus Morelia, C.P. 58089, Morelia, M\'exico\\
$^{3}$ Cavendish Laboratory, University of Cambridge, 19 J. J. Thomson Ave, Cambridge, CB3 0HE, UK\\
$^{4}$ Kavli Institute for Cosmology, University of Cambridge, Madingley Road, Cambridge, CB3 0HA, UK\\
$^{5}$ Institute of Astronomy, University of Cambrige, Cambridge, CB3 0HA, UK\\
$^{6}$ Institute for Astronomy, University of Hawaii, 2680 Woodlawn Drive, 96822 Honolulu,HI USA \\
$^{7}$ Kavli Institute for Astronomy and Astrophysics, Peking University, Beijing 100871, China \\
$^{8}$ National Observatory of Athens, P.Pendeli, Athens, Greece \\
$^{9}$ Physics Dept., Aristotle Univ. of Thessaloniki, Thessaloniki 54124, Greece \\
$^{10}$ Academy of Athens Research Center for Astronomy \& Applied Mathematics, Soranou Efessiou 4, 11-527 Athens, Greece \\
$^{11}$ European Southern Observatory, Santiago de Chile, Chile \\
$^{12}$ Observatorio Nacional, Rua Jos\'e Cristino 77, 20921-400 Rio de Janeiro, Brasil\\
}

\begin{document}

\date{v118 --- Compiled at \thistime\ hrs  on \today\ }

\pagerange{\pageref{firstpage}--\pageref{lastpage}} \pubyear{2016}

\maketitle

\label{firstpage}

\begin{abstract}
We present new high spectral resolution observations of 15 high-z ($1.3 \leq$ z $\leq 2.5$)  HII Galaxies (HIIG)  obtained with MOSFIRE at the Keck Observatory. These data, combined with already published data for another 31 high-z and 107 z  $\leq 0.15$ HIIG, are used to obtain new independent cosmological results using the distance estimator based on the established correlation between the Balmer emission line velocity dispersion and  luminosity for HIIG.
Our results are in excellent agreement with the latest cosmological concordance model ($\Lambda$CDM) published results.
From our analysis, we find a value for the mass density parameter of $\Omega_m=0.290^{+0.056}_{-0.069}$ (stat).
For a flat Universe we constrain the plane $\lbrace\Omega_m;w_0\rbrace = \lbrace 0.280^{+0.130}_{-0.100} ; -1.12^{+0.58}_{-0.32}\rbrace $  (stat). 
The joint likelihood analysis of HIIG with other complementary cosmic probes (Cosmic Microwave Background and Baryon Acoustic Oscillations) provides tighter constraints for the parameter space of the Equation of State of Dark Energy that are also in excellent agreement with those of similar  analyses using Type Ia Supernovae instead as the geometrical probe. 
\end{abstract}


\section{Introduction}

The first clear evidence for an accelerated cosmic expansion was given by the analysis of
Type Ia Supernovae (SNIa) data some 20 years ago \citep{Riess1998, Perlmutter1999}. Since then, analysis of the Cosmic Microwave Background (CMB) anisotropies \citep[e.g.][]{Jaffe2001, Pryke2002, Spergel2007, PlanckCollaboration2014, PlanckCollaboration2016a} and of Baryon Acoustic Oscillations (BAOs) \citep[e.g.][]{Eisenstein2005, Blake2011} in
combination with independent Hubble parameter measurements \citep[e.g.][]{Chavez2012, Freedman2012, Riess2016, Riess2018, Fernandez2018} have provided ample evidence for the presence of what has been called the Dark Energy (DE) component of the Universe.

A  main goal of present-day cosmology is to measure the DE Equation of State (EoS) and to explore its evolution (or lack of it) with look-back time  as well as searching for hints of any other property of DE.

The cosmological parameters $\Omega_{m}$, $\Omega_{\Lambda}$  and $w_0$ (plus $w_a$ in the evolutive case) are obtained combining the low redshift results (z $\lesssim 2$) for BAOs and SNIa (\citealt{Riess1998,Perlmutter1999,Hicken2009,Amanullah2010,Riess2011,Suzuki2012,Betoule2014,Scolnic2018}) with high redshift results (z $\sim$ 1000 Planck CMB fluctuations; e.g. \citealt{PlanckCollaboration2014, PlanckCollaboration2016a}). This is because the individual solutions are degenerate and  only combining them one can obtain competitive results.
It is important to remark that there are no determinations of cosmological parameters at intermediate redshift ($2 \lesssim$ z $\lesssim 10$), where the maximum difference in cosmological models that include an evolving DE EoS occurs \citep[cf.][]{Plionis2011}.

The \lsigG\ relation between the emission line velocity dispersion ($\sigma$) and the
Balmer line luminosity ($L[\mathrm{H}x]$, usually H$\beta$) of HII Galaxies  (HIIG) has
already been proven as a cosmological tracer
\citep[e.g.][and references therein]{Melnick2000, Siegel2005, Plionis2011,
Chavez2012, Chavez2014, Terlevich2015, Chavez2016}. It has also been shown that the
\lsigb\ relation has been used in the local Universe to significantly constrain the
value of $H_0$ \citep{Chavez2012,Fernandez2018}. 

HIIG are a very promising  tracer for the parameters of the DE EoS precisely because they can be observed, using the current available
infrared instrumentation, up to \mbox{z $\sim 4$} \citep[cf.][]{Terlevich2015, Chavez2016}. Even when their scatter
on the Hubble-Lem\^aitre diagram is about a factor of two larger than in the case of high-z SNIa, this disadvantage is
compensated by the fact that  HIIG  are observed to much larger redshifts than SNIa where, as mentioned above, the degeneracies for different DE models are largely reduced \citep[cf.][]{Plionis2011}.  \citet{Chavez2016} presented simulations that  predict substantial improvement in the constraints when increasing the sample of high-z HIIG  to 500, a goal that can be achieved in reasonable observing times with existing large telescopes and state-of-the-art instrumentation.

In this work, we use a new set of high spectral resolution observations of high-z HIIG obtained with MOSFIRE at the Keck telescope to constrain in an independent manner cosmological parameters on the important range of intermediate redshift $1.0 \lesssim$ z $\lesssim 3.0$.
We increased the observed sample with selected data from the literature and apply both  a standard $\chi ^2$  and a Markov Chain Monte Carlo (MCMC) methods to find the probability distribution of the solutions for a set of cosmological parameters and combine the results with those obtained using different probes (SNIa, BAOs, CMB). 

This paper is organized as follows, \S \ref{sec:Data} describes the data sample, our new data, its reduction process and analysis, and the extra data obtained from the literature. In  \S \ref{sec:Extinction correction} we discuss the extinction law applied to the data.

In \S \ref{sec:ResultsDiscussion} we present and discuss the results while the conclusions are given in \S \ref{sec:Conclusions}.  Serendipitous objects found in the 2D spectra, in Appendix \ref{sec:Others objects in the 2D spectra} and the reduced 2D and 1D Keck spectra are shown in Appendix \ref{sec:MOSFIRE spectra}.


\begin{table*}
	\small
	\centering
	\caption{Observed sample.}
	\label{tab:table1}
	\begin{tabular}{lcccccc} 
	\hline
Name	 & R.A. & DEC. & Field & PSF & Exp. time &  Ref. \\
& (J2000) & (J2000) & & (arcsec) &(seconds)&\\
		\hline
HDF-BX1376	&	12	36	52.962	&	62	15	45.70	& GOODS-N & 0.79\arcsec & 10,560 & 1,2	\\
HDF-BX1368	&	12	36	48.249	&	62	15	56.43	& &&	&1,2	\\
HDF-BX1409	&	12	36	47.408	&	62	17	28.77	& &	&&1,2	\\
HDF-BX305	&	12	36	37.105	&	62	16	28.71	& &	&&1,2	\\
HDF-BX1311	&	12	36	30.522	&	62	16	26.15	& &	&&1,2	\\
HDF-BX1388	&	12	36	44.832	&	62	17	15.88	& &	&&1,2	\\
3D-HST10245	&	12	36	23.280	&	62	15	28.00	& & &&	3	\\
UDS9			&	02	17	31.822	&	-05	14	40.70	& UDS	& 0.58\arcsec & 7,200 &4	\\
UDS23		&	02	17	32.882	&	-05	10	38.06	& & &&	4	\\
UDS25		&	02	17	36.514	&	-05	10	31.31	& & && 	4	\\
UDS40		&	02	17	45.146	&	-05	09	36.34	& & &&	4	\\
UDS-14655	&	02	17	33.926	&	-05	11	43.08	& & &&	5,6	\\
UDS-11484	&	02	17	43.536	&	-05	12	43.60	& & &&	5,6	\\
UDS-12435	&	02	17	38.607	&	-05	14	05.36	& & &&	6	\\
UDS-10138	&	02	17	41.606	&	-05	14	32.13	& & &&	6	\\
UDS-4501		&	02	17	33.783	&	-05	15	02.87	& & &&	5,6 	\\
UDS-109082	&	02	17	37.392	&	-05	13	07.88	& & &&	7 	\\
UDS-112374	&	02	17	32.544	&	-05	11	56.32	& & &&	7	\\
UDS-113972	&	02	17	40.353	&	-05	11	16.80	& & &&	7 \\
UDS-116891	&	02	17	39.784	&	-05	10	18.75	& & &&	7	\\
UDS-118417	&	02	17	35.301	&	-05	09	43.54	& & &&	7	\\
Lensed target2	&	02	17	37.237	&	-05	13	29.78	& & &&	8,9,10	\\
COSMOS-16207	&	10	00	43.937	&	02	22	22.53	& COSMOS&0.64\arcsec & 10,080 &	5,6	\\
COSMOS-13848	&	10	00	42.478	&	02	20	43.42	& & &&	5,6	\\
COSMOS-16566	&	10	00	40.960	&	02	20	53.86	& & &&	6	\\
COSMOS-18358	&	10	00	40.120	&	02	22	00.63	& & &&	6	\\
COSMOS-12807	&	10	00	38.295	&	02	19	59.92	& & &&	5,6	\\
COSMOS-15144	&	10	00	37.629	&	02	21	38.79	& & &&	5,6	\\
COSMOS-17118	&	10	00	36.277	&	02	21	14.89	& & &&	6	\\
COSMOS-19049	&	10	00	33.333	&	02	22	24.16	& & &&	6	\\
COSMOS13073	&	10	00	29.954	&	02	22	56.23	& & &&	11 	\\
COSMOS32915	&	10	00	35.087	&	02	22	27.30	& & &&	11 	\\
COSMOS133784	&	10	00	31.031	&	02	22	10.44	& & &&	12 	\\
COSMOS134172	&	10	00	26.463	&	02	22	26.84	& & &&	12 	\\
zCOSMOS-411737	&	10	00	32.356	&	02	21	21.00	& && &	13	\\
		\hline
		\multicolumn{7}{l}{}\\
		\multicolumn{7}{l}{References: (1): \cite{Erb2006a}; (2): \cite{Erb2006b}, (3): \cite{Lundgren2012}; (4): \cite{van_der_Wel2011};}\\
		\multicolumn{7}{l}{(5): \cite{Maseda2013};  (6): \cite{Maseda2014}; (7): \cite{Bruce2012}; (8): \cite{Brammer2012}; (9): \cite{Tu2009};}\\
		\multicolumn{7}{l}{(10): \cite{Cooray2011}; (11) \cite{Belli2014}; (12): \cite{Krogager2014}; (13): \cite{Mancini2011}.}\\
		\multicolumn{7}{l}{}\\
	\end{tabular}
\end{table*}

\section{Data sample}
\label{sec:Data}

\subsection{MOSFIRE}
\label{sec:MOSFIRE}

The new high spectral resolution near-IR spectra presented in this work were obtained using MOSFIRE \citep{McLean2010,McLean2012}, the multi-object spectrograph at the Cassegrain focus of the 10-m Keck I telescope in Mauna Kea, Hawaii, during the night of January 27th, 2016. 
Candidate objects were selected from the literature according to 3 conditions: 

i)  redshift ranges $1.2 <$ z $< 1.7$ and $1.9 <$ z $< 2.6$ in order to observe either H$\alpha$ or H$\beta$ and \mbox{[$\Oiii$]$\lambda$5007\AA} emission lines in the  H band; 

ii) high equivalent widths (EW) in their emission lines; and 

iii) candidate belong to a dense cosmological field in order to have at least 10 HIIG in the MOSFIRE field of view.

As in our group's previous related papers, the second condition was selected because stellar population synthesis models for star-forming galaxies in bursting episodes \citep{Leitherer1999}, predict that if the EW(H$\beta$) $>$ 50 \AA\ (EW(H$\alpha$) $>$ \mbox{200 \AA})  the sample is composed by systems in which a single starburst younger than 5 Myr dominates the total luminosity. At the same time, this condition minimizes the contamination by an older underlying stellar population.
To account for uncertainties in the measurements of EWs reported in the literature, we relaxed  somehow the conditions so that the published values for the candidates were 
\mbox{EW(H$\beta$) $>$ 25 \AA} or EW(H$\alpha$) $>$ \mbox{150 \AA}. 

Since the H$\beta$ line in these objects is in general weaker than the [$\Oiii$]$\lambda$5007\AA\ one, our selection criteria include also the [$\Oiii$] line.
If we assume that the EW[$\Oiii$]  behaves in the same way  in HIIG at high as at low redshifts, we can use the F[$\Oiii$]/F(H$\beta$) distribution of a sample of HIIG at low redshift in order to calculate statistically the distribution of the EW[$\Oiii$] at high redshift through the relation:
\begin{equation}
EW[\Oiii]=\frac{F[\Oiii]}{F(H\beta)}\times EW(H\beta),
\label{eq:EWOIII distribution}
\end{equation}

Using the result from \citet{Chavez2014} for a sample of 95 HIIG at low redshift having both \mbox{F[$\Oiii$]} and F(H$\beta$) data, we adopted the selection criterion \mbox{EW[$\Oiii$] $\gtrsim $ 400 \AA} for our new high redshift sample. 

Following these criteria, we  selected from \citet{Erb2006a, Erb2006b,Tu2009, Mancini2011,Cooray2011,van_der_Wel2011, Bruce2012, Brammer2012, Lundgren2012, Maseda2013, Maseda2014} a sample of 35 candidates for our MOSFIRE run in the GOODS North \citep[GOODS-N;][]{Giavalisco2004}, the Cosmic Evolution Survey \citep[COSMOS;][]{Scoville2007, Koekemoer2007} and the Ultra Deep Survey \citep[UDS;][]{Lawrence2007, Cirasuolo2007} fields. 
The observed sample is detailed in Table \ref{tab:table1}. The target name is given in the first column; the coordinates in the second and third column; the cosmological field that each object belongs to in column 4; the seeing in arcseconds during the  observations of each field in column 5; the total exposure time per field in seconds in column 6 and in column 7 the literature reference for  the sample objects.

\begin{figure*}
\begin{center}
\includegraphics[width=1.3\columnwidth]{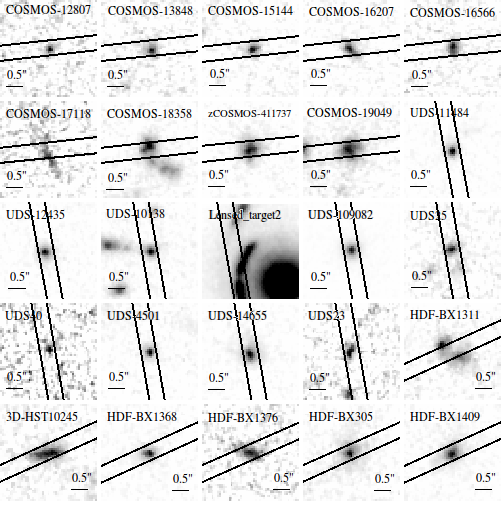}
\end{center}
\caption[Sample Image.]{\footnotesize Target images in the f160w filter (H band) from CANDELS HST Program. Superposed are the slits with the orientation and width  set for the MOSFIRE observations. The slit width of 0.48\arcsec\ represents (for a $\lambda$CDM cosmology and $H_0$=71 km/s Mpc$^{-1}$) a physical scale of $\sim$ 4 kpc.}
\label{sample image}
\end{figure*}

\subsubsection{Observations and Data Reduction}
\label{sec:Observations and Data Reduction}
MOSFIRE has a cryogenic configurable slit unit (CSU), consisting of 46 pairs of bars each one of length 7.1\arcsec. The bars can be configured in the direction perpendicular to the slits anywhere within an effective field of view of 3\arcmin x 6\arcmin.

We designed our masks using  MAGMA\footnote{MAGMA (MOSFIRE Automatic GUI-based Mask Application) is a tool that is used to design new slitmasks for use with MOSFIRE and is available from the website http://www2.keck.hawaii.edu/inst/mosfire/magma.html.}. The wavelength coverage of the spectra depends on the position of the slit in the CSU, thus it differs slightly from target to target. Therefore, the standard star must be observed in two wavelength settings (POS A and POS C) to guarantee the full coverage of the wavelength range of the science targets.

The high  resolution spectra (R = 5,340 in the H band) were obtained using a 0.48\arcsec slit width. Total exposure times per cosmological field ranged between 2 and 3 hrs.
For each run we  measured the point spread function (PSF) from Gaussian fits to the spatial profile of stars observed simultaneously in the three fields. The resulting FWHM of the PSF ranges between 0.5\arcsec and 0.8\arcsec measured in the H band (see Table \ref{tab:table1}).

To identify each target we used the images from the Cosmic Assembly Near-IR Deep Extragalactic Legacy Survey (CANDELS), Multi-Cycle Treasury Program with the NASA/ESA HST \citep{Grogin2011,Koekemoer2011} in the filter f160w corresponding to the H band. The slit orientations were 96, 10  and 114 degrees for the COSMOS, UDS and GOODS-N fields, respectively. Since all slit widths were 0.48\arcsec, the compactness of the sources can be appreciated in Figure \ref{sample image}. For $\Lambda$CDM cosmology and $H_0$ = 71 km/s Mpc$^{-1}$, the scale at \mbox{z = 1.55} is 8.3 kpc/arcsec and at \mbox{z = 2.3} is 8.1 kpc/arcsec. Thus the 0.48\arcsec slit represents a metric scale of $\sim$ 4 kpc.

The data reduction was carried out using the MOSFIRE Data Reduction Pipeline (DRP) developed by the MOSFIRE team\footnote{http://www2.keck.hawaii.edu/inst/mosfire/drp.html.}. The MOSFIRE DRP produces flat-field corrected, background subtracted, wavelength calibrated, and rectified 2D spectra for each slit on a given mask. The 2D wavelength solutions in the H band were obtained from the night sky OH emission lines for each slit.

The final product of the MOSFIRE DRP is wavelength calibrated 2D spectra. The one-dimensional (1D) spectra and the flux calibration were obtained using IRAF\footnote{IRAF (Image Reduction and Analysis Facility) is a software for the reduction and analysis of astronomical data. It is written and supported by the National Optical Astronomy Observatories (NOAO).}. For the flux calibration, we need to select either POS A or POS C of the reference star spectra.  So, depending on the redshift and slit position of  each  target  on the field of view, the wavelength coverage of the spectra differs. 

The wavelength range of POS A is from 15,200\AA\ to 17,730\AA\ for the UDS field and from 15,160\AA\ to 17,920\AA\ for the GOODS-N field while  POS C goes from 15,000\AA\ to 17,350\AA\ for all the  fields.

\subsubsection{Redshifts}
Emission lines were detected  in  25 of the 35  candidates. The sample of 25 HIIG contains 11 objects at z $\sim$ 2.3 and 14 at z $\sim$ 1.5.

To determine the redshift and width of the emission lines we fitted  H$\alpha$ for the objects at z $\sim$ 1.5 and \mbox{[O III]$\lambda\lambda$4959, 5007 \AA} and H$\beta$ for the objects at z $\sim$ 2.3. An example of the Gaussian fitting to the H$\alpha$ line for the target COSMOS-16566 is shown in Figure \ref{fig:montecarlo analysis}.

For the objects at z $\sim$ 2.3 the redshifts were measured using at least 2 emission lines. For those which have only one emission line (objects at z $\sim$ 1.5) we compared our values with those from the literature that have been measured using other lines. 
Three objects in the UDS field (UDS23, UDS25 and UDS40) were selected from their photometric redshift indicating  [O III] emission in the redshift range 1.6 $<$ z $<$ 1.8
\citep{van_der_Wel2011}. Our MOSFIRE/Keck data shows a strong line detection that we identified as H$\alpha$ at a redshift  z=1.6113 for UDS40, z=1.6658 for UDS25  and z=1.6001 for UDS23  thus confirming the photometric estimate.

As a result, 24 of the 25 measured redshifts coincide with those reported in the literature with small variations of $\pm$ 0.001. For the remaining object, UDS-109082, we found variations of +0.23 and -0.05 with respect to the published redshift; combining this difference in redshift with the fact that in our spectrum only one emission line is present, we assume that the observed line is H$\alpha$ at redshift 1.6814.

The redshifts measured for each line are shown in columns 2, 3, 4 and 5 of Table \ref{tab:redshift} and the 2D and 1D spectra and the Gaussian fits to the emission lines are shown in Apendix \ref{sec:MOSFIRE spectra}.

\subsubsection{Velocity Dispersions}

\begin{figure}
\hspace*{-0.5cm}
	\includegraphics[width=1.\columnwidth]{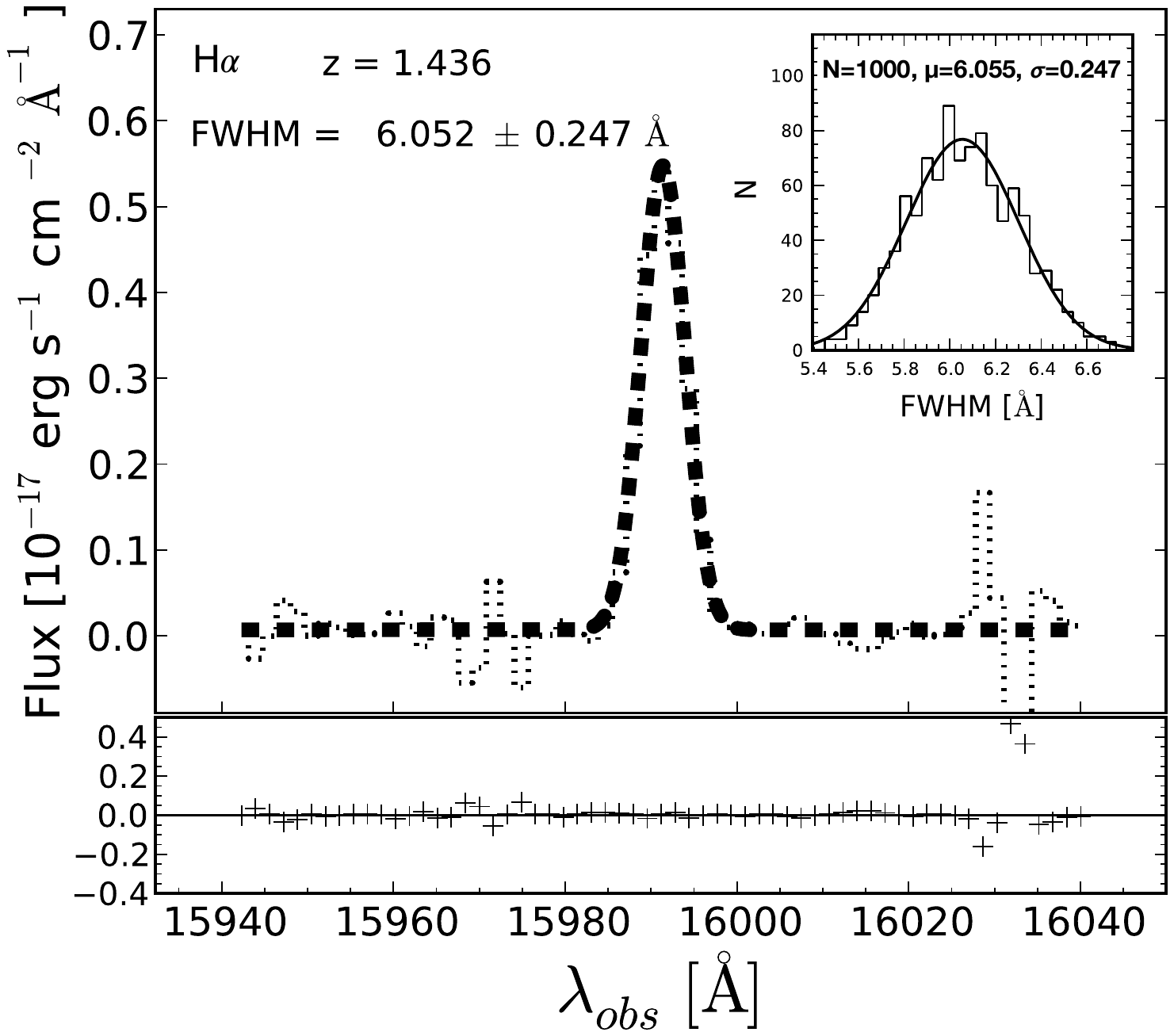}
    \caption{H$\alpha$ line for  COSMOS-16566. The dashed grey line is the spectrum, the dashed dark line is the Gaussian fit to the emission line and the box underneath shows the residuals. The inset at the upper right corner is the Montecarlo analysis performed to the line where the standard deviation of the resulting distribution is taken as the uncertainty of the measured FWHM of the Gaussian fit.}
    \label{fig:montecarlo analysis}
\end{figure}

We have measured the velocity dispersion  ($\sigma$) of the ionized gas from Gaussian fits to the emission lines. The $\sigma$ of the observed profile being,

\begin{equation}
	\sigma_{obs}=\frac{FWHM}{2\sqrt{2ln2}},
	\label{eq:sigma_obs}
\end{equation}
where FWHM is the full width at half-maximum of the emission line. The uncertainties were estimated using a Montecarlo analysis where a set of random realizations of each spectrum was generated using the r.m.s. intensity of the continuum adjacent to the emission line (see Figure \ref{fig:montecarlo analysis}). The FWHM 1$\sigma$ uncertainty was  estimated from the standard deviation of the distribution of FWHM measurements. The fit to the observed lines and the distribution of FWHM obtained from the Montecarlo simulations are also shown in figures \ref{1D and 2D spectra of Halpha} and \ref{1D and 2D spectra of O III}.

The sigma of the observed profile for each line can be broken down into four components:

\begin{equation}
\sigma^2_{obs} = \sigma^2 + \sigma_{th}^2 + \sigma_i^2 + \sigma_{fs}^2,
\label{eq:sigma}
\end{equation}
where $\sigma_{th}$, $\sigma_i$ and $\sigma_{fs}$ are the thermal, instrumental and fine structure broadening components, respectively, and $\sigma $ is the intrinsic velocity dispersion.  The thermal broadening can be calculated from the expression:
\begin{equation}
\sigma_{th}=\sqrt{\frac{kT_e}{m}},
\end{equation}
where $k$ is  the Boltzmann constant, $m$ is the mass of the ion in question and $T_e$ is the electron temperature in  degrees Kelvin, for which a reasonable value is $T_e$ = 10,000 K.

For the fine structure width, which is important for hydrogen recombination lines but not for metal lines such as [$\Oiii$], we adopted a value of \mbox{$\sigma_{fs}^2$(H$\alpha$) = 10.233 $\mathrm{km^2\ s^{-2} }$} or $\sigma_{fs}^2$(H$\beta$) = 5.767 $\mathrm{km^2\ s^{-2} }$ \citep{Garcia-Dias2008}. Finally, the instrumental broadening is measured from the sky lines width giving a value of \mbox{$\sigma_i$ = 22.68 $\mathrm{km\ s^{-1}}$}.

Summarizing, we determine the one-dimensional (1D) velocity dispersion by fitting a Gaussian profile
 to each emission line and from those values subtracting in quadrature  the thermal, instrumental and fine structure broadening using Equation \ref{eq:sigma}. The resulting values of the velocity dispersion for each emission line are given  in km s$^{-1}$ in columns 3, 4, 5 and 6 of Table \ref{tab:redshift,sigma and fluxes}.

\begin{table*}
	\small
	\centering
	\caption{Measured redshifts and EWs}
	\label{tab:redshift}
		\begin{tabular}{lcccccccc} 
			\hline
Name	 & z[O III] & z[O III] & z(H$\beta$) & z(H$\alpha$) & EW[O III] & EW[O III] & EW(H$\beta$)&EW(H$\alpha$)\\
& $\lambda$5007\AA & $\lambda$4959\AA & & & $\lambda$5007\AA & $\lambda$4959\AA & &\\
		\hline
HDF-BX1376	&	2.4293	&	2.4291	&	2.4297	&	$\ldots$	&	218$\pm49$	&	88$\pm13$	&	77$\pm24$	&	$\ldots$	\\
HDF-BX1368	&	2.4412	&	2.4412	&	2.4403	&	$\ldots$	&	342$\pm45$	&	166$\pm37$	&	67$\pm14$	&	$\ldots$	\\
HDF-BX1409	&	2.2457	&	2.2460	&	2.2463	&	$\ldots$	&	236$\pm54$	&	37$\pm9$	&	45$\pm9$	&	$\ldots$	\\
HDF-BX305	&	2.4833	&	2.4832	&	2.4843	&	$\ldots$	&	135$\pm12$	&	76$\pm17$	&	45$\pm6$	&	$\ldots$	\\
HDF-BX1311	&	2.4840	&	2.4840	&	2.4845	&	$\ldots$	&	285$\pm43$	&	154$\pm50$	&	63$\pm14$	&	$\ldots$	\\
3D-HST10245	&	$\ldots$	&	$\ldots$	&	$\ldots$	&	1.3592	&	$\ldots$	&	$\ldots$	&	$\ldots$	&	172$\pm48$	\\
UDS40	&	$\ldots$	&	$\ldots$	&	$\ldots$	&	1.6113	&	$\ldots$	&	$\ldots$	&	$\ldots$	&	$\ldots$	\\
UDS25	&	$\ldots$	&	$\ldots$	&	$\ldots$	&	1.6658	&	$\ldots$	&	$\ldots$	&	$\ldots$	&	243$\pm102$	\\
UDS23	&	$\ldots$	&	$\ldots$	&	$\ldots$	&	1.6001	&	$\ldots$	&	$\ldots$	&	$\ldots$	&	194$\pm140$	\\
UDS-14655	&	2.2983	&	2.2988	&	2.2989	&	$\ldots$	&	443$\pm115$	&	124$\pm49$	&	61$\pm9$	&	$\ldots$	\\
UDS-11484	&	2.1855	&	2.1856	&	2.1858	&	$\ldots$	&	672$\pm118$	&	109$\pm28$	&	67$\pm32$	&	$\ldots$	\\
UDS-109082	&	$\ldots$	&	$\ldots$	&	$\ldots$	&	1.6814	&	$\ldots$	&	$\ldots$	&	$\ldots$	&	249$\pm141$	\\
Lensed target2	&	2.2868	&	2.2867	&	2.2874	&	$\ldots$	&	72$\pm13$	&	30$\pm5$	&	27$\pm12$	&	$\ldots$	\\
UDS-12435	&	$\ldots$	&	$\ldots$	&	$\ldots$	&	1.6114	&	$\ldots$	&	$\ldots$	&	$\ldots$	&	224$\pm67$	\\
UDS-10138	&	2.1510	&	2.1511	&	2.1511	&	$\ldots$	&	411$\pm161$	&	97$\pm26$	&	63$\pm18$	&	$\ldots$	\\
UDS-4501	&	2.2973	&	2.2973	&	$\ldots$	&	$\ldots$	&	576$\pm243$	&	106$\pm20$	&	$\ldots$	&	$\ldots$	\\
COSMOS-16207	&	$\ldots$	&	$\ldots$	&	$\ldots$	&	1.6482	&	$\ldots$	&	$\ldots$	&	$\ldots$	&	256$\pm52$	\\
COSMOS-13848	&	$\ldots$	&	$\ldots$	&	$\ldots$	&	1.4433	&	$\ldots$	&	$\ldots$	&	$\ldots$	&	213$\pm122$	\\
COSMOS-16566	&	$\ldots$	&	$\ldots$	&	$\ldots$	&	1.4366	&	$\ldots$	&	$\ldots$	&	$\ldots$	&	248$\pm89$	\\
COSMOS-18358	&	$\ldots$	&	$\ldots$	&	$\ldots$	&	1.6491	&	$\ldots$	&	$\ldots$	&	$\ldots$	&	166$\pm24$	\\
COSMOS-12807	&	$\ldots$	&	$\ldots$	&	$\ldots$	&	1.5820	&	$\ldots$	&	$\ldots$	&	$\ldots$	&	230$\pm90$	\\
COSMOS-15144	&	$\ldots$	&	$\ldots$	&	$\ldots$	&	1.4120	&	$\ldots$	&	$\ldots$	&	$\ldots$	&	248$\pm72$	\\
COSMOS-17118	&	$\ldots$	&	$\ldots$	&	$\ldots$	&	1.6554	&	$\ldots$	&	$\ldots$	&	$\ldots$	&	244$\pm106$	\\
COSMOS-19049	&	$\ldots$	&	$\ldots$	&	$\ldots$	&	1.3688	&	$\ldots$	&	$\ldots$	&	$\ldots$	&	169$\pm32$	\\
zCOSMOS-411737	&	2.4451	&	2.4453	&	2.4454	&	$\ldots$	&	334$\pm48$	&	69$\pm10$	&	67$\pm7$	&	$\ldots$	\\
		\hline
		\multicolumn{9}{l}{Redshifts measured using Gaussian fits with the PYTHON routine mpfit (https://www.scipy.org/citing.html)}\\
		\multicolumn{9}{l}{Equivalent widths measured using the IRAF task splot.}\\
		\end{tabular}
\end{table*}

\subsubsection{Fluxes}

The star observed (HD225023) is a photometric (not  spectrophotometric) standard. In order to calibrate in flux we used its observed spectrum combined with a black-body curve corresponding to its spectral type (A0 star with a magnitude H=6.985 in the CIT/CTIO system by \citet{Elias1982}.
We applied the extinction curve for Mauna Kea, Hawaii, in the bands J, H, K', Ks and K, respectively for \mbox{2.2 mm} of precipitable water vapour.

To verify our calibration, we applied it to the observed star spectrum. We obtained for POS A an 
H band  flux at effective wavelength \mbox{1.65 $\mu$m}  of \mbox{F$_{1.65\mu m} = 1.846\times10^{-13}$ erg s$^{-1}$ cm$^{-2}$ \AA$^{-1}$} for the UDS field, \mbox{F$_{1.65\mu m} = 1.829\times10^{-13}$ erg s$^{-1}$ cm$^{-2}$ \AA$^{-1}$} for the COSMOS field, \mbox{F$_{1.65\mu m} = 1.824\times10^{-13}$ erg s$^{-1}$ cm$^{-2}$ \AA$^{-1}$} for the GOODS-N field. For the POS C corresponding to the COSMOS, UDS and GOODS-N fields we obtained  \mbox{F$_{1.65\mu m} = 1.869\times10^{-13}$ erg s$^{-1}$ cm$^{-2}$ \AA$^{-1}$}.

We converted the m$_{\textsc{H}}$=6.985 to flux using the equation:\\
\begin{equation}
	F_\nu = F_010^{-6.985/2.5},
	\label{eq:F_nu}
\end{equation}
where F$_0=980$ Jy corresponds to the CIT/CTIO  system. Transforming F$_\nu$ to F$_\lambda$ using the equation:\\
\begin{equation}
	 F_\lambda=F_\nu\frac{c}{\lambda^2},
	\label{eq:F_lambda}
\end{equation}
with an effective wavelength of \mbox{1.65 $\mu m$}, we obtained \mbox{F$_{1.65\mu m}=1.730\times10^{-13}$ erg s$^{-1}$ cm$^{-2}$ \AA$^{-1}$}. We conclude that the science targets calibrated with the standard star observed in the POS A for the UDS, COSMOS and GOODS-N fields have a flux uncertainty due to the calibration method of 7\%, 6\% and
 5\%, respectively. The science targets calibrated with the standard star observed in the POS C for the 3 cosmological fields have a flux uncertainty due to the calibration method of 8\% with respect to their total flux.

The emission line fluxes were measured using the IRAF task \textit{splot} and their uncertainties were estimated from the expression:
\begin{equation}
	\epsilon_{Flux} = \sqrt{\epsilon^2_F + \epsilon^2_{cal F}},
	\label{eq:sigma_Flux}
\end{equation}
where $\epsilon_{cal F}$ is the uncertainty due to the flux calibration process and $\epsilon_{F}$ was calculated from the expression \citep{Tresse1999}:
\begin{equation}
	\epsilon_{F}=\epsilon_cD\sqrt{2N_{pix} + EW/D},
	\label{eq:sigma_F}
\end{equation}
where $\epsilon_c$ is the mean standard deviation per pixel of the continuum at each side of the line, D is the spectral dispersion in \AA/pixel, N$_{pix}$ is the number of pixels covered by the line and EW is the rest equivalent width in \AA\ estimated also by using the IRAF task \textit{splot}. The EW uncertainties were estimated as \citep{Tresse1999}:
\begin{equation}
	\epsilon_{EW} = \frac{EW}{F}\epsilon_cD\sqrt{EW/D + 2N_{pix} + (EW/D)^2/N_{pix}},
	\label{eq:sigma_F}
\end{equation}
where F is the flux in ergs s$^{-1}$ cm$^{-2}$.

The EWs are shown in columns 6, 7 ,8 and 9 of Table \ref{tab:redshift} and the 
measured fluxes  are listed in columns 7, 8, 9 and 10 of Table \ref{tab:redshift,sigma and fluxes}.
Our measurements are in good agreement with three of the four objects with published values 
 (see Table \ref{tab:Fluxes comparison}). For reasons discussed in Section \ref{sec:Objects no considered in the analysis} and in Apendix \ref{sec:Others objects in the 2D spectra}, the discordant object has been excluded from the analysis, as indicated in Table \ref{tab:redshift,sigma and fluxes}.

\begin{table*}
	\centering
	\caption{Redshift, velocity dispersions and fluxes of the relevant lines.}
	\label{tab:redshift,sigma and fluxes}
		\resizebox{\textwidth}{!}{
		\begin{tabular}{lccccccccc} 
			\hline
Name	 & z$^a$ & $\sigma$[O III]$^{b}$ & $\sigma$[O III]$^{b}$ & $\sigma$(H$\beta)^b$ & $\sigma$(H$\alpha)^b$ & F[O III]$^{c}_{\textit{obs}}$ & F[O III]$^{c}_{\textit{obs}}$ &	F(H$\beta)^c_{\textit{obs}}$ & F(H$\alpha)^c_{\textit{obs}}$\\
& & $\lambda5007$\AA\ & $\lambda4959$\AA\ &  &  &  $\lambda5007$\AA & $\lambda4959$\AA\  &  & \\
		\hline
${}^\dagger$HDF-BX1376	&	2.4294	&	58.1$\pm3.5$	&	51.2$\pm4.9$	&	43.6$\pm11.9$	&	$\ldots$	&	6.48$\pm0.60$	&	2.03$\pm0.19$	&	1.33$\pm0.23$	&	$\ldots$	\\
HDF-BX1368	&	2.4409	&	122.3$\pm4.1$	&	124.1$\pm8.0$	&	42.6$\pm4.1$	&	$\ldots$	&	13.83$\pm0.88$	&	5.44$\pm0.55$	&	2.10$\pm0.24$	&	$\ldots$	\\
HDF-BX1409	&	2.2460	&	102.7$\pm7.7$	&	62.7$\pm7.0$	&	145.6$\pm19.3$	&	$\ldots$	&	5.46$\pm0.64$	&	0.96$\pm0.17$	&	1.30$\pm0.21$	&	$\ldots$	\\
HDF-BX305	&	2.4836	&	76.9$\pm3.1$	&	109.0$\pm23.7$	&	125.0$\pm16.9$	&	$\ldots$	&	6.21$\pm0.45$	&	2.39$\pm0.36$	&	1.95$\pm0.22$	&	$\ldots$	\\
HDF-BX1311	&	2.4842	&	97.4$\pm4.0$	&	85.3$\pm16.4$	&	147.0$\pm24.0$	&	$\ldots$	&	9.04$\pm0.95$	&	2.99$\pm0.46$	&	2.08$\pm0.32$	&	$\ldots$	\\
${}^\dagger$3D-HST10245	&	1.3592	&	$\ldots$	&	$\ldots$	&	$\ldots$	&	73.9$\pm6.4$	&	$\ldots$	&	$\ldots$	&	$\ldots$	&	4.03$\pm0.50$	\\
${}^\dagger$UDS40	&	1.6113	&	$\ldots$	&	$\ldots$	&	$\ldots$	&	58.8$\pm10.2$	&	$\ldots$	&	$\ldots$	&	$\ldots$	&	1.50$\pm0.23$	\\
${}^\dagger$UDS25	&	1.6658	&	$\ldots$	&	$\ldots$	&	$\ldots$	&	56.7$\pm3.9$	&	$\ldots$	&	$\ldots$	&	$\ldots$	&	4.77$\pm0.74$	\\
${}^\dagger$UDS23	&	1.6001	&	$\ldots$	&	$\ldots$	&	$\ldots$	&	44.7$\pm5.3$	&	$\ldots$	&	$\ldots$	&	$\ldots$	&	1.14$\pm0.26$	\\
UDS-14655	&	2.2986	&	62.4$\pm2.2$	&	51.6$\pm3.5$	&	33.7$\pm6.6$	&	$\ldots$	&	11.32$\pm1.16$	&	3.07$\pm0.48$	&	1.50$\pm0.17$	&	$\ldots$	\\
${}^\dagger$UDS-11484	&	2.1856	&	62.5$\pm1.5$	&	75.4$\pm8.4$	&	67.7$\pm17.2$	&	$\ldots$	&	18.83$\pm1.54$	&	6.06$\pm0.81$	&	2.08$\pm0.54$	&	$\ldots$	\\
${}^\dagger$UDS-109082	&	1.6814	&	$\ldots$	&	$\ldots$	&	$\ldots$	&	58.2$\pm10.7$	&	$\ldots$	&	$\ldots$	&	$\ldots$	&	2.58$\pm0.52$	\\
Lensed target2	&	2.2869	&	58.7$\pm5.3$	&	41.6$\pm18.1$	&	52.3$\pm10.6$	&	$\ldots$	&	3.47$\pm0.42$	&	1.41$\pm0.19$	&	0.61$\pm0.20$	&	$\ldots$	\\
${}^\dagger$UDS-12435	&	1.6114	&	$\ldots$	&	$\ldots$	&	$\ldots$	&	67.5$\pm5.5$	&	$\ldots$	&	$\ldots$	&	$\ldots$	&	4.87$\pm0.63$	\\
UDS-10138	&	2.1511	&	68.0$\pm3.8$	&	71.8$\pm8.1$	&	84.5$\pm12.6$	&	$\ldots$	&	11.41$\pm1.56$	&	4.49$\pm0.69$	&	2.25$\pm0.39$	&	$\ldots$	\\
UDS-4501	&	2.2973	&	77.4$\pm3.2$	&	73.3$\pm7.5$	&	$\ldots$	&	$\ldots$	&	9.89$\pm1.18$	&	4.44$\pm0.53$	&	1.15$\pm0.33$	&	$\ldots$	\\
${}^\dagger$COSMOS-16207	&	1.6482	&	$\ldots$	&	$\ldots$	&	$\ldots$	&	42.4$\pm1.5$	&	$\ldots$	&	$\ldots$	&	$\ldots$	&	6.87$\pm0.61$	\\
${}^\dagger$COSMOS-13848	&	1.4433	&	$\ldots$	&	$\ldots$	&	$\ldots$	&	28.0$\pm5.4$	&	$\ldots$	&	$\ldots$	&	$\ldots$	&	1.28$\pm0.21$	\\
${}^\dagger$COSMOS-16566	&	1.4366	&	$\ldots$	&	$\ldots$	&	$\ldots$	&	41.4$\pm2.3$	&	$\ldots$	&	$\ldots$	&	$\ldots$	&	3.65$\pm0.41$	\\
${}^\dagger$COSMOS-18358	&	1.6491	&	$\ldots$	&	$\ldots$	&	$\ldots$	&	66.2$\pm4.0$	&	$\ldots$	&	$\ldots$	&	$\ldots$	&	10.99$\pm0.97$	\\
${}^\dagger$COSMOS-12807	&	1.5820	&	$\ldots$	&	$\ldots$	&	$\ldots$	&	53.3$\pm4.8$	&	$\ldots$	&	$\ldots$	&	$\ldots$	&	1.88$\pm0.23$	\\
${}^\dagger$COSMOS-15144	&	1.4120	&	$\ldots$	&	$\ldots$	&	$\ldots$	&	45.2$\pm2.2$	&	$\ldots$	&	$\ldots$	&	$\ldots$	&	6.94$\pm0.71$	\\
${}^\dagger$COSMOS-17118	&	1.6554	&	$\ldots$	&	$\ldots$	&	$\ldots$	&	31.8$\pm3.3$	&	$\ldots$	&	$\ldots$	&	$\ldots$	&	3.56$\pm0.44$	\\
COSMOS-19049	&	1.3688	&	$\ldots$	&	$\ldots$	&	$\ldots$	&	117.9$\pm10.1$	&	$\ldots$	&	$\ldots$	&	$\ldots$	&	6.11$\pm0.63$	\\
zCOSMOS-411737	&	2.4453	&	81.2$\pm2.6$	&	72.9$\pm6.4$	&	75.6$\pm5.4$	&	$\ldots$	&	10.53$\pm0.76$	&	2.92$\pm0.28$	&	2.24$\pm0.20$	&	$\ldots$	\\
			\hline
			\multicolumn{10}{l}{Flux in 10$^{-17}$ erg s$^{-1}$ cm$^{-2}$ and velocity dispersion in km/s.}\\
			\multicolumn{10}{l}{Typical uncertainties in redshift are $\sim$ 10$^{-4}$.}\\			
			\multicolumn{10}{l}{$^a$ Mean redshift.}\\
			\multicolumn{10}{l}{$^b$ Gaussian fits done using the PYTHON routine mpfit.}\\
			\multicolumn{10}{l}{$^c$ Measurements done using the IRAF task splot.}\\
			\multicolumn{10}{l}{${}^\dagger$ HIIG selected for the cosmological analysis.}\\
		\end{tabular}}
\end{table*}

\begin{table}
	\centering
	\caption{Observed [O III] fluxes; This work vs. literature.}
	\label{tab:Fluxes comparison}
	\begin{tabular}{lcc} 
		\hline
Name	 & Flux[O III]$\lambda 5007^a$ & Flux[O III]$\lambda 5007^b$ \\
& this work & literature\\
		\hline
UDS-14655	&	11.3$\pm1.2$	&	21.9$\pm3.1$	\\
UDS-11484	&	18.8$\pm1.5$	&	21.5$\pm2.8$	\\
UDS-10138	&	11.4$\pm1.6$	&	10.9$\pm0.6$	\\
UDS-4501	        &	9.9$\pm1.2$	&	13.7$\pm2.8$	\\
		\hline
		\multicolumn{3}{l}{Flux in units of 10$^{-17}$ erg s$^{-1}$ cm$^{-2}$.}\\
		\multicolumn{3}{l}{$^a$ Measured  using the IRAF task splot.}\\
		\multicolumn{3}{l}{$^b$ Fluxes taken from \cite{Maseda2013,Maseda2014}.}\\
	\end{tabular}
\end{table}
\subsection{Excluded objects}
\label{sec:Objects no considered in the analysis}
Following \citet{Chavez2014} and \citet{Terlevich2015} we have selected only  those HIIG which have high equivalent width in their emission lines and a logarithmic velocity dispersion, $\log \sigma + {\rm error} \leq 1.84$.

We have excluded two additional objects, HDF-BX1368 and  UDS-14655 on the grounds of
having  H$\beta$ contaminated by a sky emission line. In principle it would be possible to use $\sigma$[O III] instead of $\sigma$(H$\beta$) as in \citet{Terlevich2015}. However, due to the sky contamination it is not possible to accurately determine the H$\beta$ fluxes to combine with the $\sigma$[O III] \citep[cf][]{Melnick2017}.
A third excluded object, Lensed target2,  belongs to the UDS-01 Lensing system \citep[SL2SJ02176-0513;][]{Tu2009, Cooray2011, Brammer2012}, which is an uncommon quite complex lens system showing two multiply-imaged systems at different redshifts lensed by a foreground massive galaxy at z$_{lens}$ = 0.656: a bright cusp arc at z$_{arc}$ = 1.847 and an additional double-image system at z$_{dbl}$ = 2.29 estimated by \citet{Brammer2012}. It is interesting that we detect the HeI$\lambda$5876 \AA\ emission line  confirming the reported redshift of the bright cusp arc of z = 1.847 (see Appendix \ref{sec:Others objects in the 2D spectra}) and  H$\beta$ and [$\Oiii$]$\lambda\lambda$4959,5007 \AA\ which confirm the reported redshift of the double-image system of z = 2.29. We labelled this galaxy as Lensed target 2 since it is the second gravitationally lensed object in the lensing system (see Table \ref{tab:redshift,sigma and fluxes} and Figure \ref{sample image}). As a confirmed complex gravitational lens we decided to remove it from the analysis.

All this reduces the MOSFIRE  sample to the 15 objects indicated with a dagger symbol in Table \ref{tab:redshift,sigma and fluxes}.

\subsection{Literature sample}
\label{sec:Literature sample}

In order to increase the sample we have added published HIIG data  following the same selection criteria. 

a) At low z ($0.01 \leq$ z $\leq 0.15$) 107  HIIG  extensively analysed in \citet{Chavez2014}.

b) At intermediate and high-z ($0.6 \leq$ z $\leq 2.5$) we included 25 HIIG from \citet{Erb2006a}, \citet{Masters2014} and \citet{Maseda2014} plus our 6 HIIG observed with VLT-XShooter publised in \citet{Terlevich2015}.\\

The total number of HIIG used for the remainder of this paper is  thereby 153.

\section{Extinction correction}
\label{sec:Extinction correction}
The extinction correction was performed using two different laws: the one by \citet[][]{Calzetti2000}, which has been widely used for massive starburst galaxies; and that of \citet[][]{Gordon2003}, that corresponds to the Large Magellanic Cloud (LMC) supershell near the prototypical Giant HII Region (GHIIR) 30 Doradus in the LMC.

The extinction corrected fluxes were determined, as is the usual practice, from the expression:
\begin{equation}
F(\lambda) = F_{obs}(\lambda)10^{0.4\textsc{Av}k(\lambda)/\textsc{Rv}},
\label{eq:flux correction}
\end{equation}
where $k(\lambda) = A(\lambda)/E(B - V )$ is given by the extinction law used. We adopt $k(H\beta) = 3.33$ and $k(H\alpha) = 2.22$ or $k(H\beta) = 4.60$ and $k(H\alpha) = 3.32$ for \citet{Gordon2003} and \citet{Calzetti2000}, respectively.

Given the redshifts of the objects observed with MOSFIRE, and as we only have H band data, we cannot measure the Balmer decrement directly. Instead the extinction (Av) was derived from the published \mbox{E(B - V)} whenever available, using the value of Rv = 2.77 given by \citet{Gordon2003} or \mbox{Rv = 4.05} given by \citet{Calzetti2000} extinction curves. For those objects where the reddening was not available the mean \mbox{Av/Rv = 0.2208} or 0.1454 from our local sample was adopted for \citet{Gordon2003} or \citet{Calzetti2000}, respectively.

\begin{figure}
\centering
	\includegraphics[width=1.0\columnwidth]{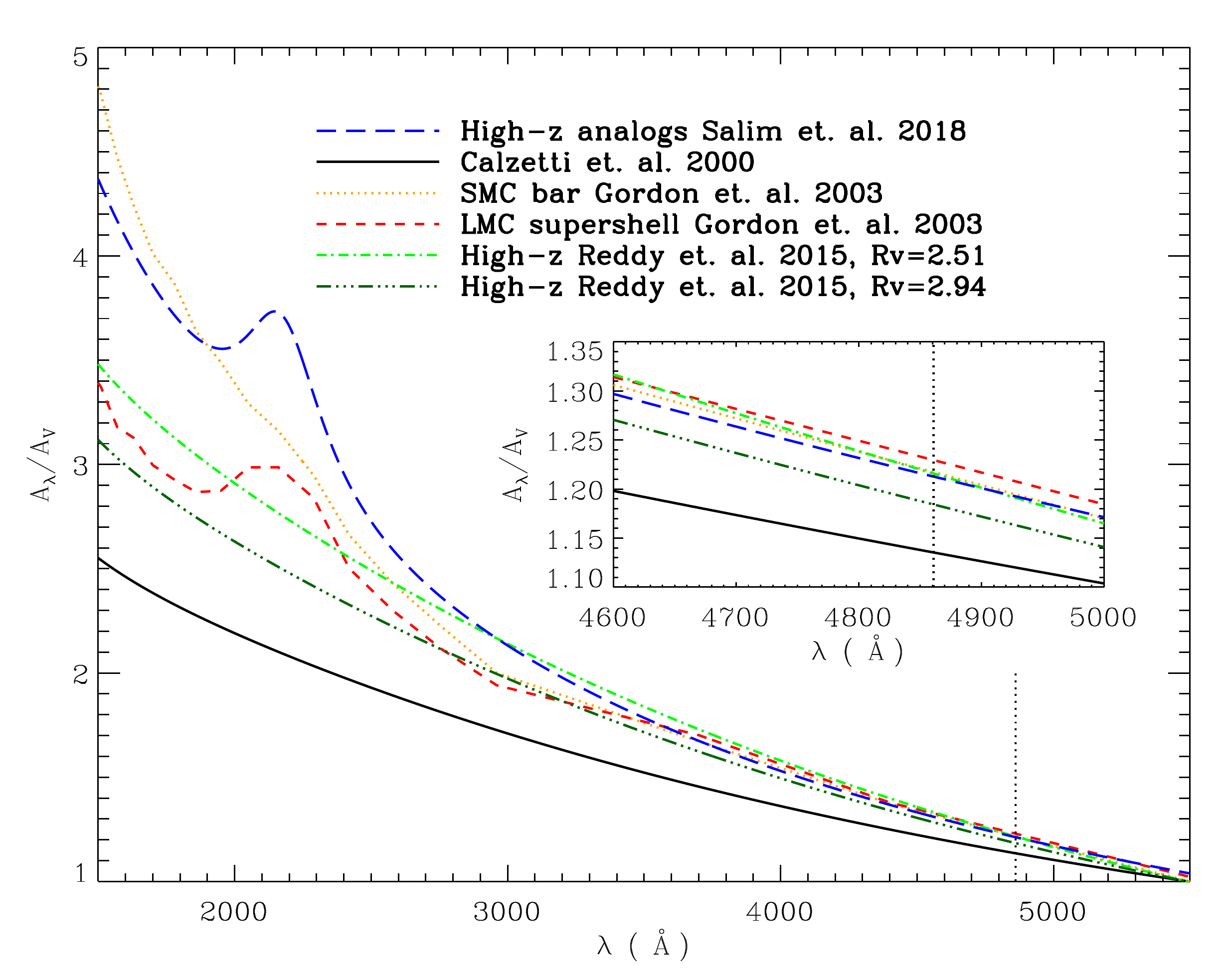}
\vspace*{-0.5cm}	
    \caption{Dust attenuation curves for different samples of star-forming galaxies. The inset show a zoom between 4600\AA\ and 5000\AA. The vertical dotted line indicates the position of $H\beta$.}
    \label{fig:ExtinctionLaw}
\end{figure}

In Figure \ref{fig:ExtinctionLaw} we present the dust attenuation curves for different samples of star-forming galaxies. It is clear from the figure  that the dust attenuation curve derived from analogs of high redshift star forming galaxies given by \citet{Salim2018} (long-dashed blue line) and from star forming galaxies at z$\sim$2 given by \citet{Reddy2015} (dot-dashed green line) agrees quite well with the Large and Small Magellanic Clouds  slopes given by \citet{Gordon2003} (dotted orange and dotted short-dashed red lines). Therefore, for our sample of HIIG at high redshift we prefer the results obtained using the \citet{Gordon2003} extinction curve for the supershell in the Large Magellanic Cloud. However, to facilitate the comparison with our previous work we also present the results using \citet{Calzetti2000} law.

\section{Results and Discussion}
\label{sec:ResultsDiscussion}

\subsection{Constraining cosmological parameters with HIIG}

To calculate the parameters of the \lsigG\ relation in a unified way including HIIG and GHIIR, we define the following likelihood function:
\begin{equation}
        \mathcal{L}_{HII} \propto \exp{(- \frac{1}{2} \chi^2_{HII})},
\label{eq:lkh}
\end{equation}
where:
\begin{equation}
        \chi^2_{HII} =  \sum_{n}\frac{(\mu_o(\log f, \log \sigma | \alpha, \beta) - \mu_{\theta}(z | \theta))^2}{\epsilon^2}.
\label{eq:chi}
\end{equation}

where $\mu_o$ is the distance modulus calculated from a set of observables as:
\begin{equation}
         \mu_o = 2.5(\alpha + \beta \log \sigma  - \log f - 40.08),
\end{equation}
where $\alpha$ and $\beta$ are the \lsigG\ relation's intercept and slope, respectively, $\log \sigma$ is the logarithm of the measured velocity dispersion and $\log f$ is the
logarithm of the measured flux. 

For HIIG the theoretical distance modulus, $\mu_{\theta}$, is given as:
\begin{equation}
         \mu_{\theta} = 5 \log d_L (z, \mathbf{\theta}) + 25,
\label{eq:thdm}
\end{equation}
where z is the redshift, $d_L$ is the luminosity distance in Mpc and $\theta$ is a given set of cosmological parameters. 

For GHIIR, the value of $ \mu_{\theta}$ is inferred from primary indicators. The GHIIR sample is described in detail in \cite{Fernandez2018}.

Finally $\epsilon^2$, the weights in the likelihood function,  can be given as:
  \begin{equation}
	  \epsilon^2 = \epsilon^2_{\mu_{o}, stat} + \epsilon^2_{\mu_{\theta}, stat} + \epsilon^2_{sys},
\label{eq:epsilon}         
 \end{equation}
 where $\epsilon_{stat}$ are the statistical uncertainties  given as:
 \begin{equation}
	 \epsilon^2_{\mu_{0}, stat} = 6.25(\epsilon_{\log f}^2 + \beta^2\epsilon_{\log \sigma}^2 + \epsilon_{\beta}^2\log \sigma^2 + \epsilon_{\alpha}^2),
  \end{equation}
where $\epsilon_{\log f}$,  $\epsilon_{\log \sigma}$ and $\epsilon_{\alpha}$  are the uncertainties associated with the logarithm of the flux, the logarithm of the velocity dispersion and the intercept of the \lsigG\ relation respectively and $\epsilon_{\mu_{\theta}, stat}$  in equation \ref{eq:epsilon} is the uncertainty associated with the distance modulus as propagated from the redshift uncertainty in the case of HIIG and for the case of GHIIR as given by the cepheid measurements uncertainty. Finally, $\epsilon_{sys}$ are the systematic uncertainties as described in \citet{Chavez2016} mentioned for completeness but  not included in the present analysis (see also section \ref{sec:ConstrainingOm})

Using the likelihood function given in Equation \ref{eq:lkh} in the MultiNest Bayesian inference algorithm \citep[cf.][]{Feroz2008, Feroz2009, Feroz2013}, we estimate the parameters of the \lsigG\ relation for the sample of 107 local  ($0.01 \leq$  z $\leq 0.15$) HIIG discussed in \S \ref{sec:Literature sample} and 36 GHIIR at  z $\leq 0.01$  described in \cite{Fernandez2018} for which distances have been estimated from primary indicators, both samples corrected for extinction using the \citet{Gordon2003} extinction curve (see \S \ref{sec:Extinction correction}). The resulting parameters are:

\begin{eqnarray}
        \alpha &=& 33.268\pm 0.083 \\
        \beta &=&  5.022\pm 0.058 .
\label{eq:parama}
\end{eqnarray}

\begin{figure}
\begin{center}
\hspace*{-0.8cm}
\includegraphics[width=1.1\columnwidth]{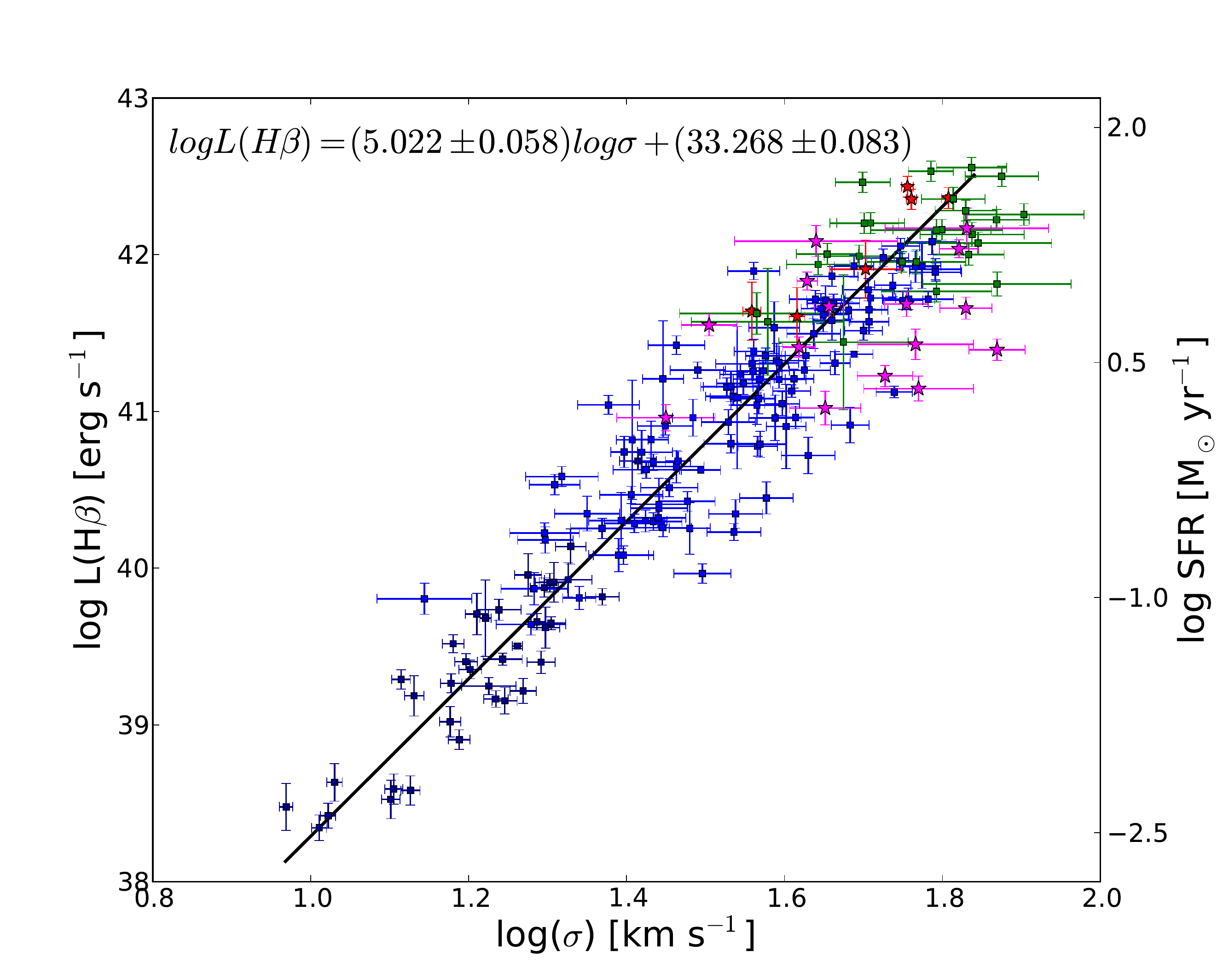}
\end{center}
\vspace*{-0.3cm}
\caption[L-sigma relation.]{\footnotesize \lsigG\ relation for the
  combined local (107 HIIG and 36 GHIIR) and high redshift (46 HIIG)
  samples using the \citet{Gordon2003} extinction curve. The fit
  corresponds to only the local sample of 143 objects, i.e., does not
  include the 46 high-z HIIG. Dark blue squares: GHIIR. Blue squares:
  local HIIG. Pink stars: our high-z MOSFIRE observations: Red stars:
  our high-z XShooter observations \citep{Terlevich2015}. Green
  squares: data from the literature (see \S \ref{sec:Literature sample}). The parameters of the
  fit are indicated at the top.}
\label{fig:L-sigma relation}
\end{figure}

The \lsigG\ relation shown in Figure \ref{fig:L-sigma relation}, includes  the new data for high-z objects presented in \S \ref{sec:MOSFIRE} and those analysed previoulsy in \citet{Terlevich2015} and \citet{Chavez2016}.

In order to compare with  \citet{Chavez2016} results, we also used their $\alpha$ and $\beta$  parameters derived after correcting the data for extinction using  the \citet{Calzetti2000}  curve:
\begin{eqnarray}
        \alpha &=& 33.11 \pm 0.145 \\
        \beta &=& 5.05 \pm 0.097 .
\label{eq:paramb}
\end{eqnarray}

The luminosity distance $d_{L}$ of the sources tracing the Hubble expansion is employed to calculate the theoretical distance moduli (see Equation \ref{eq:thdm}). We define, for convenience, an extra parameter independent of the Hubble constant as:
\begin{equation}
D_L(z,\theta)= (1+z) \int_{0}^{z}{\frac{dz^{'}}{E(z^{'}, \theta)}}\;.
\end{equation}
i.e., $d_L=c D_L/H_0$. $E(z, \theta)$ for a flat Universe is given by:
\begin{equation}\label{eq:Ez}
        E^2(z, \theta) = \Omega_{r}(1+z)^4 + \Omega_{m}(1+z)^3 + \Omega_{w} (1+z)^{3y}\exp\left( \frac{-3 w_a z}{1+z}\right)
\end{equation}
with $y=(1 + w_0 + w_a)$.
The parameters $w_0$ and $w_a$ refer to the DE EoS, the general form of which is:
\begin{equation}
p_{w} =  w(z) \rho_{w} \;,
\end{equation}
with $p_{w}$ the pressure and $\rho_{w}$ the density of the
postulated DE fluid.
Different DE models have been proposed and many are
parametrized using a Taylor expansion around the present epoch:
\begin{equation}
        w(a) = w_0 + w_a(1-a)\Longrightarrow w(z)=w_0+w_a \frac{z}{1 + z}\;,
\end{equation}
\citep[CPL model;][]{Chevallier2001, Linder2003, Peebles2003,
  Dicus2004, Wang2006}.
The cosmological constant is just a special case of DE, given
for $(w_0,w_a)=(-1,0)$, while the so called wCDM models
are such that $w_a=0$ but $w_0$ can take values $\neq -1$.

From a set of cosmological parameters, in the general case considered here given as
 $\theta = \{\Omega_m, w_0, w_a \}$, and
the redshift, we can obtain the
theoretical distance modulus of a source as:
\begin{equation}
\mu_{\theta} = 5 \log D_L (z, \mathbf{\theta}) + \mu_{0},
\label{eq:mu2}
\end{equation}
where $\mu_{0}=42.384-5\log h$ and $h \equiv H_0/100$.
Inserting Equation \ref{eq:mu2} into Equation \ref{eq:chi}
we find after some simple algebra that:
\begin{equation}\label{eq:expand-xi2}
\chi^2(\theta)=A(\theta)-2B(\theta)\mu_0+C\mu_0^2\;,
\end{equation}
where,
\begin{eqnarray}\label{eq:xi2-A-B-C}
 A(\theta) &=& \sum\limits_{i=1}^{N}\frac{[\mu_{\rm o}(z_{i})-
5{\rm log}D_{L}(z_{i},\theta)]^2}{\epsilon^2}\;,\nonumber \\
 B(\theta) &=&\sum\limits_{i=1}^{N}\frac{\mu_{\rm o}(z_{i})-5{\rm log}D_{L}(z_{i},\theta)}
{\epsilon^2}\;,\nonumber \\
 C &=& \sum\limits_{i=1}^{N} \frac{1}{\epsilon^2}\;. \nonumber
\end{eqnarray}
For $\mu_{0}=B/C$, Equation \ref{eq:expand-xi2}
has a minimum at:
\begin{equation}
{\tilde \chi}^{2}(\theta)=A(\theta)-\frac{B^2(\theta)}{C}.
\end{equation}
Therefore, instead of using
$\chi^{2}$ we now minimise ${\tilde \chi}^{2}$
which is independent of $\mu_{0}$ and thus of the value of the Hubble constant \citep[cf.][]{Nesseris2005}.

\begin{figure*}
\begin{center}
\hspace*{-1.5cm}
\includegraphics[width=2.5\columnwidth]{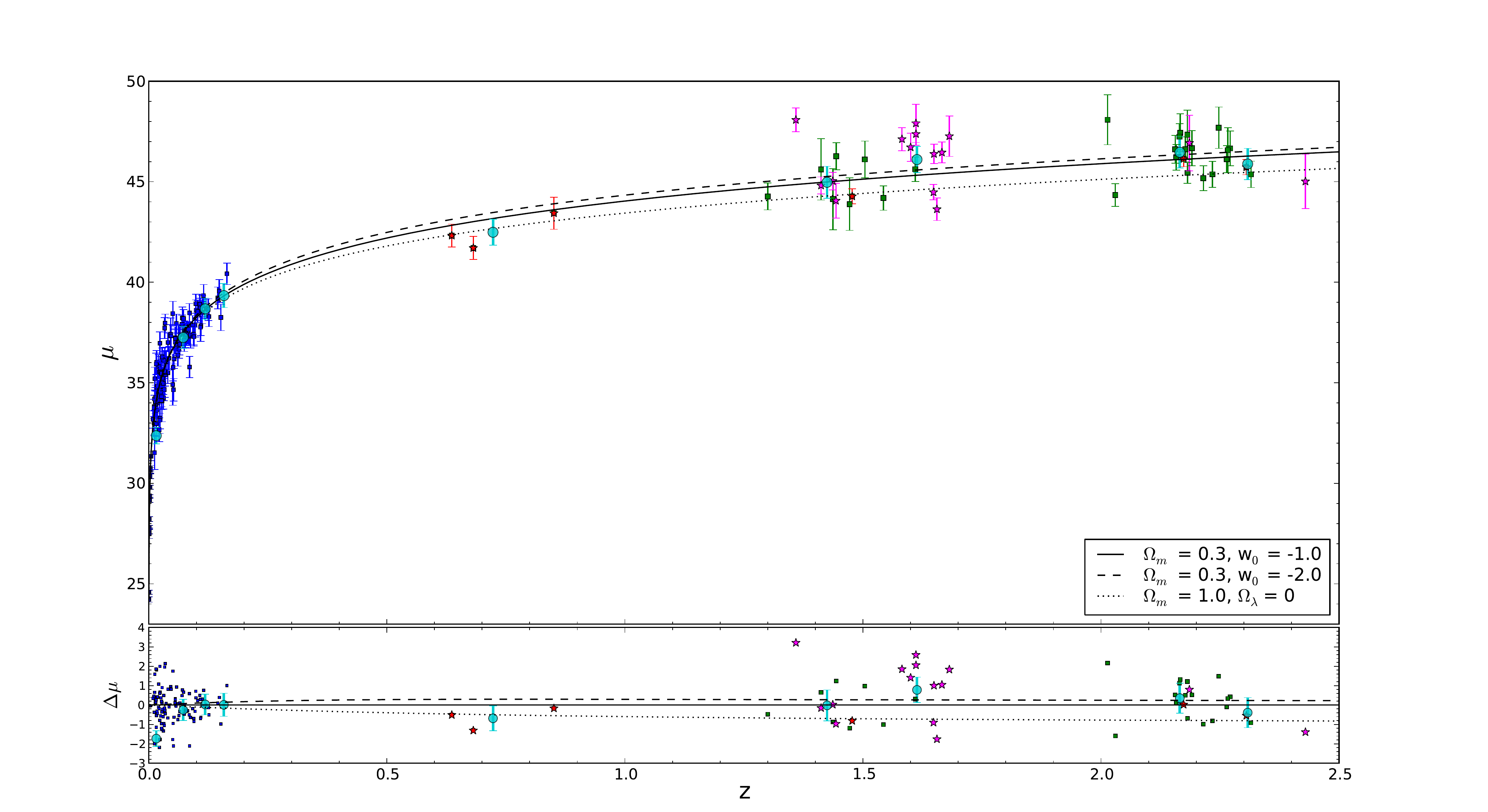}
\end{center}
\caption[Hubble diagram.]{\footnotesize Hubble-Lem\^aitre diagram connecting our local and high redshift samples up to z$\sim$2.5 for three different cosmologies shown in the inset. Dark blue squares: GHIIR; blue squares: local HIIG; pink stars: our high-z MOSFIRE observations; red stars: our high-z XShooter observations; green squares: data from the literature and sky blue circles: the mean in redshift bines. Residuals are plotted in the bottom panel. Note the huge dynamical range covered by the distance modulus with a single distance estimator ($L-\sigma$).}
\label{fig:Hubble Diagram}
\end{figure*}

\begin{figure}
\centering
\hspace*{-0.7cm}
        \includegraphics[width=1.1
        \columnwidth]{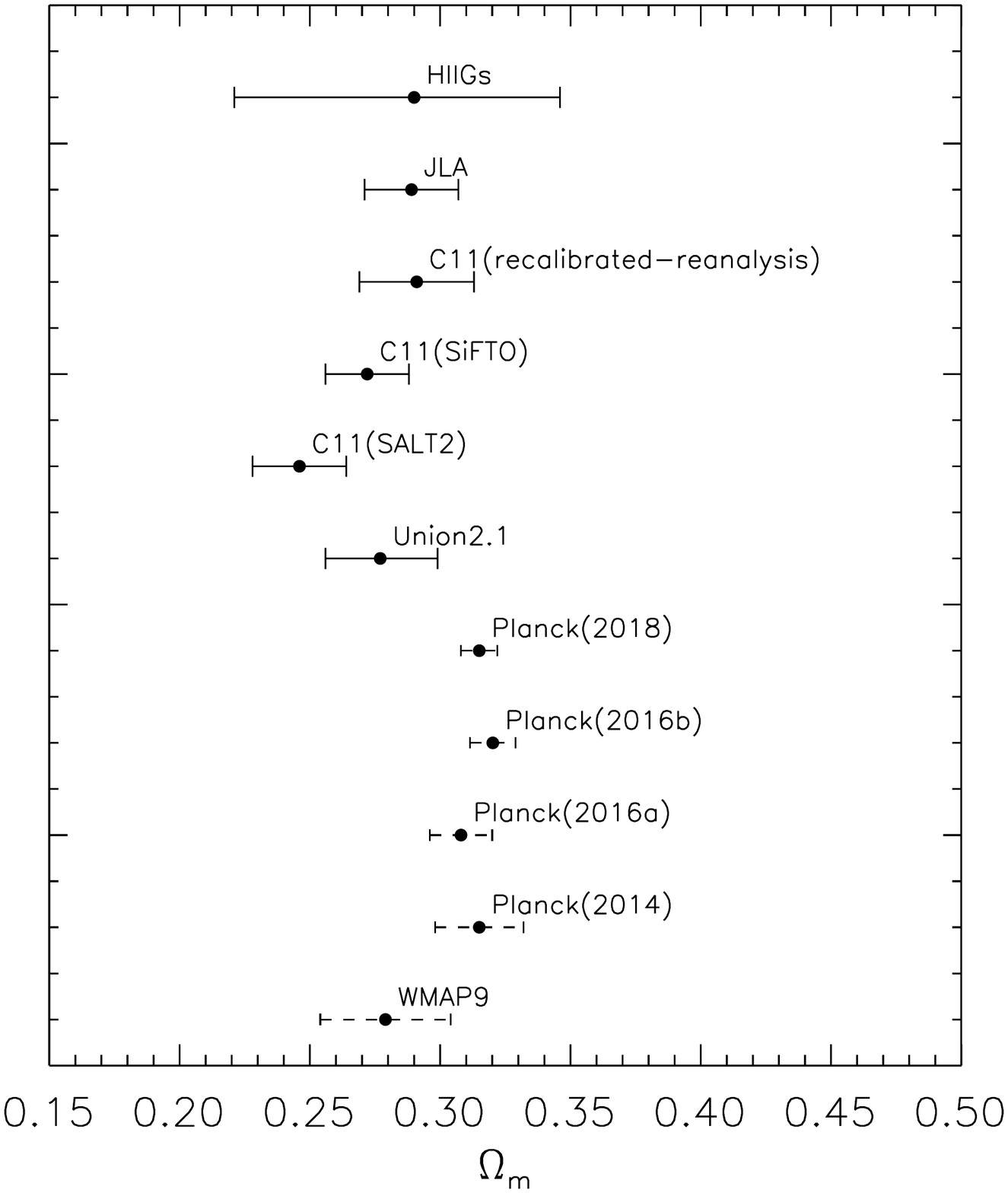}
\vspace*{-0.5cm}
    \caption{Comparison of the result of this work with the most recent estimates of $\Omega_m$ from the literature. The dashed error bars include statistical + systematic uncertainties and the continuous error bars, only statistical uncertainties. Our result for $\Omega_m$ is in the top line.}
    \label{fig:valoresOM}
\end{figure}

\subsubsection{Constraining $\Omega_m$}
\label{sec:ConstrainingOm}

Applying the method described above to the joint local and high-z sample of
153 HIIG and using both of the two fitting approaches, the
$\chi^2$-minimization procedure and the MultiNest MCMC, 
we find $\Omega_m=0.276^{+0.066}_{-0.054}$ and
$\Omega_m=0.290^{+0.056}_{-0.069}$  respectively under the priors shown in Table
\ref{tab:priors} and for the extinction curves of \citet{Gordon2003}.
Using the \citet{Calzetti2000} extinction curves we find a lower value of
$\Omega_m=0.219^{+0.072}_{-0.057}$ (based on the $\chi^2$-minimization procedure).
Note that the uncertainties are statistical and do not include the contribution of
systematic errors.

In Figure \ref{fig:Hubble Diagram} we present the Hubble diagram of our data using the \citet{Gordon2003} extinction curve and three different cosmologies, for comparison, as described in the inset in the figure.

It is important to underline that the  difference in the value of $\Omega_m$, determined using the two different methods, MCMC and $\chi^2$-minimization, is well within $1\sigma$ and can be  attributed to the different error-weighting schemes that they employ.

It is interesting to note that the relatively large values of the reduced $\chi^2$ of the fits, i.e.,\ $\sim 1.7$ and $\sim 1.1$, respectively, for the two different extinction curves used (see Table
\ref{tab:par}), could well be attributed to the fact that we have not added in quadrature the systematic uncertainty in $\mu$. For a value $\sim 0.22$, added in quadrature to the Equation \ref{eq:epsilon}, the reduced $\chi^2$ drops to $\sim 1$. One important such systematic, which is not taken into account, is related to the second parameter (size of the HII region) in the \lsigG\
correlation (see for example \citet{Chavez2014}).

\subsubsection{Comparison with previous $\Omega_m$ results}
The comparison of our MCMC estimated $\Omega_m = 0.290^{+0.056}_{-0.069}$
(stat) with other literature results is summarized in  Figure
\ref{fig:valoresOM}. 
In Planck Collaboration XVI
\citeyearpar[Sect. 5.4]{PlanckCollaboration2014} the light-curve
parameters and covariance matrices obtained for the SiFTO light-curve
model \citep{Conley2008} lead to an $\Omega_m$ value significantly
lower than that obtained from the SALT2 \citep{Guy2007} analysis.
Interesting is the comparison of the SiFTO and SALT2 analyses obtained
when the systematic uncertainties are not taken into account. 

For example \citet{Betoule2014} find $\Omega_m=0.289 \pm 0.018$ (stat)
that differs from the value published in \citep[hereafter
  C11]{Conley2011} of $\Omega_m=0.246 \pm 0.018$ (stat) and
$\Omega_m=0.272 \pm 0.016$ (stat) by 2.4$\sigma$ and 1.1$\sigma$ using
the SALT2 and SiFTO light-curve models, respectively.
The differences between SALT2 and SiFTO analyses are probably caused
by  different weights assigned to SNIa on the Hubble diagram
rather than differences in the models. The difference is reduced by
the recalibrated-reanalysis of the C11 sample by \citet{Betoule2014},
who find $\Omega_m=0.291 \pm 0.022$ (stat), reducing the discrepancy
between C11 and the Planck Collaboration XVI \citeyearpar{PlanckCollaboration2014} value of $\Omega_m=0.315 \pm 0.017$ (stat + sys).

Our determination of $\Omega_m = 0.290^{+0.056}_{-0.069}$ (stat) is in agreement with the SNIa measurements of $\Omega_m=0.289 \pm 0.018$ (stat), based on 740 SNIa (\citet{Betoule2014}), and of $\Omega_m=0.277^{+0.022}_{-0.021}$ (stat), based on the 580 SNIa of the Union 2.1 sample (\citet{Suzuki2012}), where the highest redshift is 1.415. Our value is also consistent with the most recent CMB measurement from WMAP9 \citep{Bennett2013}, which provides a value of $\Omega_m=0.279 \pm 0.025$, as well as with the Planck Collaboration measurement \citeyearpar{PlanckCollaboration2016a,PlanckCollaboration2016b,PlanckCollaboration2018}  of $\Omega_m=0.308 \pm 0.012$; $\Omega_m= 0.320 \pm 0.009$ and $\Omega_m= 0.315 \pm 0.007$ respectively, where both statistical and systematic uncertainties are included.

\begin{table}
\caption{Priors for Constrained Parameters.}
\label{tab:priors}
\centering
\begin{tabular}{ l l }
\hline \hline
Parameter & Prior \\
\hline
\multicolumn{2}{c}{Cosmological Parameters}\\
\hline
$h$ & Uniform [0.5, 1.0] \\
$\Omega_m$ & Uniform [0.0, 1.0] \\
$w_0$ & Uniform [-2.0, 0.0] \\
$w_a$ & Uniform [-4.0, 2.0] \\
$w_b$ & Uniform [0.0, 0.05] \\
\hline
\multicolumn{2}{c}{HIIG Nuisance Parameters}\\
\hline
$\alpha$ & Uniform [25.0, 35.0] \\
$\beta$ & Uniform [0.0, 8.0] \\
\hline
\multicolumn{2}{c}{SNIa Nuisance Parameters}\\
\hline
$\alpha$ & Uniform [0.0, 0.3] \\
$\beta$ & Uniform [2.0, 4.0] \\
$M_{B}^1$ & Uniform [-18.0, -21.0] \\
$\Delta_M$ & Uniform [-2.0, 0.0] \\
\hline \hline
\end{tabular}
\end{table}

\subsubsection{Comparison with previous  results in the $\lbrace\Omega_m,w_0\rbrace$ plane}
\begin{figure}
\centering
\hspace*{-0.7cm}
        \includegraphics[width=1.15
        \columnwidth]{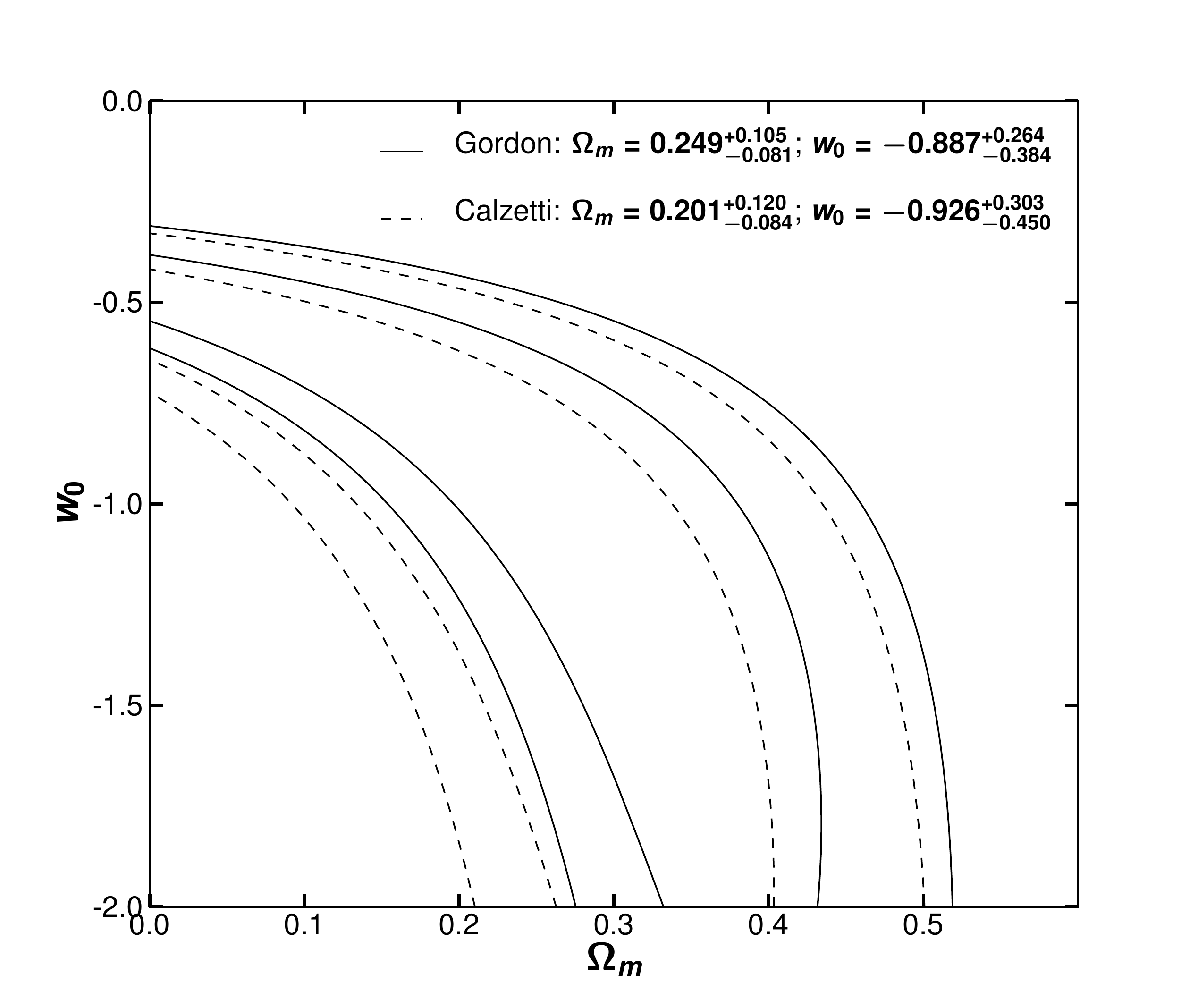}
\vspace*{-0.5cm}
    \caption{$\chi^2$ likelihood contours for $\Delta\chi^2=\chi^2_{tot}-\chi^2_{tot,min}$ equal to 2.30 and 6.17 corresponding to the 1$\sigma$ and 2$\sigma$ confidence levels in the $\{\Omega_m, w_0\}$ plane for our 153 HIIG at intermediate and high redshift. a)  solid line: \citet{Gordon2003} extinction correction ($\alpha = 33.268 \pm 0.083$ and $\beta =  5.022 \pm 0.058$), b) dotted line: \citet{Calzetti2000} extinction correction ($\alpha = 33.11 \pm 0.145$ and $\beta =  5.05 \pm 0.097 $).}
    \label{fig:chi2-Likelihood contours}
\end{figure}

We compare our current with our previous results in the $\{\Omega_m, w_0\}$ plane \citep{Terlevich2015,Chavez2016}, based on the $\chi^2$-minimzation procedure, in Figure 
\ref{fig:chi2-Likelihood contours} and in Table \ref{tab:par}. 

Figure \ref{fig:chi2-Likelihood contours} shows the 1$\sigma$ and 2$\sigma$ confidence levels 
in the $\lbrace\Omega_m,w_0\rbrace$ plane based on the  $\chi^2$-minimzation procedure, for the \lsigG\ parameters corresponding to the 
\citet{Calzetti2000} (dashed contours) and the \citet{Gordon2003} (continuous contours) extinction correction curves, for which we obtain  ${\Omega_m, w_0}=\lbrace 0.201^{+0.120}_{-0.084}, -0.926^{+0.303}_{-0.450} \rbrace$ and $\lbrace 0.249^{+0.105}_{-0.081}, -0.887^{+0.264}_{-0.384} \rbrace$, respectively. Values are listed in Table \ref{tab:par}. 

The  addition of just 15 new high-z HIIG to the sample presented in our previous work \citep[cf.][]{Chavez2016} has produced a significant improvement, exemplified by the fact that now we obtain
$\lbrace\Omega_m, w_0\rbrace$ central values that are in agreement with other analyses and with errors that are reasonably small considering the modest size of our high-z HIIG sample.

\begin{figure}
\begin{center}
\includegraphics[width=1.0\columnwidth]{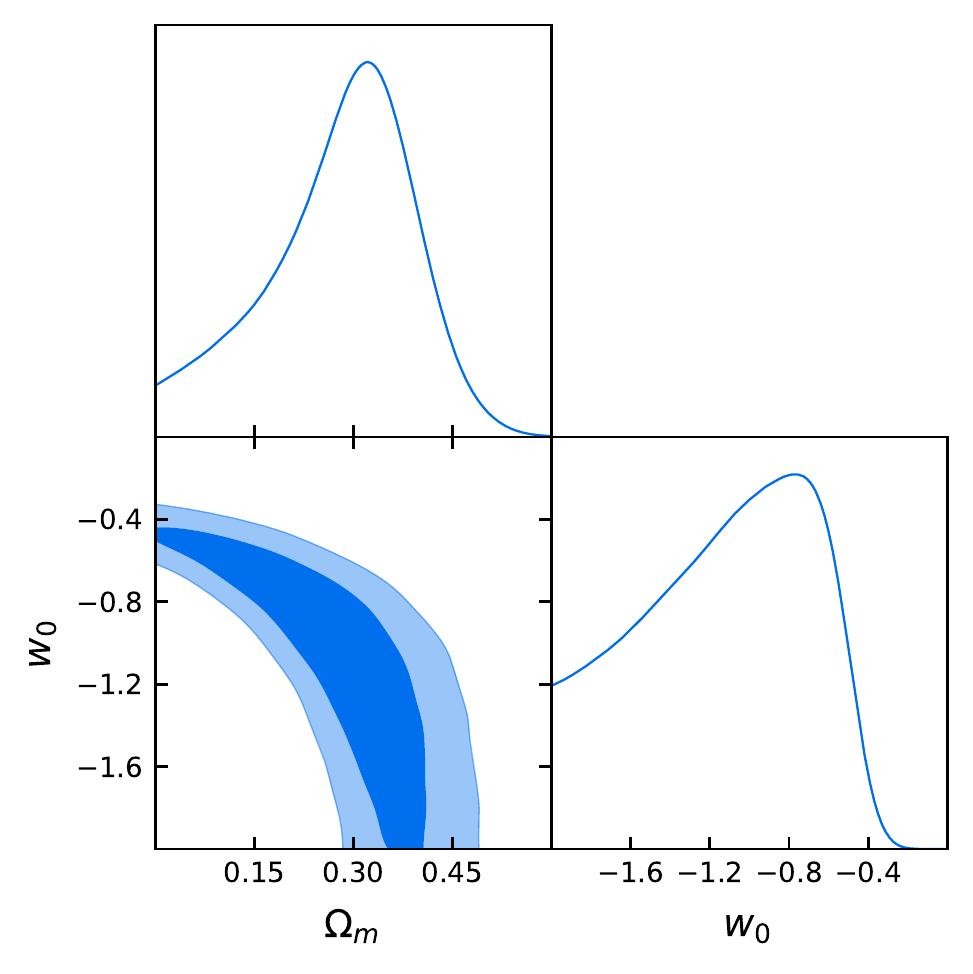}
\end{center}
\caption{Likelihood contours  corresponding to the 1$\sigma$ and
  2$\sigma$ confidence levels in the $\{\Omega_m, w_0\}$ space for the
  HIIG sample corrected for extinction with \citet{Gordon2003}. The
  distance estimator parameters are: $\alpha = 33.268 \pm 0.083$ and
  $\beta =  5.022 \pm 0.058$.}
\label{fig:h2a}
\end{figure}

Figure \ref{fig:h2a} and Table \ref{tab:par} show the MCMC $\{\Omega_m, w_0\}$ constraints based on our sample of 153 HIIG using the priors shown in Table \ref{tab:priors}. When we constrain the $\lbrace\Omega_m,w_0\rbrace$ plane, we obtain $\Omega_m=0.280^{+0.130}_{-0.100}$ and $w_0=-1.12^{+0.58}_{-0.32}$ (stat). 
Our results are in agreement, within 1$\sigma$, with the SNIa measurements of $\{\Omega_m, w_0 \}= \{0.281^{+0.067}_{-0.092},-1.011^{+0.208}_{-0.231}\}$ (stat), based on the Union 2.1 sample of 580 SNIa (\citet{Suzuki2012}) and of $\{ 0.350\pm0.035, -1.251\pm0.144 \}$ (stat), based on the most recent sample of 1048 SNIa (\citet{Scolnic2018}). The smaller uncertainties in the SNIa based results reflect
the much larger samples, i.e., 580 and 1048 SNIa versus 153 HIIG.

\begin{figure*}
\begin{center}
\includegraphics[width=1.5\columnwidth]{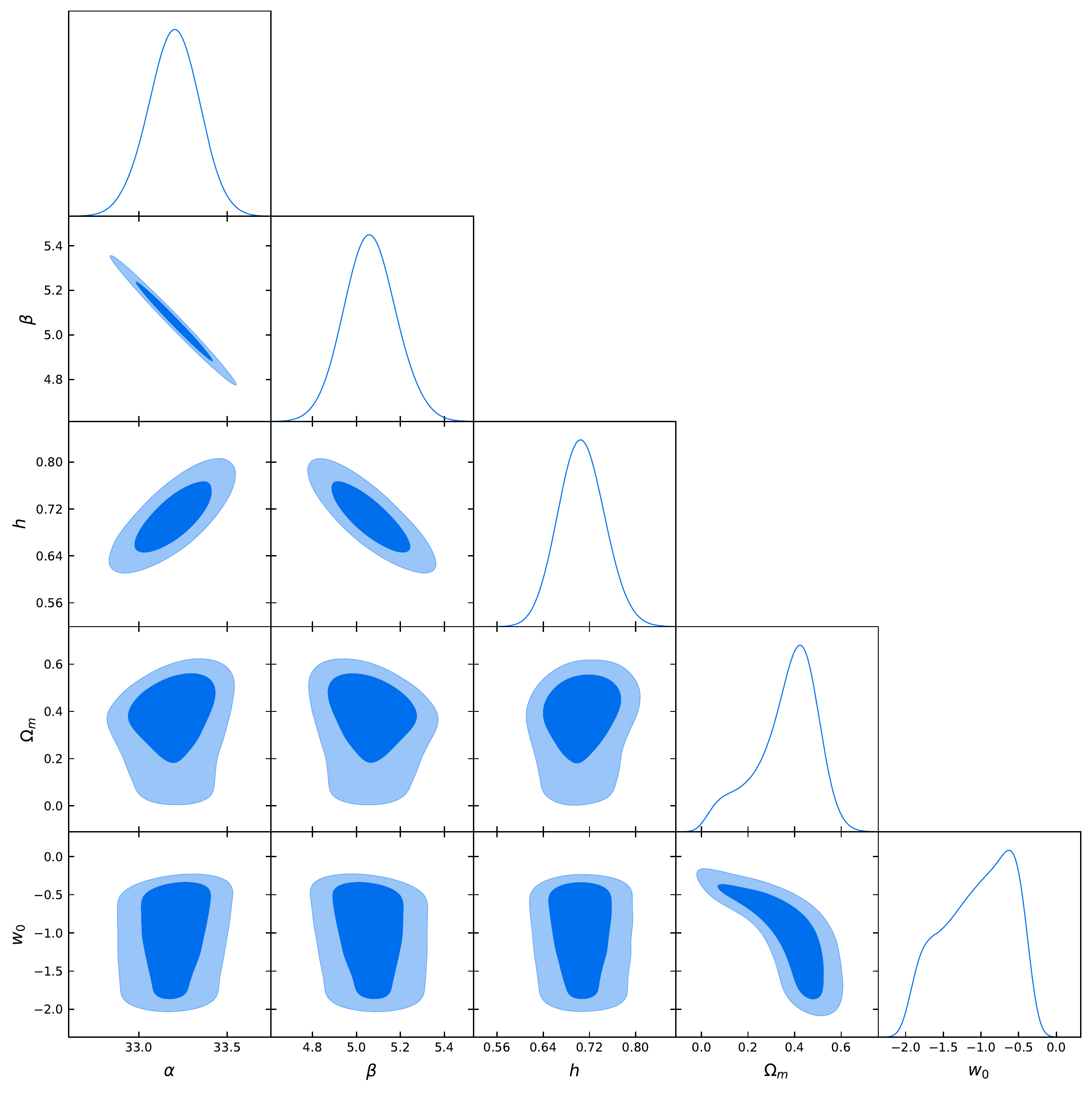}
\end{center}
\caption{Likelihood contours  corresponding to the 1$\sigma$ and
  2$\sigma$ confidence levels in the $\{\alpha, \beta, h,\Omega_m, w_0\}$ space for the
	HIIG sample corrected for extinction with \citet{Gordon2003}. The resulting constraints are $\{\alpha, \beta, h,\Omega_m, w_0\} = \{33.20\pm 0.14, 5.06\pm 0.12, 0.706\pm 0.040, 0.363^{+0.16}_{-0.081}, -1.03^{+0.62}_{-0.33} \}$.}
\label{fig:h2b}
\end{figure*}

\subsubsection{HII galaxies and GHIIR simultaneous constraints}
Using simultaneously GHIIR and HIIG data, a global  fit  of  all  the
free  parameters, nuisance and cosmological,  provides
the following results $\alpha=33.20\pm 0.14$, $\beta=5.06\pm 0.12$, $h = 0.706\pm 0.040$,  $\Omega_{m}=0.363^{+0.16}_{-0.081}$ and $w_0 = -1.03^{+0.62}_{-0.33}$ ; while in Figure \ref{fig:h2b} we plot the $1\sigma$ and $2\sigma$ contours in various planes.

Remarkably, although the errors are quite large, the corresponding
mean values are in excellent agreement with those of \citet{Scolnic2018}.
This confirms our proposal that the Hubble relation based on the HIIG could provide an efficient tool to understand the mechanism of late (dark energy) cosmic acceleration.
We argue that  new data from VLT/KMOS (Gonz\'alez-Mor\'an et al. in preparation) and from 
GTC/MEGARA (at present in an observing queue)  will allow us to measure the DE EoS parameter as well as to check whether $w$ depends on time.

\subsection{CMB constraints}
The information about the position of the acoustic peaks of the CMB, can be quantified by three observables: $\{l_a, R, \omega_{b}\}$, where $l_a$ is the acoustic scale related to the comoving sound speed horizon, $R$ is the shift parameter \citep{Bond1997, Nesseris2007}, the ratio of the position of the first peak to that of a reference model, and 
\begin{equation}
\omega_{b} \equiv \Omega_b h^2 ,
\label{eq:wb}
\end{equation}
where $\Omega_b$ is the current baryon density parameter. The first two quantities are related to the angular diameter distance, which in flat space can be given by:
\begin{equation}
d_A(z, \theta)= \frac{c}{H_0} \frac{1}{1+z} \int_0^z{\frac{dz'}{E(z', \theta)}}
\end{equation}
and to the comoving sound speed horizon:
\begin{equation}
r_s(z, \theta)= \frac{c}{H_0}  \int_0^{1/(1+z)}{da \frac{c_s(a)}{a^2E(a)}},
\label{eq:csh}
\end{equation}
where $c_s(a)$ is the sound speed of the baryon-photon fluid before recombination:
\begin{equation}
c_s(a) = \frac{1}{\sqrt{3+ 3(3\Omega_{b}/4\Omega_{\gamma})a}},
\end{equation}
where $\Omega_{\gamma}$ is the current photon energy density parameter.

The acoustic scale is given by:
\begin{equation}
l_a = (1+z_{\star}) \frac{\pi d_A(z_{\star}, \theta)}{r_s(z_{\star}, \theta)},
\label{eq:la}
\end{equation}
the shift parameter can be given as:
\begin{equation}
R(\theta)=\sqrt{\Omega_{m}}\int_{0}^{z_{\star}} \frac{dz}{E(z, \theta)},
\label{eq:R}
\end{equation}
and $z_{\star}$ is given by the fitting formula of \citet{Hu1996}:
\begin{equation}
z_{\star} = 1048[1 + 0.00124(\Omega_{b}h^2)^{-0.738}] [1 + g_1(\Omega_{m}h^2)^{g_2}],
\end{equation}
\begin{equation}
g_1 = \frac{0.0783 (\Omega_{b}h^2)^{-0.238}}{1 + 39.5(\Omega_{b}h^2)^{0.763}},
\end{equation}
\begin{equation}
g_2 = \frac{0.560}{1 + 21.1(\Omega_{b}h^2)^{1.81}}.
\end{equation}

In \citet{Chavez2016} we used only the measured shift parameter, according to
Planck data \citep{Shafer2014} as $R=1.7499\pm 0.0088$ at the
redshift of decoupling (i.e.,  at the last scattering surface). Here we use updated values for $\{l_a, R, \omega_{b}\}$ derived from the Planck 2015 likelihoods assuming a flat Universe \citep[][]{PlanckCollaboration2016a}. The parameters values derived from observations are given by cf. \citet[][]{Wang2016}:
\begin{equation}
\begin{split}
	\mathbf{P_o} & = \{l_a, R, \omega_{b}\} \\ 
		     & = \{301.77 \pm 0.09, 1.7482 \pm 0.0048, 0.02226 \pm 0.00016 \}, 
\end{split}
\end{equation}
with an inverse covariance matrix, $\mathbf{C^{-1}}$, given as:

\[
\begin{bmatrix}
	$1.47529862e+02$ & $-9.50243997e+02$ &  $6.75330855e+03$ \\
       $-9.49518029e+02$ & $8.87656028e+04$  &  $1.66515286e+06$ \\
       $6.78491359e+03$  & $1.66491606e+06$  &  $7.46953427e+07$
\end{bmatrix}
\]

The final likelihood function is:
\begin{equation}
\label{eq:cmb}
	\mathcal{L}_{CMB} \propto \exp{\left[- \frac{1}{2} \mathbf{\Delta C^{-1} \Delta^{\dagger}} \right]} \;,
\end{equation}
where, ${\bf \Delta} = {\bf P}_{\theta} - {\bf P}_{o}$, and ${\bf P}_{\theta} = \{l_a, R, \omega_{b}\}$ is the vector of modeled parameters as described in Equations \ref{eq:la}, \ref{eq:R} and \ref{eq:wb}; likewise ${\bf C^{-1}}$ is the inverse of the corresponding covariance matrix and $\bf \Delta^{\dagger}$ the transpose of $\bf \Delta$.

In Table \ref{tab:par} we show the maximum likelihood  results for the plane $\{\Omega_m, w_0, w_b\}$ using the updated values of $\{l_a, R, \omega_{b}\}$ as described above. For the parameter modeling we use a value of $h = 0.6774$, which is consistent with Planck latest determination \citep[][]{PlanckCollaboration2016a}. We use the  MultiNest algorithm with the priors listed in Table \ref{tab:priors} to calculate the probability distribution.

\subsection{BAO Constraints}
The Baryonic Acoustic Oscillation (BAO) scale, is a feature produced in the last
scattering surface by the competition between the pressure of the
coupled baryon-photon fluid and gravity. The resulting sound
waves leave an overdensity
signature at a certain length scale of the matter distribution. This
length scale is related to the comoving distance that a sound
wave can travel until recombination and in practice it manifests
itself as a feature in the correlation function of galaxies on large
scales ($\sim 100 \;h^{-1}$ Mpc).
In recent years, measurements of BAOs have proven to be an extremely useful
``standard ruler''.
The BAOs were clearly identified, for the first time,
as an excess in the clustering pattern of the SDSS luminous red galaxies
\citep{Eisenstein2005} and of the 2dFGRS galaxies \citep{Cole2005}.
Since then a large number of dedicated surveys have been used to
measure BAOs, among which the  WiggleZ Dark Energy Survey
\citep{Blake2011}, the 6dFGS \citep{Beutler2011} and
the SDSS Baryon Oscillation Spectroscopic Survey (BOSS) of SDSS-III
\citep{Eisenstein2011,Anderson2014,Aubourg2015}.

Here we use BAOs measurements from the Reconstructed 6-degree Field Galaxy Survey \citep[][]{Carter2018}, the WiggleZ Dark Energy Survey \citep[][]{Kazin2014} and the SDSS-III Baryon Oscillation Spectroscopic Survey \citep[][]{GilMarin2016} as compiled in \citet[][see the first six data points in their Table III]{Anagnostopoulos2019}, which following their notation, is given in terms of the parameter $\lambda(z_i, \theta)$ defined as:
\[
	\lambda(z_i, \theta) =
  \begin{cases}
	  d_V r_{d,fid}/r_d, & 1 \leq i \leq 4 \\
	  d_A/r_d, & 5 \leq i \leq 6 \\
  \end{cases}
\]
with $z_i$ the redshift at which the signature of the acoustic oscillations has been
measured and
\begin{eqnarray}
	d_A =& \frac{d_L(z, \theta)}{(1 + z)^2}, \\
	d_V =& \left[ \frac{c d_A^2 z (1 + z)^2}{H(z, \theta)} \right]^{1/3},
\end{eqnarray}
additionally, $r_d$ is the standard ruler, which in $\Lambda$CDM cosmology is equal to the comoving sound horizon $r_s$ as given in Equation \ref{eq:csh}.

The corresponding likelihood function is given by:
\begin{equation}
	\mathcal{L}_{BAO} \propto \exp{\left[- \frac{1}{2} \mathbf{\Delta C^{-1} \Delta^{\dagger}} \right]}\;.
\end{equation}
where, $\mathbf{\Delta} = {\bf P}_{z, \theta} - {\bf P}_{o}$, is the difference between the vector of modeled parameters, ${\bf P}_{z, \theta} = \{ \lambda(z_1, \theta), ..., \lambda(z_{10}, \theta)  \}$, as defined above, and the vector of observed data, ${\bf P}_o = \{ \lambda(z_1)_o, ..., \lambda(z_{10})_o  \}$.

The constraints for BAOs, using the MCMC MultiNest algorithm, a value of $h = 0.6774$ and the priors shown in Table \ref{tab:priors} in the plane $\{\Omega_m, w_0\}$, are given in Table \ref{tab:par}.

\subsection{Type Ia Supernovae Constraints}
For SNIa we apply a similar methodology where the likelihood function is given as:
\begin{equation}
         \mathcal{L}_{SNIa} \propto \exp{\left[- \frac{1}{2} (\mu_o - \mu_m (z | \theta))  \mathbf{C^{-1}}  (\mu_o - \mu_m(z | \theta))^{\dagger} \right]},
\end{equation}
where $\mathbf{C}$ is the covariance matrix of $\mu_o = m_{B}^{\star} - (M_B - \alpha X_1 + \beta C)$, with $m_{B}^{\star}$ the observed peak magnitude in the rest-frame $B$ band and $\alpha$, $\beta$ and $M_B$ are nuisance parameters. It has been shown that the absolute magnitude $M_B$ and the parameter $\beta$ depend on the host galaxy parameters \citep{Sullivan2011, Johansson2013}.  Following \citet{Betoule2014} we parametrize this dependence as:
\[
  M_B =
  \begin{cases}
                                   M_B^1 & \text{if $M_{\star} < 10^{10} M_{\odot}$} \\
                                   M_B^1 + \Delta_M & otherwise
  \end{cases}
\]

We analyze the JLA compilation of 740 SNIa presented in \citet[][]{Betoule2014}. We show constraints in the parameter space $\{  \alpha, \beta, M_B^1, \Delta_M, \Omega_m, w_0\}$ in Table \ref{tab:par} where  $\{  \alpha, \beta, M_B^1, \Delta_M\}$ are nuisance parameters, following \citet[][]{Betoule2014} we use a fiducial value of $h = 0.7$ for the analysis. These constraints have been calculated using the MCMC MultiNest algorithm with the priors shown in Table \ref{tab:priors}.

\begin{figure*}
\begin{center}$
\begin{array}{cc}
  \subfloat[HIIG + CMB + BAO] {\includegraphics[width=.5\textwidth]{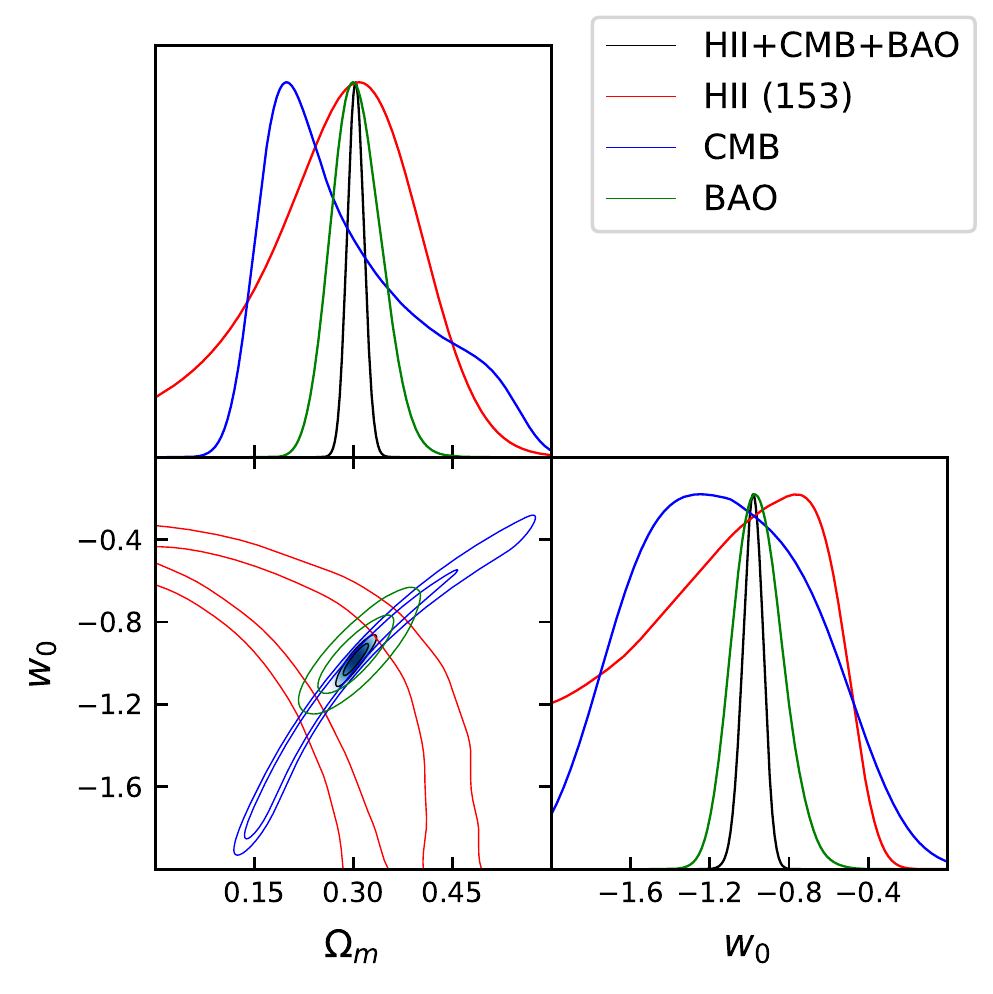}} &
  \subfloat[SNIa + CMB + BAO] {\includegraphics[width=.5\textwidth]{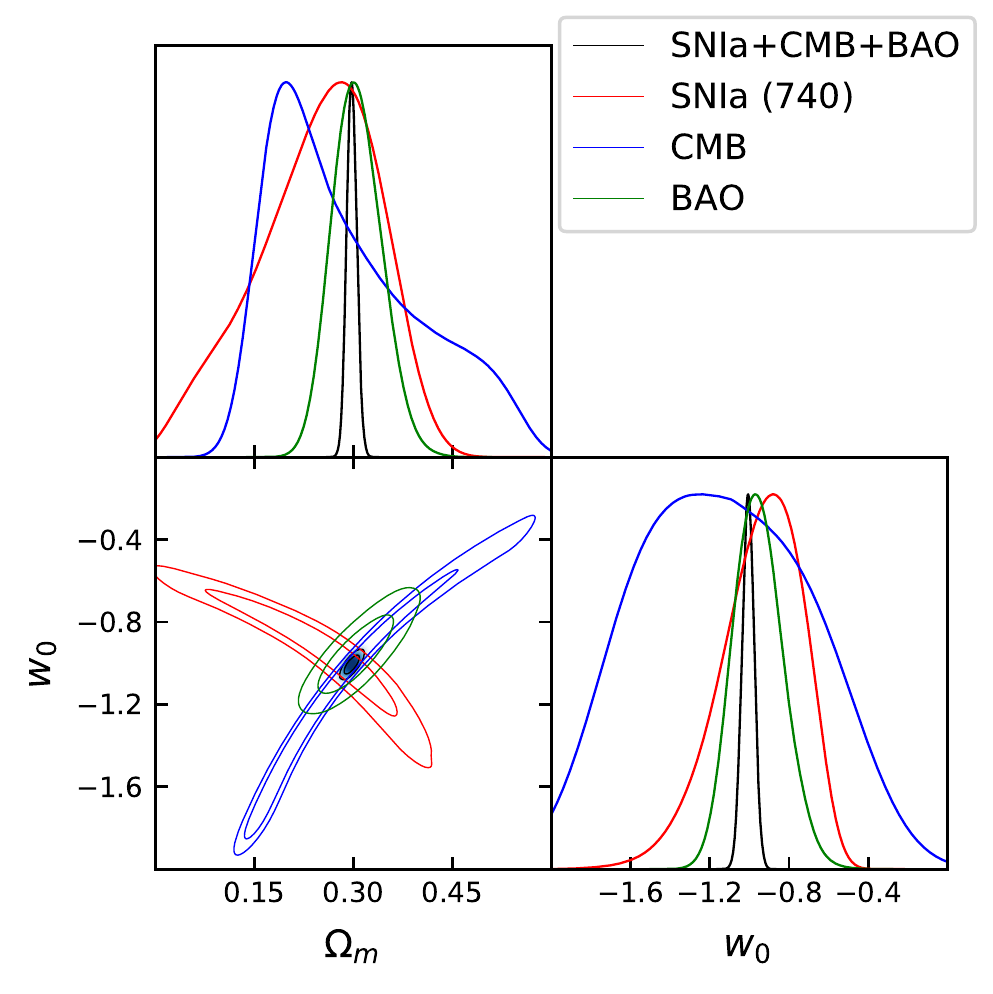}}
\end{array}$
\end{center}
\caption{Likelihood contours
        corresponding to the 1$\sigma$ and 2$\sigma$ confidence levels in the $\{\Omega_m, w_0\}$ space for a) the joint sample of HIIG, CMB\citep[][]{Wang2016} and BAO\citep[][]{Anagnostopoulos2019} and b) the joint sample of SNIa\citep[JLA][]{Betoule2014}, CMB and BAO. For HIIG and SNIa we show the number of objects used in the analyis in parentheses at the inset. Only statistical uncertainties are shown.}
\label{fig:j1}
\end{figure*}

\begin{figure*}
\begin{center}$
\begin{array}{cc}
  \subfloat[wCDM] {\includegraphics[width=.5\textwidth]{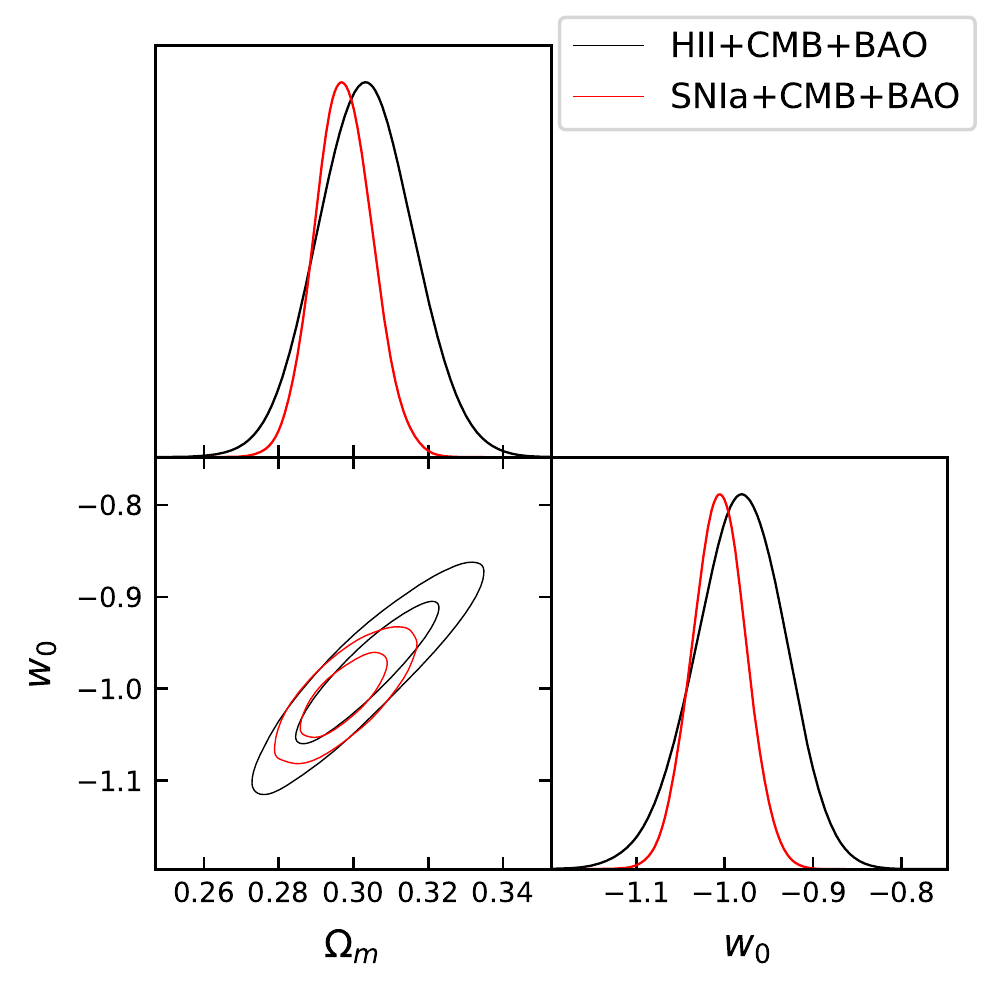}} &
  \subfloat[CPL] {\includegraphics[width=.5\textwidth]{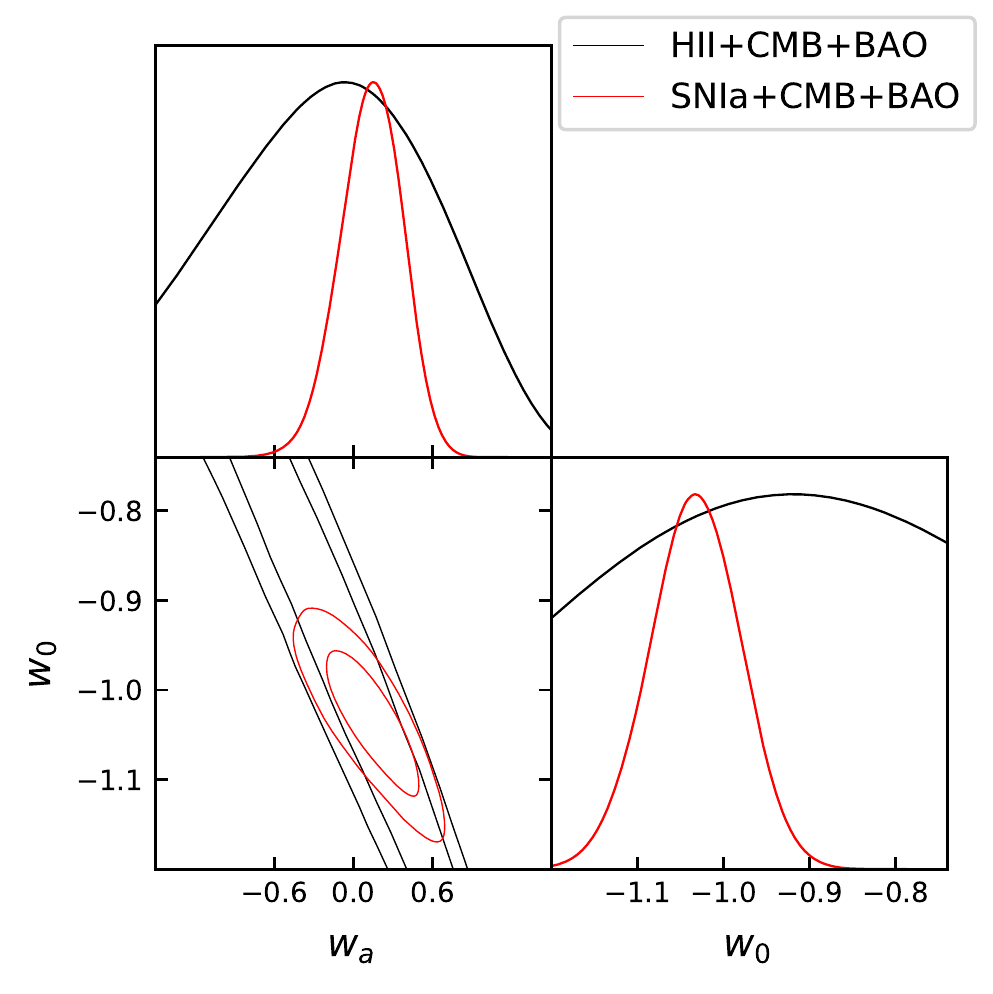}}
\end{array}$
\end{center}
	\caption{Comparison of the joint likelihood contours of the HIIG/CMB/BAO (black contours) and of the SNIa/CMB/BAO (red contours) probes. Left-hand panel: wCDM DE EoS parametrization (\{$h$, $\Omega_m$, $w_0$\}). Right-hand panel: CPL DE EoS parametrization (\{$h$, $\Omega_m$, $w_0$, $w_a$\}).}
\label{fig:j2}
\end{figure*}

\subsection{Joint analysis}

With a view to place tighter constraints on the parameter space of the DE EoS, the different cosmological probes described above were combined through a joint likelihood analysis given by the product of the individual likelihoods according to:
\begin{equation}
{\cal L}_{Tot}({\theta})= \prod_{i=1}^{n} {\cal L}_i({\theta})
\label{eq:overalllikelihood}
\end{equation}
where $n$ is the total number of cosmological probes
used\footnote{Likelihoods are normalised to their maximum
values.}. 

In Figure \ref{fig:j1} panel (a) we show the joint analysis for HIIG  \cite[with][extinction correction]{Gordon2003}, CMB and BAOs; in panel (b) the joint analysis for SNIa, CMB and BAOs for the space $\{\Omega_m, w_0\}$. In Table \ref{tab:par} we also show the results for the space $\{\Omega_m, w_0, w_a\}$. In all cases the solutions are calculated using the MCMC MultiNest algorithm with the priors shown in Table \ref{tab:priors}.

From Figure \ref{fig:j1} and Table \ref{tab:par} it is clear that the solution space of HIIG/CMB/BAO, although less constrained, is certainly compatible with the solution space of SNIa/CMB/BAO, which also can be appreciated from the results shown in Table \ref{tab:par}.

In Figure \ref{fig:j2} we show the joint likelihood contours for HIIG/BAO/CMB (black contours) and SNIa/BAO/CMB (red contours) probes for wCDM and CPL DE EoS parametrizations. It is clear that, for both parametrizations, the HIIG/BAO/CMB and SNIa/BAO/CMB joint probes agree with each other although the later produces better constraints, which is expected given the much larger number of SNIa (740) compared with the number of HIIG (153) analyzed.

\begin{table*}
	\caption{Results obtained for the different samples and their combinations, only statistical uncertainties are considered. 
	Values in parenthesis have been left fixed. Results for both extinction corrections are shown for the HIIG;  (c) \citet{Calzetti2000} extinction correction ($\alpha = 33.11 \pm 0.145$ and $\beta =  5.05 \pm 0.097 $) or (g) \citet{Gordon2003} extinction correction ($\alpha = 33.268 \pm 0.083$ and $\beta =  5.022 \pm 0.058 $).}
\label{tab:par}
\resizebox{\textwidth}{!}{
\begin{tabular}{lccccccccccc}
\hline
\hline
	Samples       & $\alpha$ & $\beta$ & $M_B^1$ & $\Delta_{M}$ & $\omega_b$ & $h$ & $\Omega_m$ & $w_0$ & $w_a$  & N & $\chi^2$ \\
\hline
HIIG$_c$ (ChiSq)  & --- & --- & --- & --- & --- & --- & $0.219^{+0.072}_{-0.057}$ & (-1.0) & (0.0) & 153 & 171.35\\
HIIG$_c$ (ChiSq)  & --- & --- & --- & --- & --- & --- & $0.201^{+0.120}_{-0.084}$ & $-0.926^{+0.303}_{-0.450}$ & (0.0) & 153 & 171.34\\
\hline
HIIG$_g$ (ChiSq)  & --- & --- & --- & --- & --- & --- & $0.276^{+0.066}_{-0.054}$ & (-1.0) & (0.0) & 153 & 256.22\\
HIIG$_g$ (ChiSq)  & --- & --- & --- & --- & --- & --- & $0.249^{+0.105}_{-0.081}$ & $-0.887^{+0.264}_{-0.384}$ & (0.0) & 153 & 256.19\\
\hline
HIIG$_g$ (MCMC)   & --- & --- & --- & --- & --- & --- & $0.290^{+0.056}_{-0.069}$ & (-1.0) & (0.0) & 153 & 254.99\\
HIIG$_g$ (MCMC)   & --- & --- & --- & --- & --- & --- & $0.280^{+0.130}_{-0.100}$ & $-1.12^{+0.58}_{-0.32}$ & (0.0) & 153 & 255.14\\
\hline
CMB 		  & --- & --- & --- & --- & $0.02225\pm 0.00016$ & (0.6774) & $0.3107\pm 0.0029$ & $-0.954^{+0.020}_{-0.018}$ & (0.0) & 3 & 0.02 \\
\hline
BAO  		  & --- & --- & --- & --- & (0.02225) & (0.6774) & $0.306^{+0.034}_{-0.041}$ & $-0.95^{+0.12}_{-0.14}$ & (0.0) & 6 & 1.76 \\
\hline
SNIa & $0.1411\pm 0.0064$  & $3.148\pm 0.077$ & $-19.041\pm 0.017$ & $-0.060\pm 0.012$ & --- & (0.7) & $0.247^{+0.11}_{-0.064}$ & $-0.94^{+0.23}_{-0.16}$ & (0.0) & 740 & 724.38\\
\hline
HIIG$_g$+CMB+BAO  & --- & --- & --- & --- & (0.02225) & $0.686\pm 0.015$ & $0.303\pm 0.013$ & $-0.983^{+0.054}_{-0.048}$ & (0.0) & 162 & 257.53\\
HIIG$_g$+CMB+BAO  & --- & --- & --- & --- & (0.02225) & $0.681\pm 0.025$ & $0.309^{+0.021}_{-0.024}$ & $-0.88^{+0.30}_{-0.34}$ & $-0.36^{+1.1}_{-0.73}$ & 162 & 257.56\\
\hline
SNIa+CMB+BAO & $0.1407\pm 0.0072$ & $3.265\pm 0.088$ & --- & $-0.064\pm 0.014$ & (0.02225) & $0.6930\pm 0.0086$ & $0.2975\pm 0.0077$ & $-1.007\pm 0.031$ & (0.0) & 749 & 589.61\\
SNIa+CMB+BAO & $0.1411\pm 0.0072$ & $3.270\pm 0.088$ & --- & $-0.063\pm 0.014$ & (0.02225) & $0.690\pm 0.011$ & $0.3001\pm 0.0090$ & $-1.033\pm 0.053$ & $0.13^{+0.25}_{-0.22}$ & 749 & 589.25\\
\hline
\hline
\end{tabular}}
\end{table*}

\section{Conclusions.}
\label{sec:Conclusions}

We have used the \lsigG\ distance indicator for HIIG to derive independently cosmological parameters. To this end, we present observations of a sample of 25 HIIG in the redshift range 1.3 $\leq$ z $\leq$ 2.5 obtained with the Keck-MOSFIRE spectrograph of which 15 were selected for this work. 
These were combined with another 6 high-z galaxies observed by us with VLT-XShooter \citep{Terlevich2015} and a compilation of 25 more objects from the literature. In total, we use the data for 46 high redshift and 107 local HIIG \citep{Chavez2014}, making a grand total of 153 HIIG  covering the redshift range 0.01 $\leq$ z $\leq$ 2.5. From the analysis of these data we have found the following:\\

1 - Using two fitting approaches, the $\chi^2$-minimization procedure and the MultiNest MCMC, we find $\Omega_m=0.276^{+0.066}_{-0.054}$ and $\Omega_m=0.290^{+0.056}_{-0.069}$ (stat) respectively, applying to the data   the extinction curve of \citet{Gordon2003}. The values of $\Omega_m$, determined using the two fitting approaches are within $1\sigma$ being the difference probably related to the disimilar error-weighting schemes that the two fittings methods employ.\\

2 - The constraints in $\Omega_m$ provided by the HIIG probe are shown to be consistent, although with larger errors, with those of the SNIa \citep{Suzuki2012} of $\Omega_m = 0.277^{+0.022}_{-0.021}$ (stat) for the Union 2.1 sample of 580 SNIa and with \citep{Betoule2014} of $\Omega_m=0.289 \pm 0.018$ (stat) for 740 SNIa  that took more than a decade to compile.\\

3 - To facilitate the comparison with our previous work \citep[cf.][]{Terlevich2015, Chavez2016} based on the $\chi^2$-minimization procedure, we present  our results using the \citet{Calzetti2000} extinction curve, finding ${\Omega_m, w_0}=\lbrace 0.201^{+0.120}_{-0.084}, -0.926^{+0.303}_{-0.450} \rbrace$, which have produced a significant improvement since now we obtain $\lbrace\Omega_m, w_0\rbrace$ central values that are in agreement with other analysis and with errors that are reasonably small considering the modest increase of the sample  by the inclusion of 15 new high-z HIIG.\\

4 - The constraints in the $\lbrace\Omega_m,w_0\rbrace$ plane provided by the HIIG are $\Omega_m=0.280^{+0.130}_{-0.100}$ and $w_0=-1.12^{+0.58}_{-0.32}$ (stat) using the \citet{Gordon2003} extinction curve. These are in agreement with the values obtained by \citet{Suzuki2012}: $\Omega_m=0.281^{+0.067}_{-0.092}$ and $w_0=-1.011^{+0.208}_{-0.231}$ (stat) for the Union 2.1 sample. \\

5 - A global fit of all the free parameters using simultaneously GHIIR and HIIG data provides the following results $\alpha=33.20\pm 0.14$, $\beta=5.06\pm 0.12$, $h = 0.706\pm 0.040$,  $\Omega_{m}=0.363^{+0.16}_{-0.081}$ and $w_0 = -1.03^{+0.62}_{-0.33}$. Although the errors are relatively large, the corresponding mean values are in excellent agreement with those of the most recent SNIa work by \citet{Scolnic2018} $\Omega_m=0.350\pm0.035$ and $w_0=-1.251\pm0.144$ (stat) for a sample of 1048 SNIa.\\

6 - The joint HIIG analysis with other cosmological probes (CMB and BAOs) further test the effectiveness of using HIIG as  tracers of the Hubble expansion. This analysis reduces dramatically the solution space in comparison with our previous results, providing quite stringent constraints on the $\lbrace\Omega_m,w_0\rbrace$ plane.\\

7 - In order to test the consistency of the derived cosmological constraints using HIIG compared with those of SNIa, we show the joint likelihood contours for HIIG/BAO/CMB and SNIa/BAO/CMB probes for wCDM and CPL DE EoS parametrizations. It is clear that, for both parametrizations, the HIIG/BAO/CMB and SNIa/BAO/CMB joint probes are in agreement with each other, although the later produces better constraints, which is expected given the much larger number of SNIa (740) compared with the number of HIIG (153) analysed.\\

It is very encouraging that even with the current small number of HIIG and after a total of only 7 observing nights, our analysis provides independent constraints on the cosmological parameters, which are in agreement with those of SNIa. This confirms our proposal that the Hubble relation for HIIG could provide an efficient tool to understand the mechanism of late cosmic acceleration.

\section*{Acknowledgements}
The authors wish to recognize and acknowledge the very significant cultural role and reverence that the summit of Maunakea has always had within the indigenous Hawaiian community.  We are most fortunate to have the opportunity to conduct observations from this mountain. ALGM and RC are grateful to the Mexican Research Council (CONACYT) for suporting this research under studentship 419392 and under grant 263561, respectively. ALGM is also grateful to Luca Ricci, support astronomer of the Keck Telescope, for his help during the observations and data reduction. RT, ET, MP and SB are grateful to the Kavli Institute for Cosmology in Cambridge  for its hospitality during a  visit in March that allowed us to conclude this work.\\  

\bibliography{bib/bibpaper2019}
\label{lastpage}

\newpage

\appendix

\section{Interlopers in the 2D spectra}
\label{sec:Others objects in the 2D spectra}

Some of  the 2D spectra show emission lines belonging to serendipitous  objects, for example the one labeled Lensed target2. In our spectrum we observe the He1$\lambda$5876\AA\ emission line confirming the reported redshift of z=1.847 given in \citep{Brammer2012}.\\

Another interesting object appears in the 2D spectrum of ZCOSMOS-411737; here we  detected  H$\alpha$ and [NII]$\lambda \lambda $6548,6584\AA\  lines at z=1.52.\\

In the 2D spectrum of COSMOS-19049, we  detected emission lines corresponding to 4 objects at different redshifts. The first one presents  [O III]$\lambda\lambda$5007,4959\AA\ and H$\beta$  at z=2.089. The second one shows only [O III]$\lambda\lambda$5007,4959\AA\ at z=2.463. The third one present only one emission line which we propose that it is H$\alpha$ at z=1.405 although we do not rule out the possibility of it being [O III]$\lambda$5007\AA\ at z=2.153. In this case   [O III]$\lambda$4959\AA\  would fall on a sky emission line, and H$\beta$ would not have been detected if it was very faint. For the target object (COSMOS-19049 at z=1.369) we detect H$\alpha$,  [N II] $\lambda\lambda$6548,6584\AA\ and [S II]$\lambda\lambda$6717,6731\AA\ .\\

In the 2D spectra of 3D-HST104245, UDS-109082, UDS23 and UDS-4501 we clearly  see the continuum of other sources.

A faint emission line  at $\lambda_{obs}$=16573.898 \AA\ is seen in the 2D spectrum of UDS-14655.  This line could be H$\alpha$ at  z=1.525; however this identification is uncertain because it looks different to other  H$\alpha$  emitting objects at such  redshift. \\

In the 2D spectrum of UDS-113972 we see an intense continuum and wide emission lines  H$\alpha$, [N II] $\lambda\lambda$6548,6584\AA\ and [S II]$\lambda\lambda$6717,6731\AA\ corresponding to an object at z=1.368.

\section{MOSFIRE data}
\label{sec:MOSFIRE spectra}
In this appendix we show for each HIIG the fits to the emission lines. We fitted H$\alpha$ for the objects at z $\sim$ 1.5 and \mbox{[O III]$\lambda\lambda$4959,5007 \AA} and H$\beta$ for the objects at z $\sim$ 2.3.\\

In the first and third rows (for objects at z $\sim$ 1.5) or in the first row (for objects at z $\sim$ 2.3), we present the 1D and 2D spectra.\\

In the second and fourth rows (for objects at z $\sim$ 1.5) or in the second, third and fourth columns (for objects at z $\sim$ 2.3), we show the fits to the H$\alpha$ or  [$\Oiii$]$\lambda\lambda$4959,5007 \AA\ and H$\beta$ emission lines, respectively. The residuals from the fits are shown in the lower panel. 
In the inset at the upper right corner we present the FWHM distribution from the Montecarlo simulations performed to estimate their errors.

\begin{figure*}
\vspace*{-0.3cm}
\begin{center}$
\begin{array}{cc}
\hspace*{0.25cm}
   {\includegraphics[width=.378\textwidth]{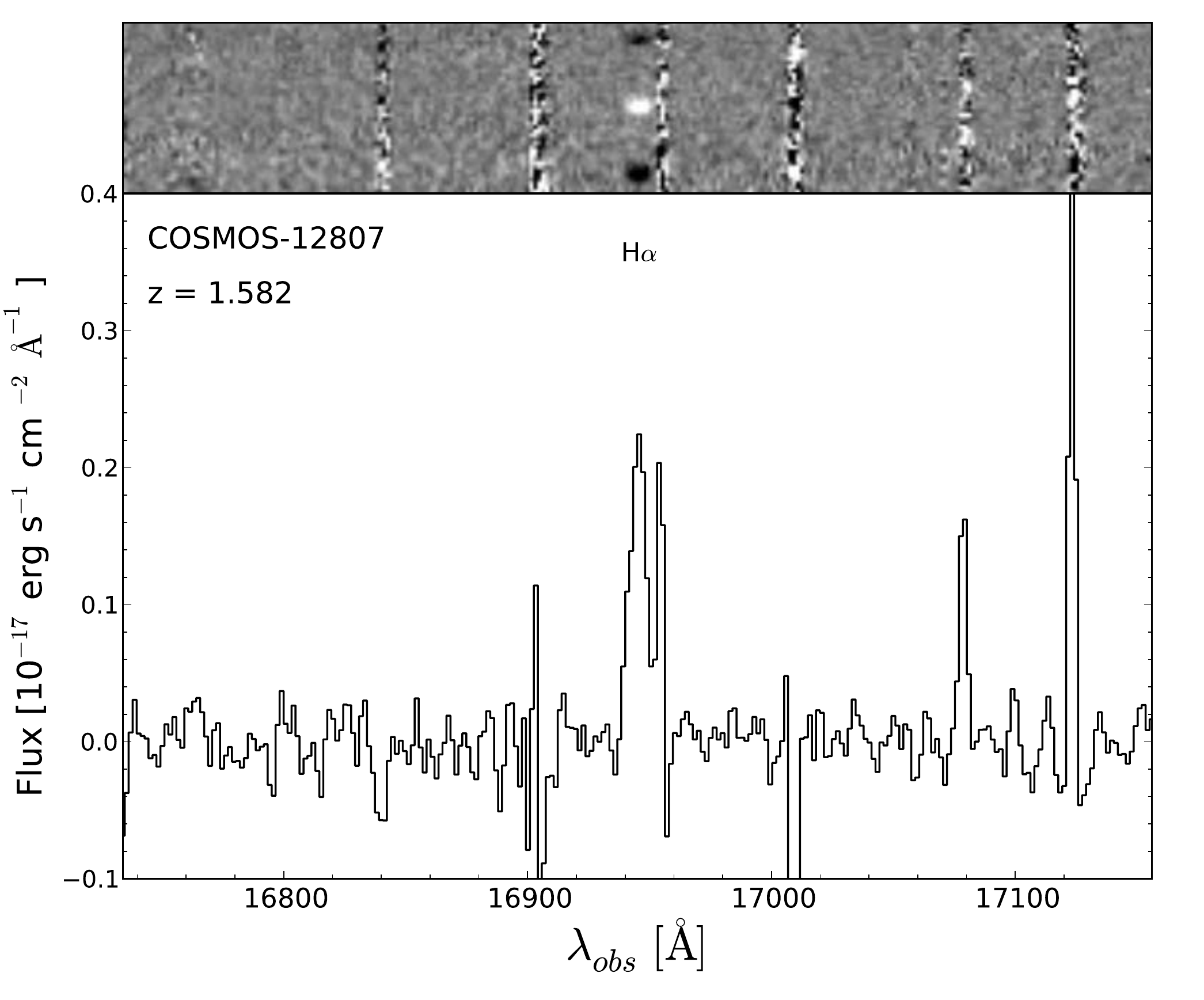}} &   
   {\hspace*{0.6cm}\includegraphics[width=.373\textwidth]{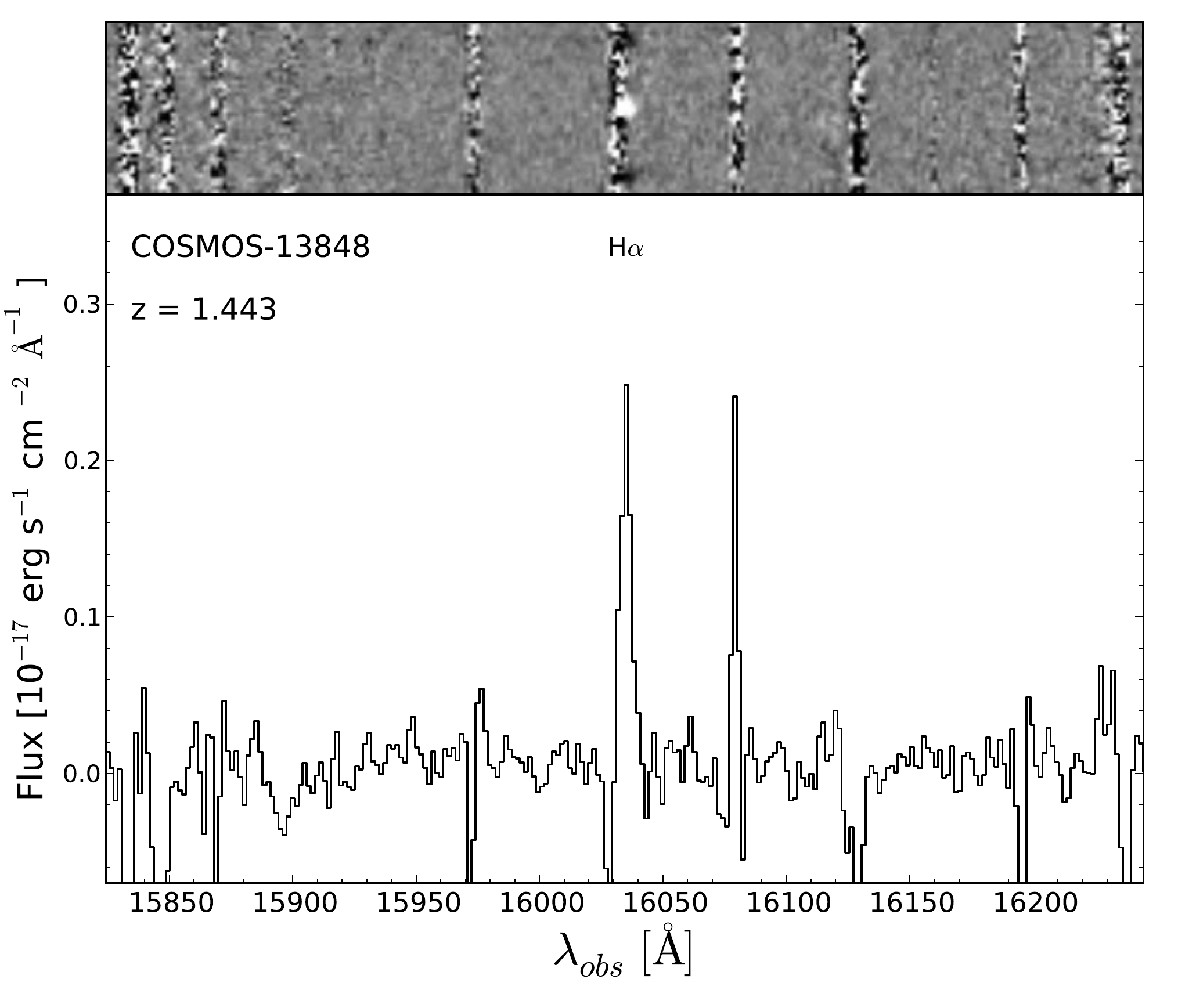}}
\end{array}$
\end{center}

\vspace*{-0.7cm}
\begin{center}$
\begin{array}{cc}
   {\hspace*{-0.15cm}\includegraphics[width=.395\textwidth]{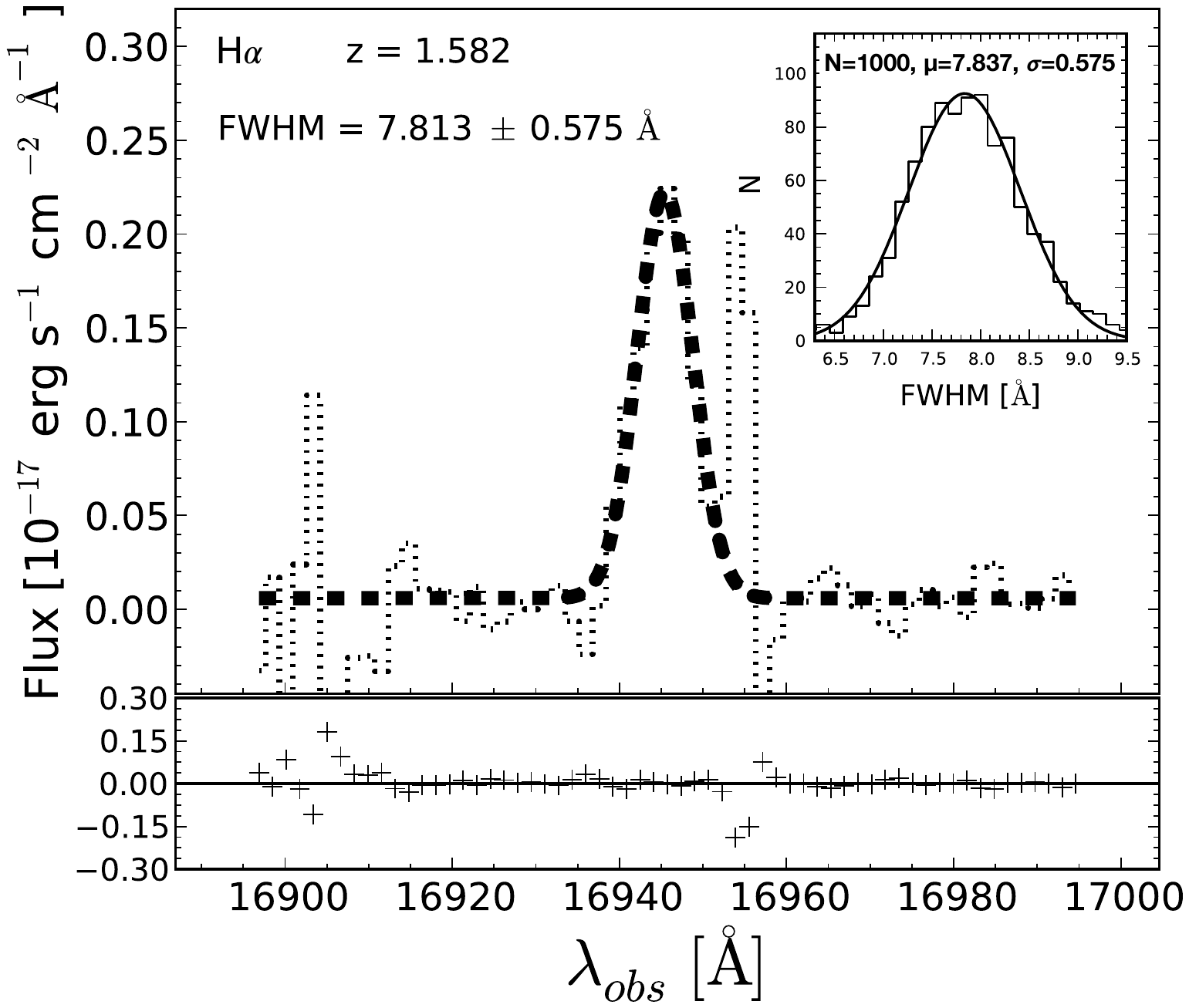}}  &
   {\hspace*{0.2cm}\includegraphics[width=.395\textwidth]{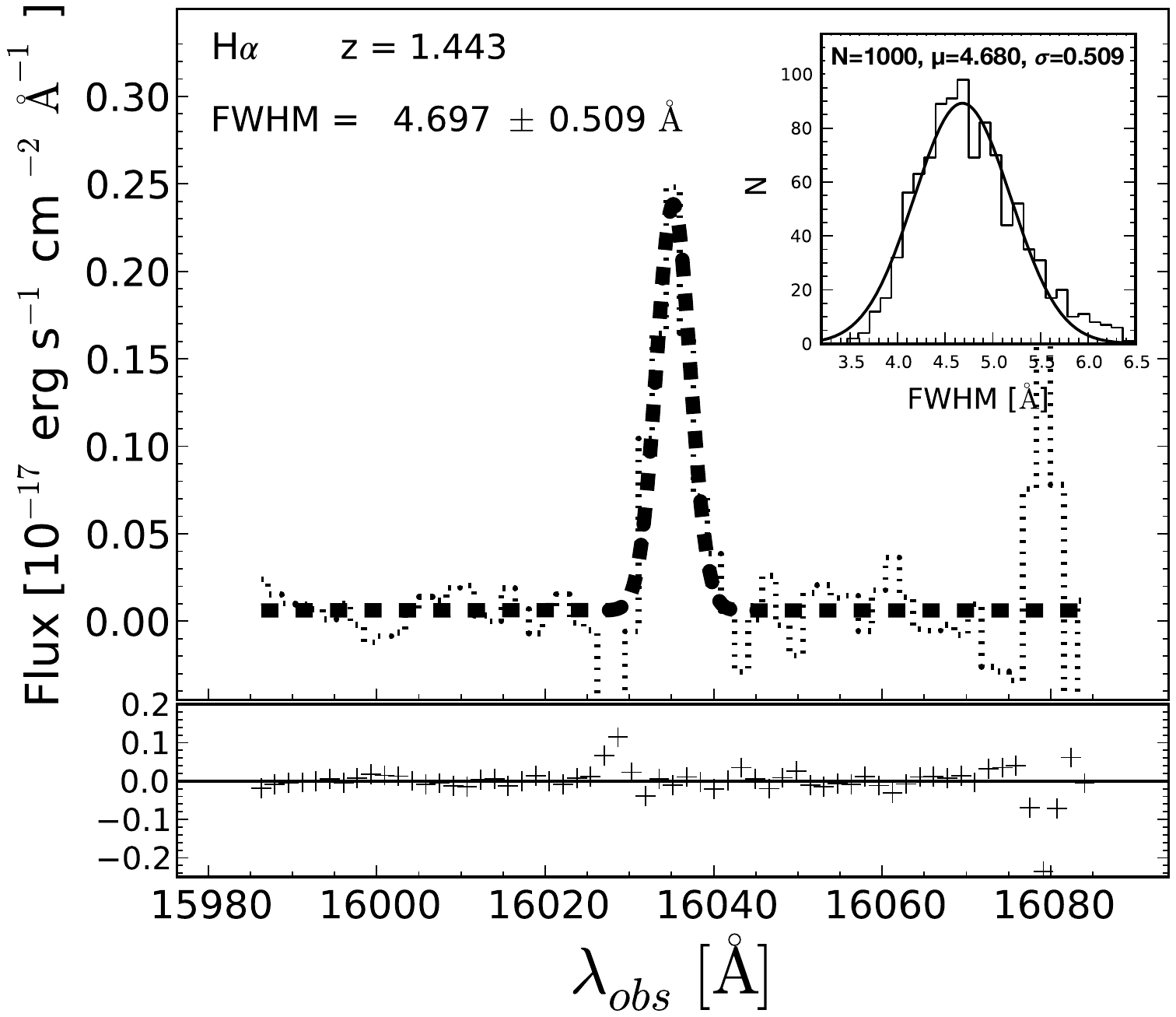}}
\end{array}$
\end{center}
\end{figure*}

\begin{figure*}
\vspace*{-0.7cm}
\begin{center}$
\begin{array}{cc}
\hspace*{0.25cm}
   {\includegraphics[width=.378\textwidth]{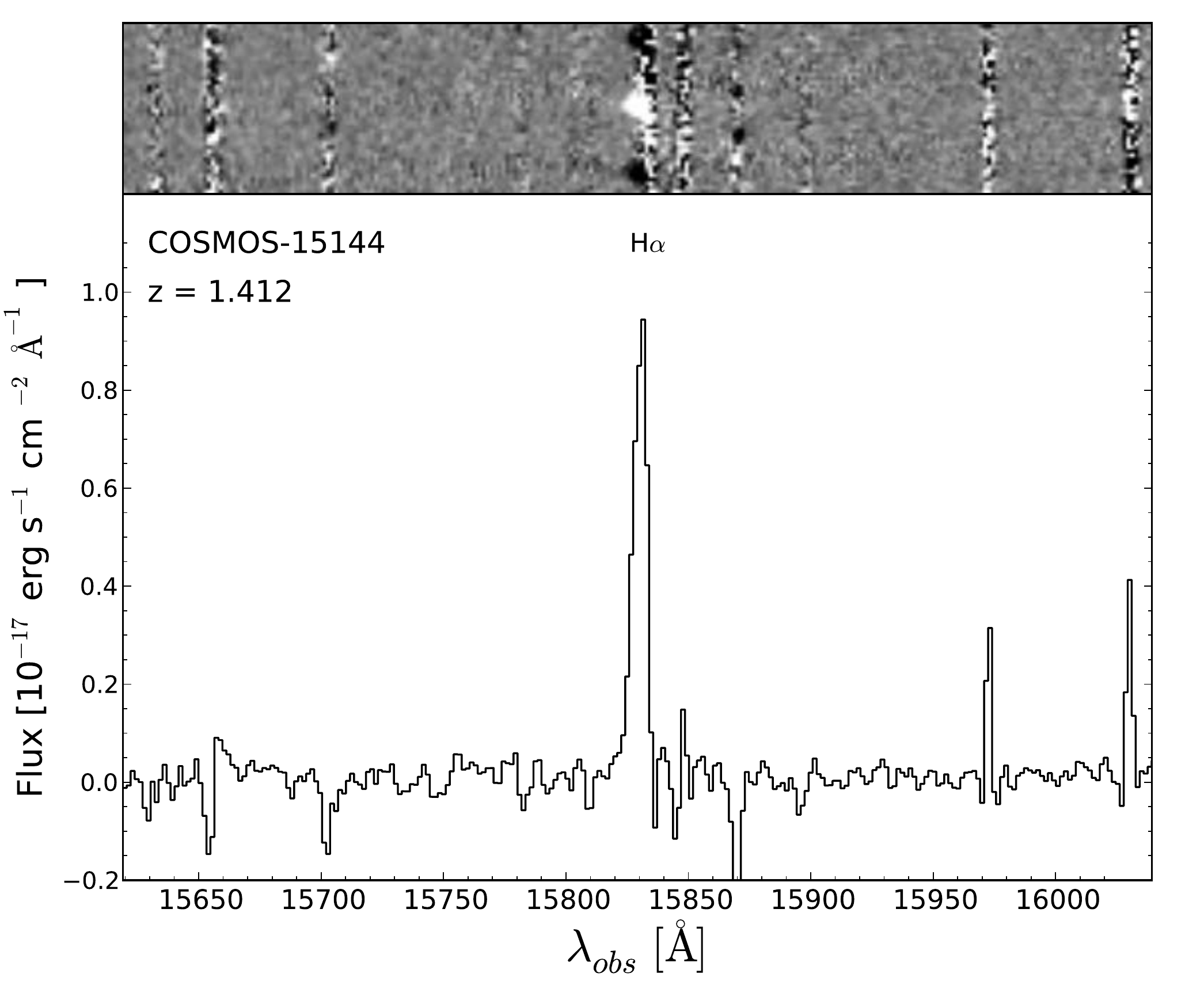}} &
   {\hspace*{0.4cm}\includegraphics[width=.378\textwidth]{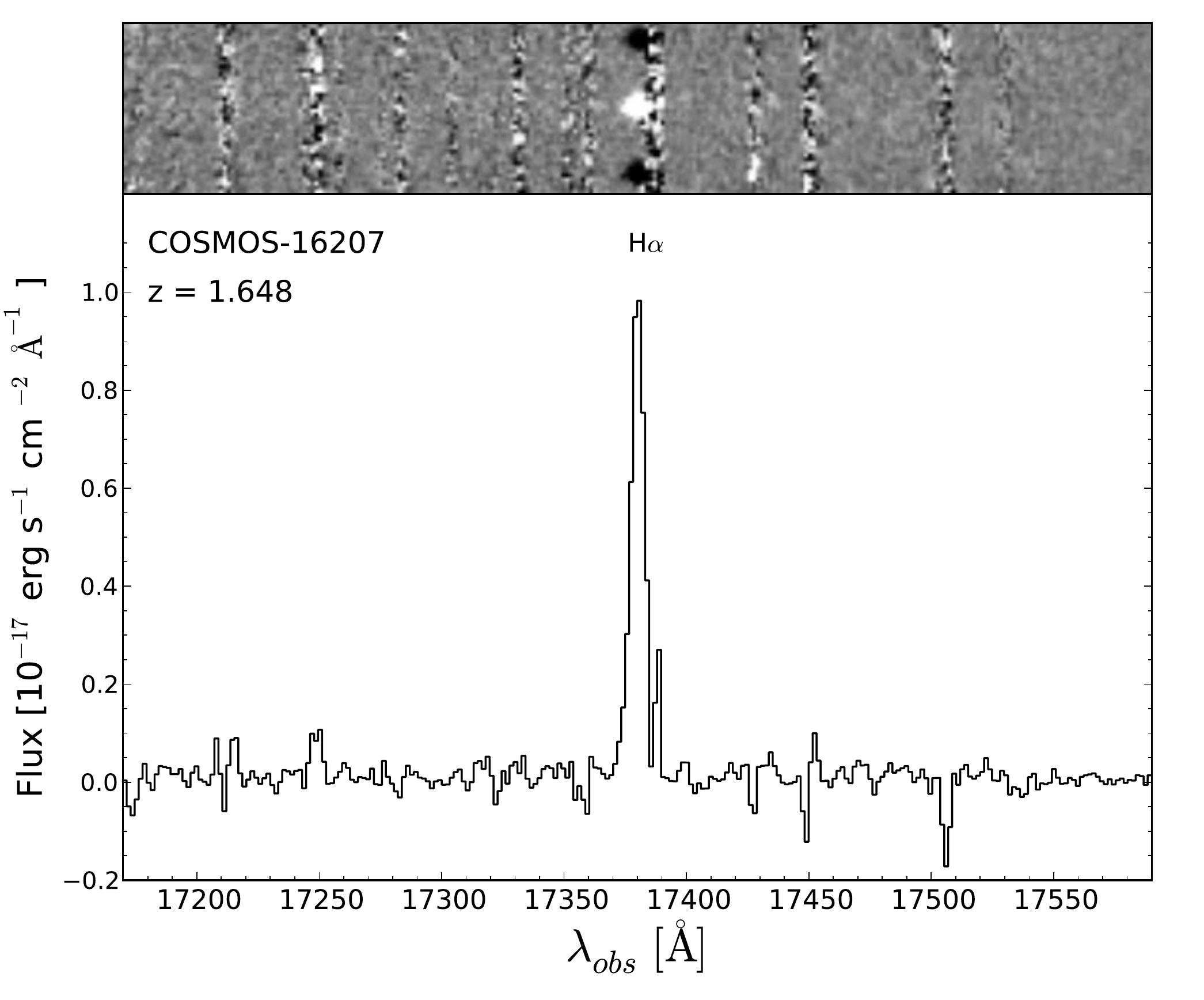}}
\end{array}$
\end{center}

\vspace*{-0.7cm}
\begin{center}$
\begin{array}{cc}
   {\hspace*{0.01cm}\includegraphics[width=.385\textwidth]{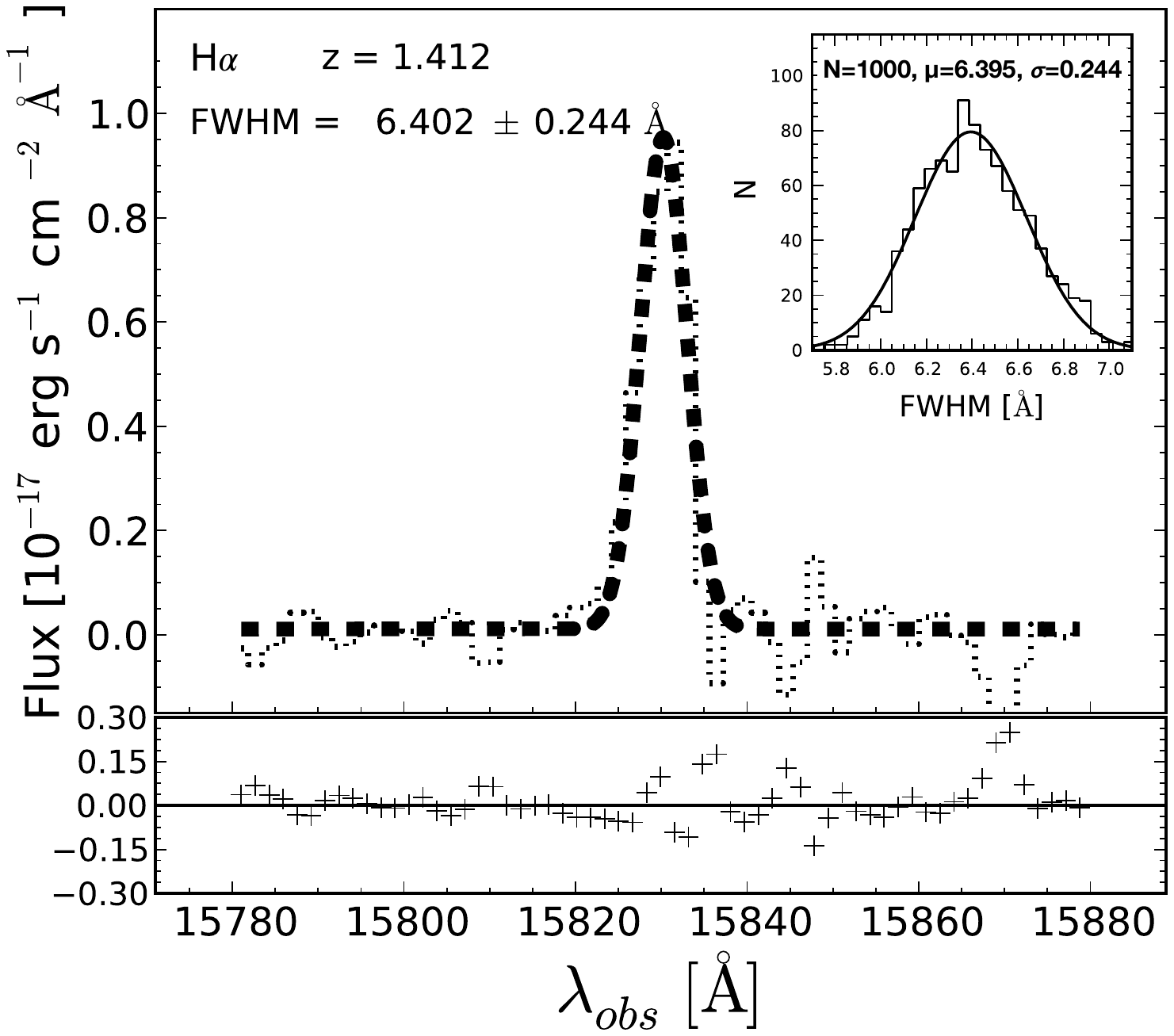}}  &
   {\hspace*{0.3cm}\includegraphics[width=.388\textwidth]{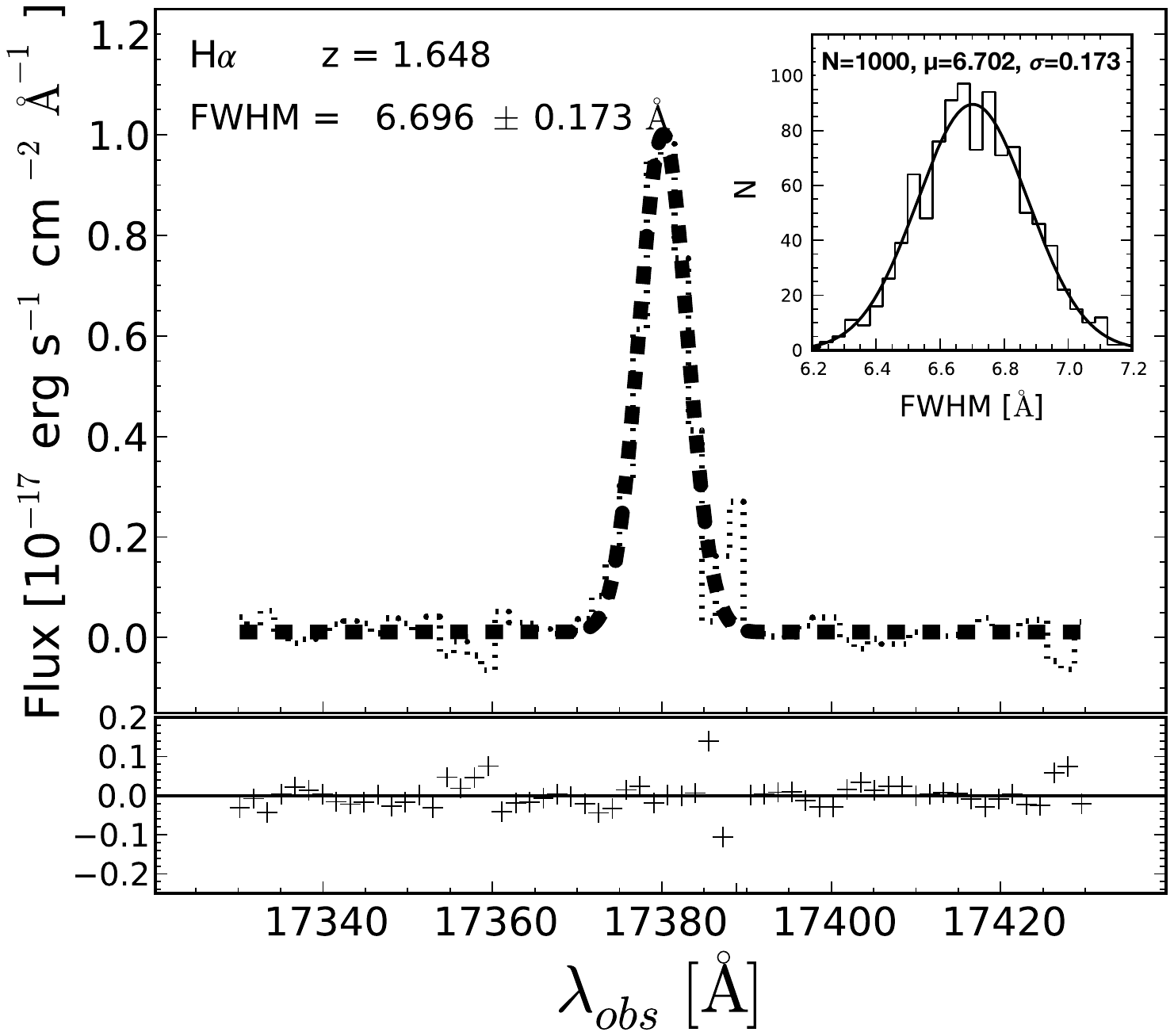}}
\end{array}$
\end{center}
\end{figure*}

\clearpage
\begin{figure*}
\vspace*{-0.3cm}
\begin{center}$
\begin{array}{cc}
\hspace*{0.4cm}
   {\includegraphics[width=.376\textwidth]{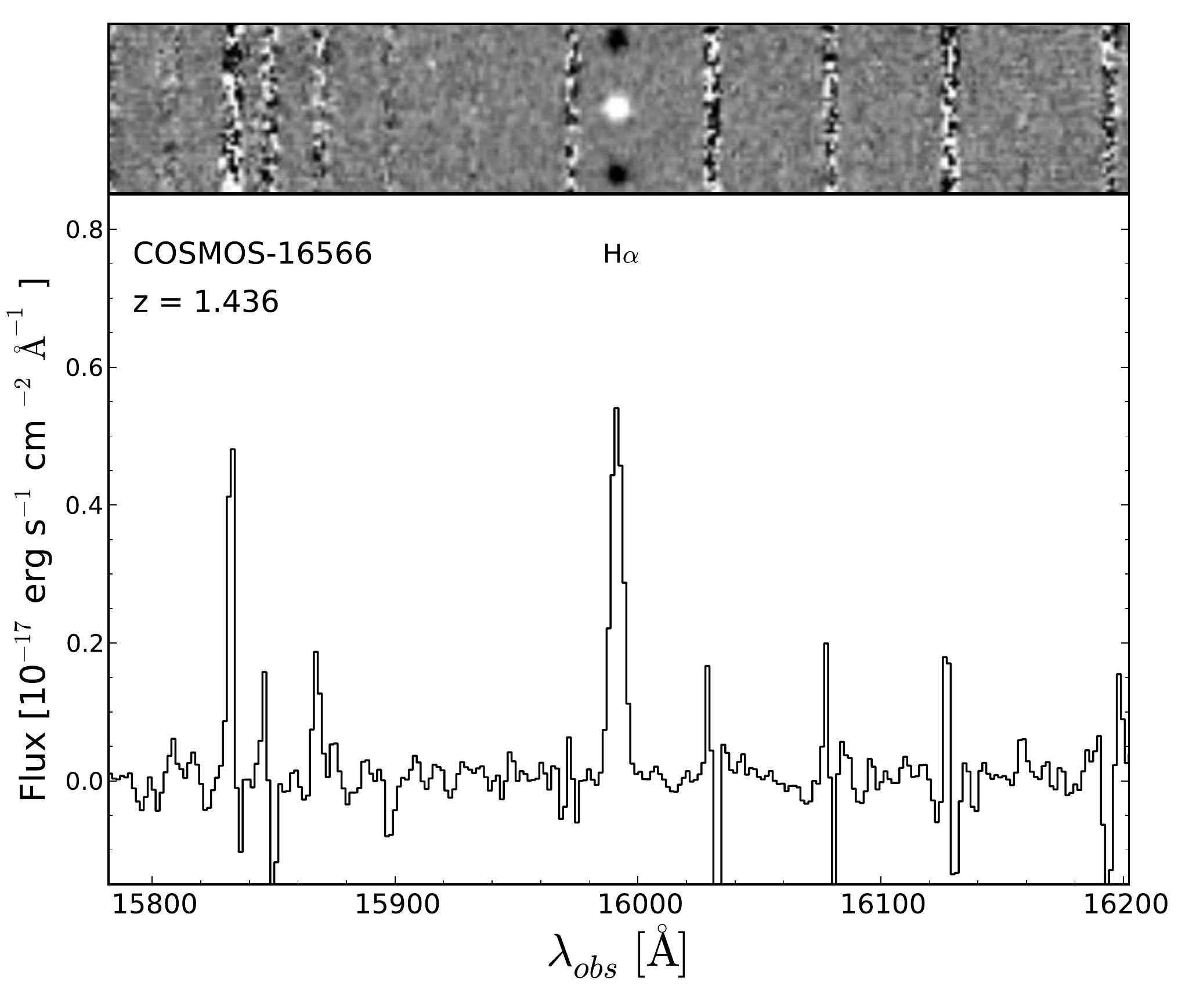}} &
   {\hspace*{0.4cm}\includegraphics[width=.378\textwidth]{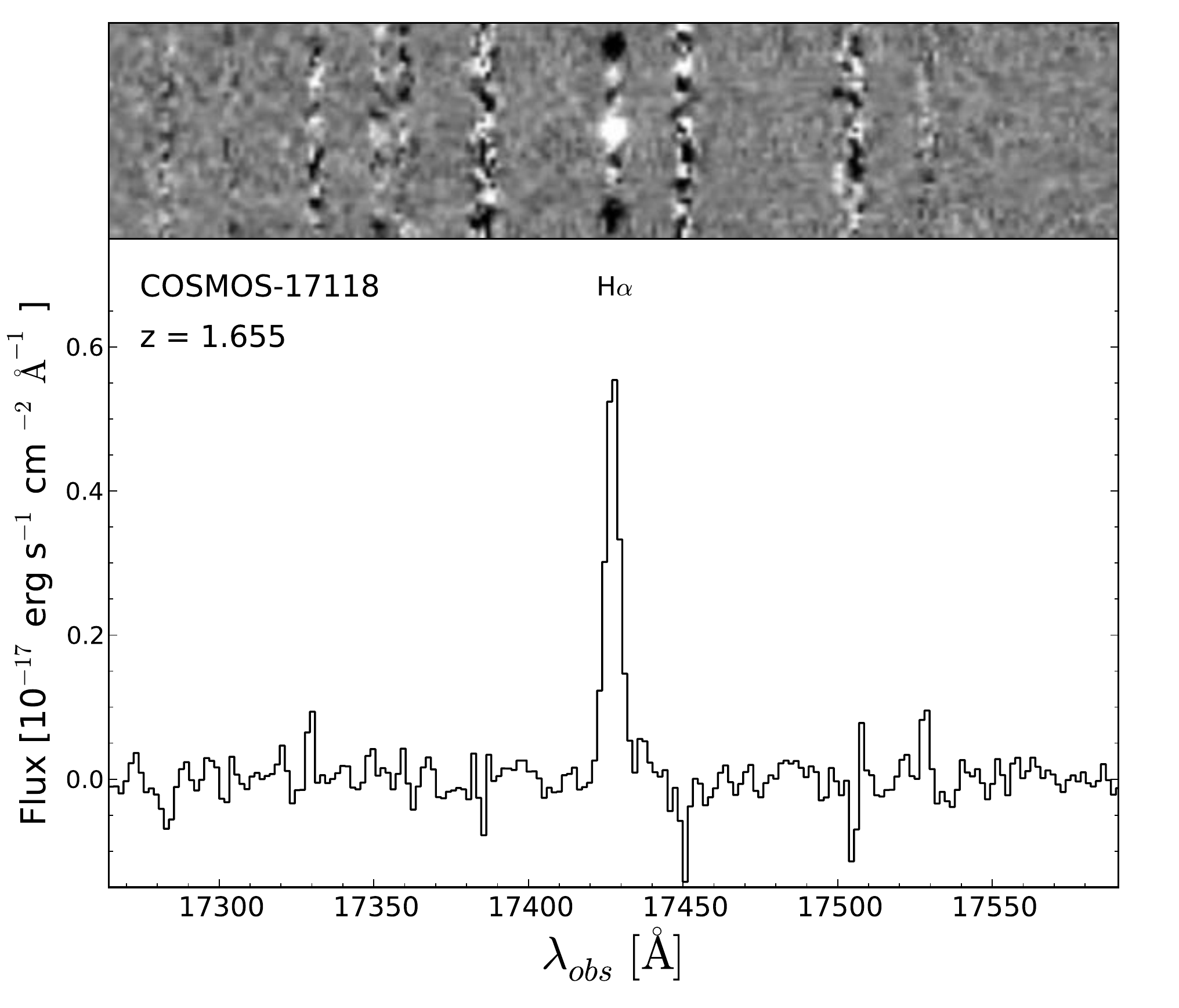}} 
\end{array}$
\end{center}

\vspace*{-0.7cm}
\begin{center}$
\begin{array}{cc}
   {\hspace*{-0.0cm}\includegraphics[width=.385\textwidth]{fig/profiles_line/COSMOS/COSMOS-16566/COSMOS16566_Ha_obs_FINAL.pdf}} 
   {\hspace*{0.6cm}\includegraphics[width=.386\textwidth]{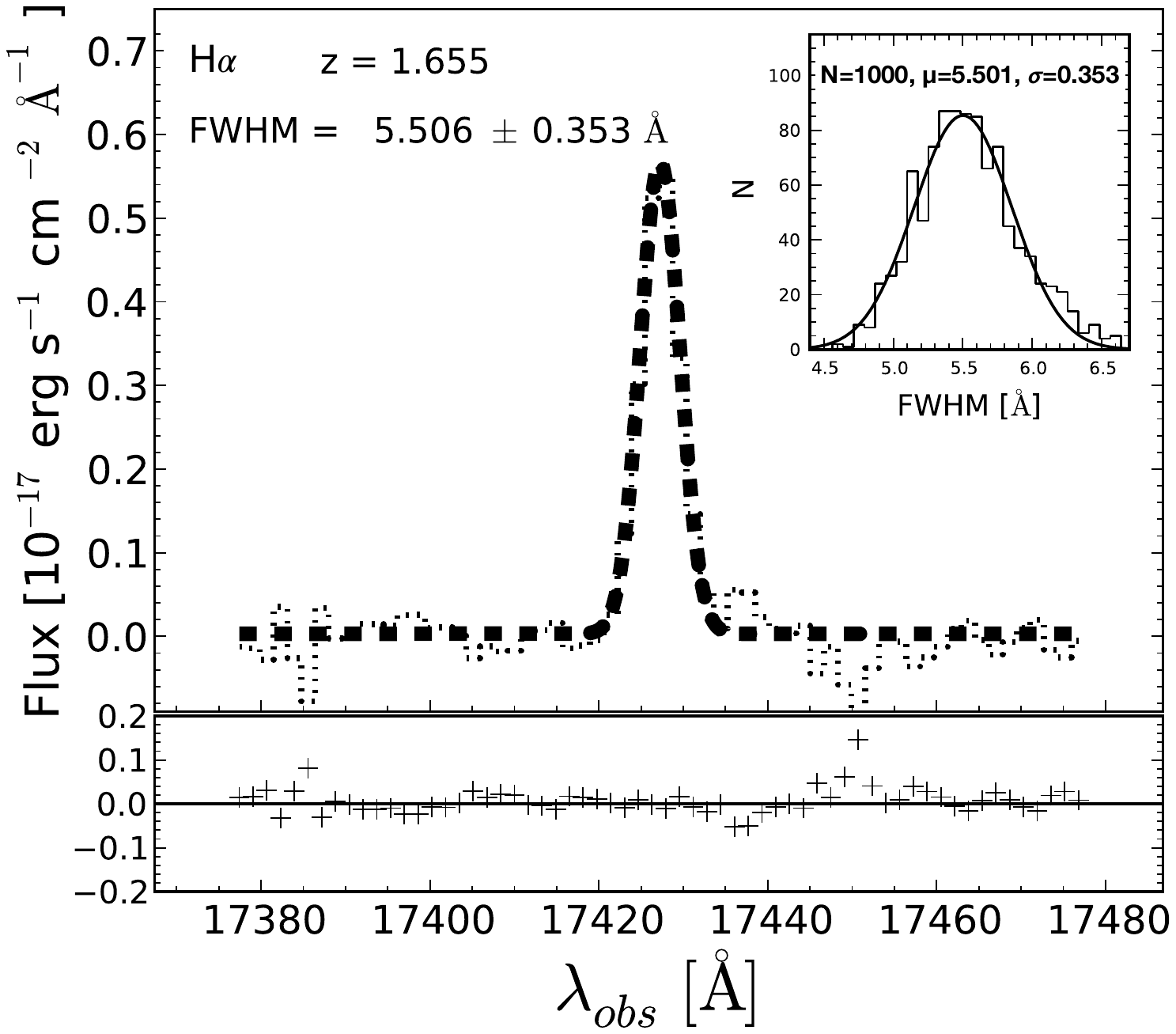}}
\end{array}$
\end{center}
\end{figure*}

\begin{figure*}
\vspace*{-0.7cm}
\begin{center}$
\begin{array}{cc}
\hspace*{0.3cm}
   {\includegraphics[width=.383\textwidth]{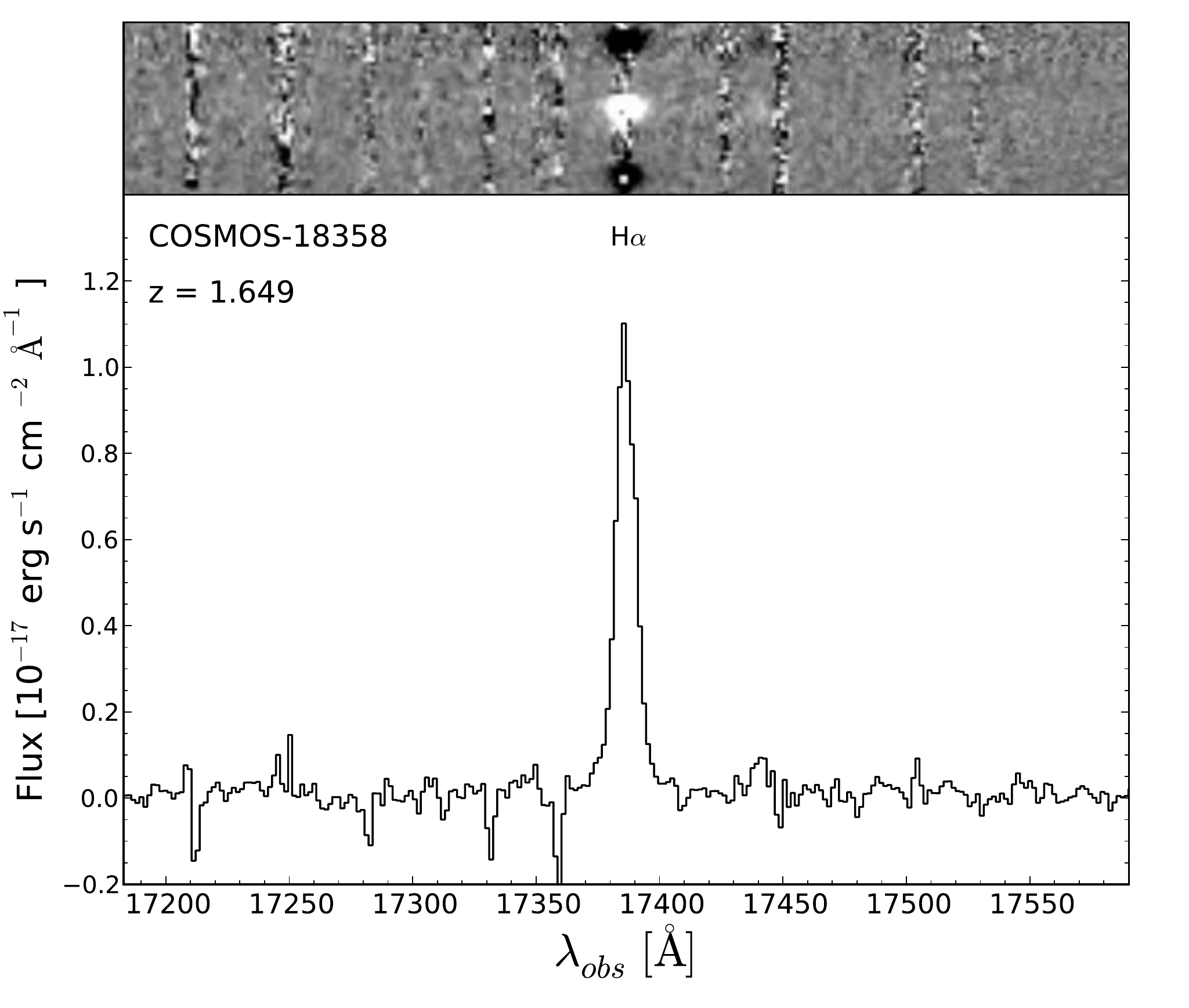}} &
   {\hspace*{0.3cm}\includegraphics[width=.383\textwidth]{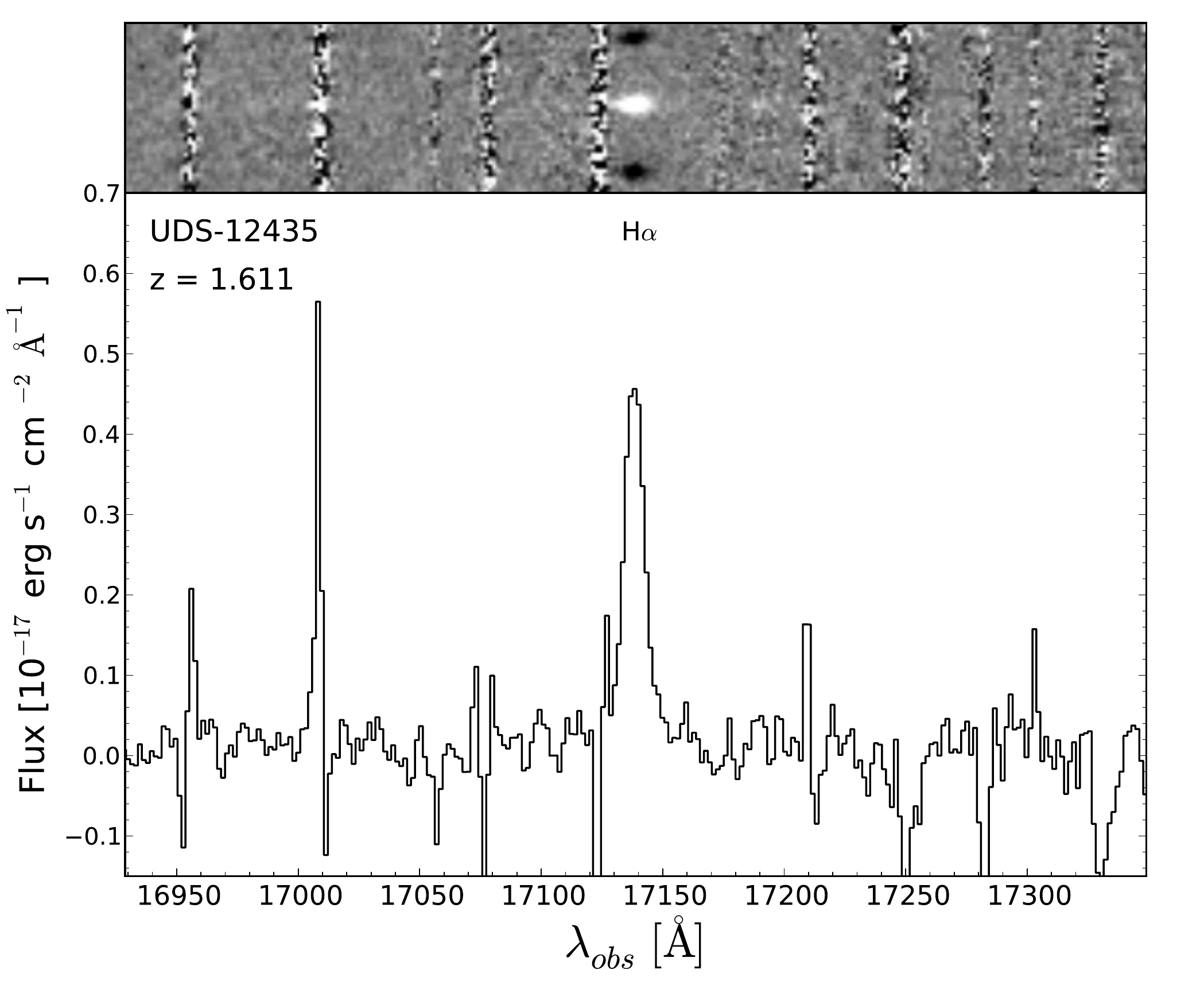}}
\end{array}$
\end{center}

\vspace*{-0.7cm}
\begin{center}$
\begin{array}{cc}
   {\hspace*{-0.08cm}\includegraphics[width=.385\textwidth]{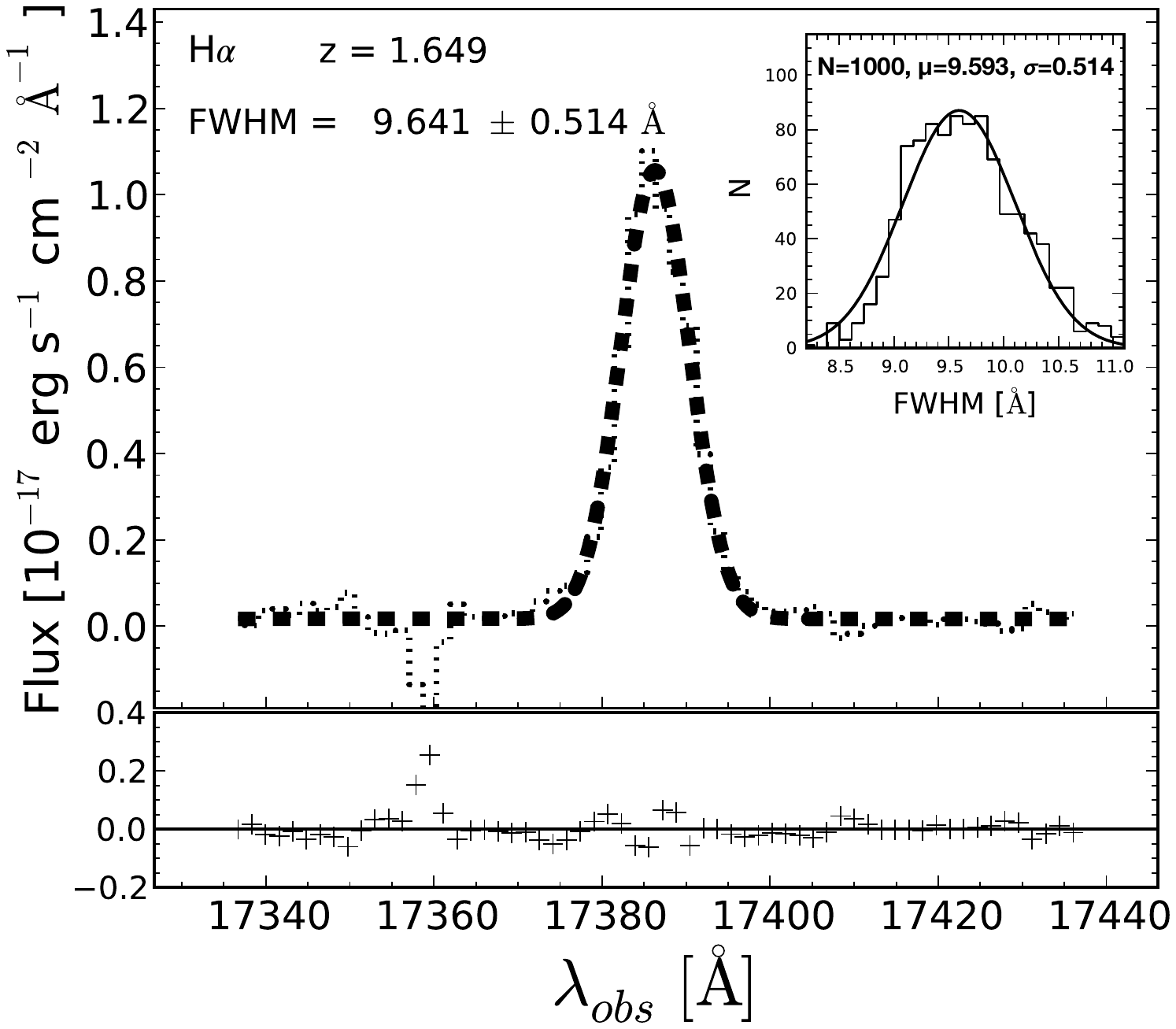}}  &
   {\hspace*{0.3cm}\includegraphics[width=.38\textwidth]{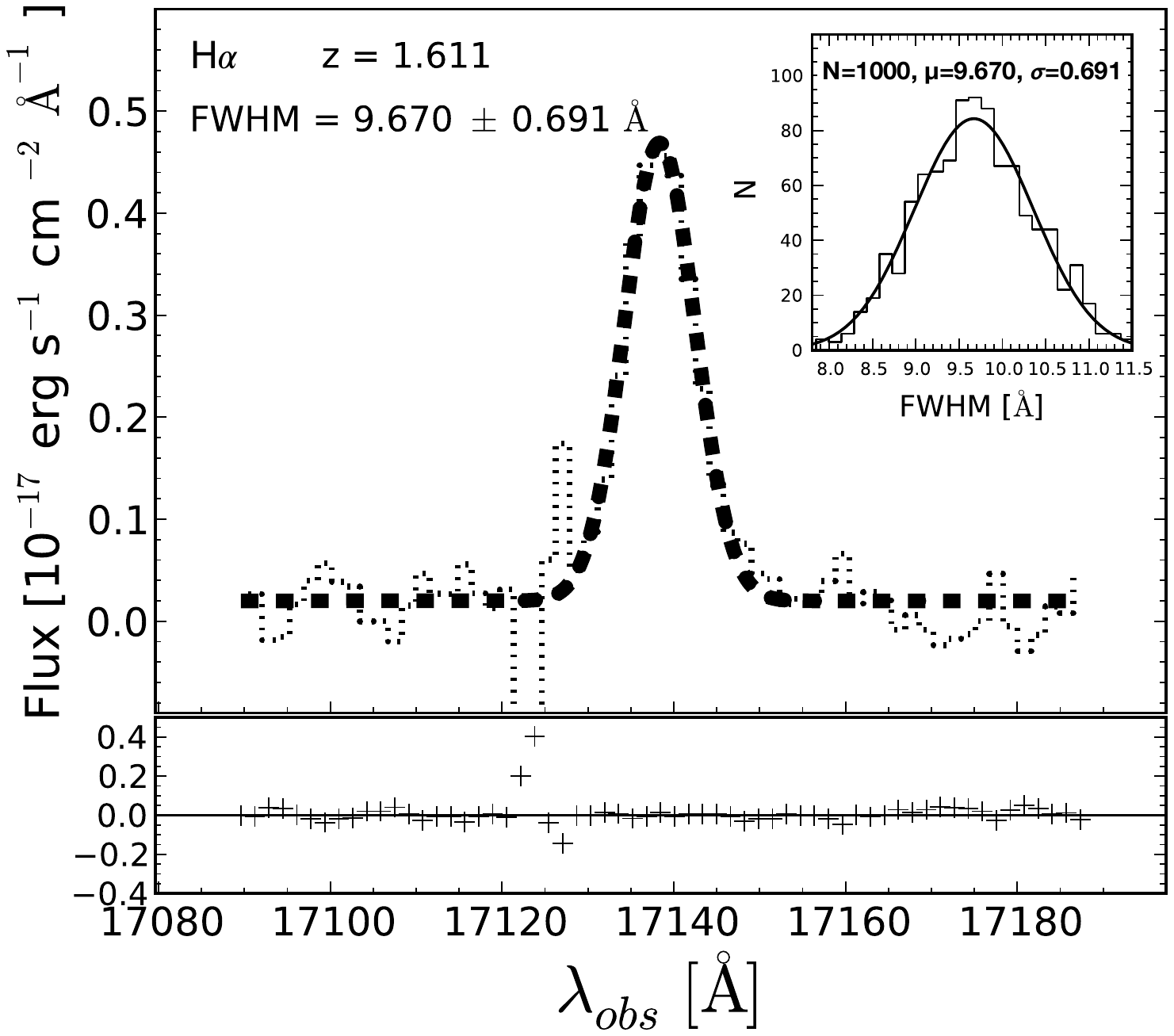}}
\end{array}$
\end{center}
\end{figure*}

\clearpage
\begin{figure*}
\vspace*{-0.2cm}
\begin{center}$
\begin{array}{cc}
\hspace*{0.25cm}
   {\includegraphics[width=.374\textwidth]{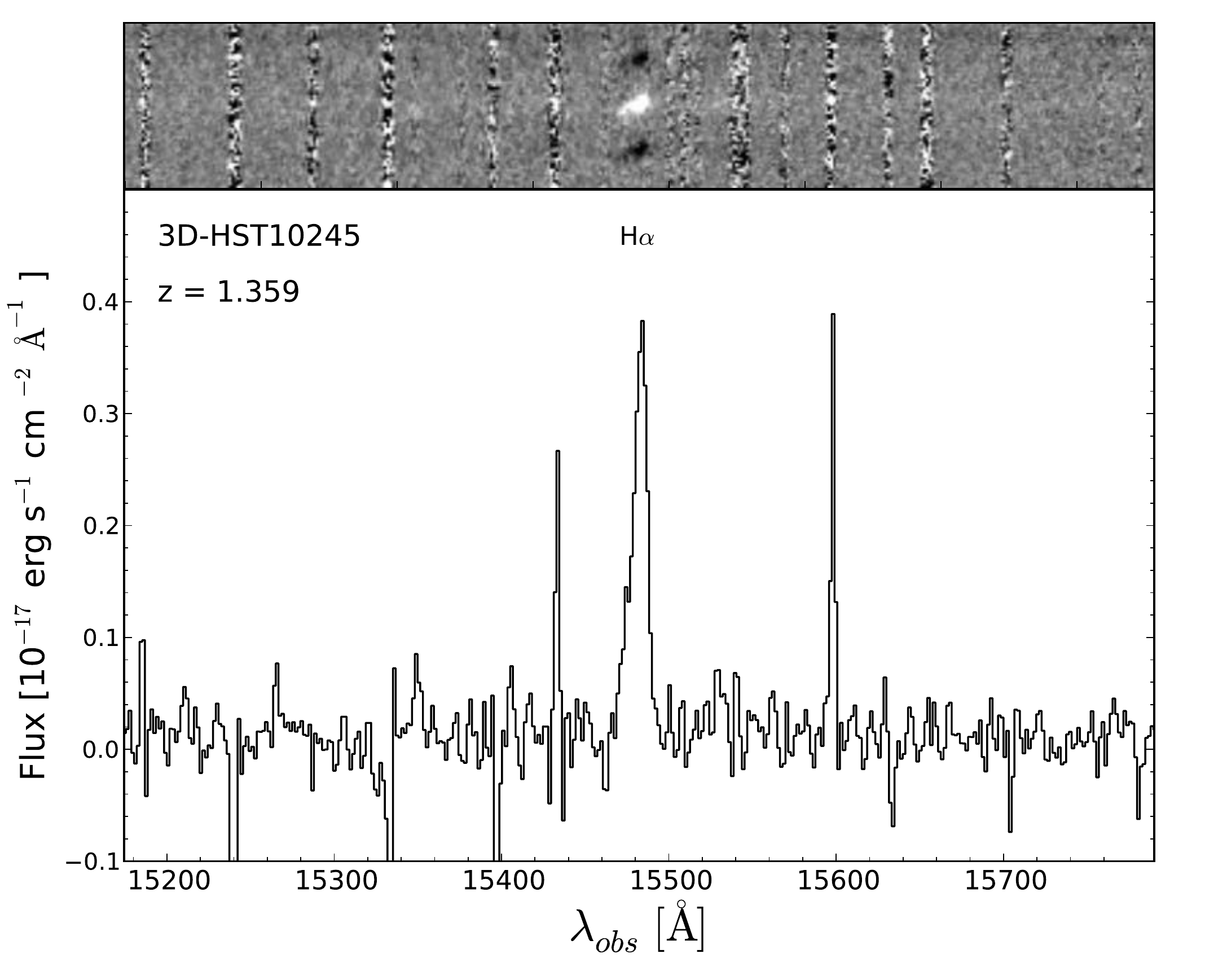}} &
   {\hspace*{0.4cm}\includegraphics[width=.374\textwidth]{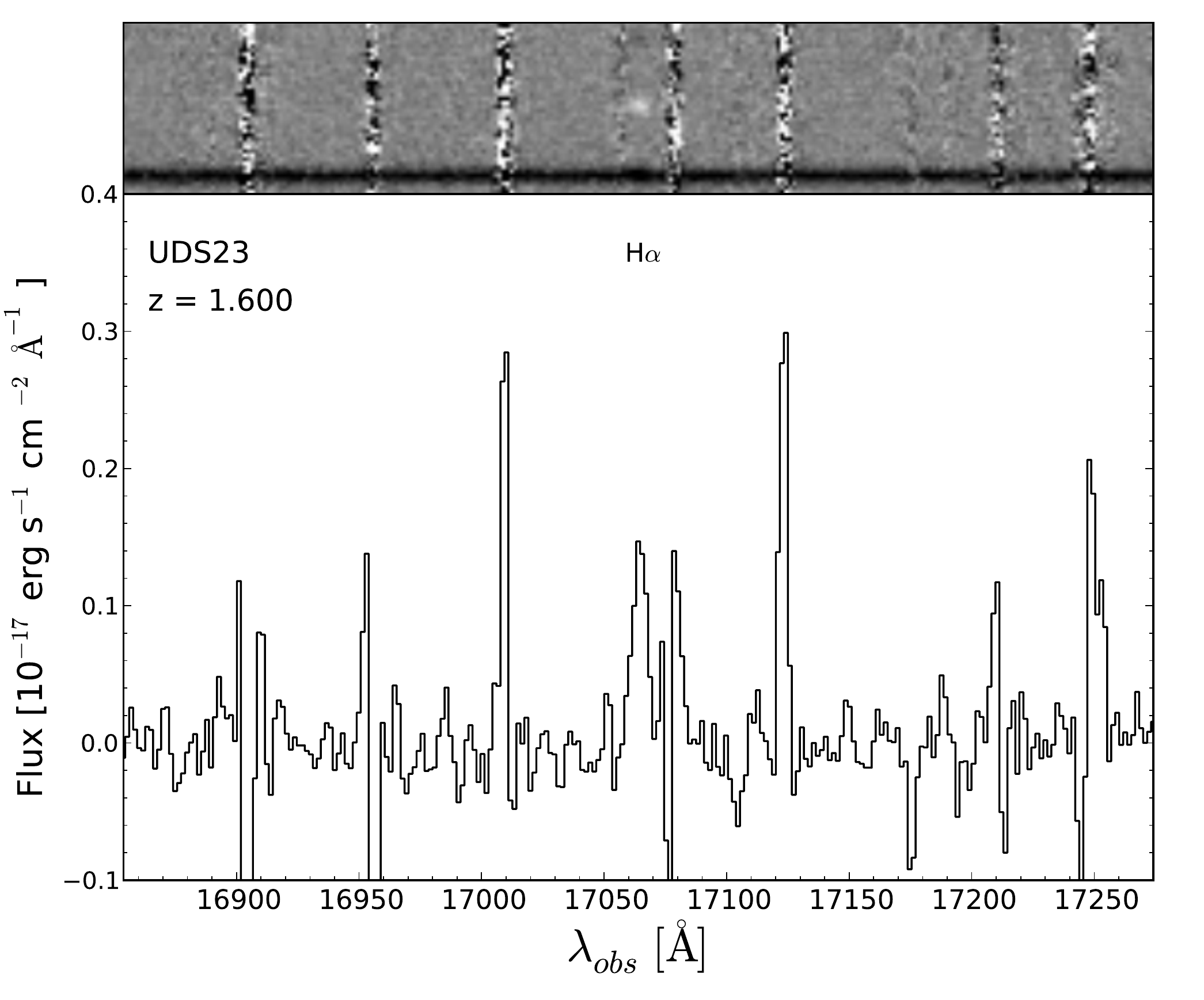}}
\end{array}$
\end{center}

\vspace*{-0.7cm}
\begin{center}$
\begin{array}{cc}
   {\hspace*{-0.03cm}\includegraphics[width=.383\textwidth]{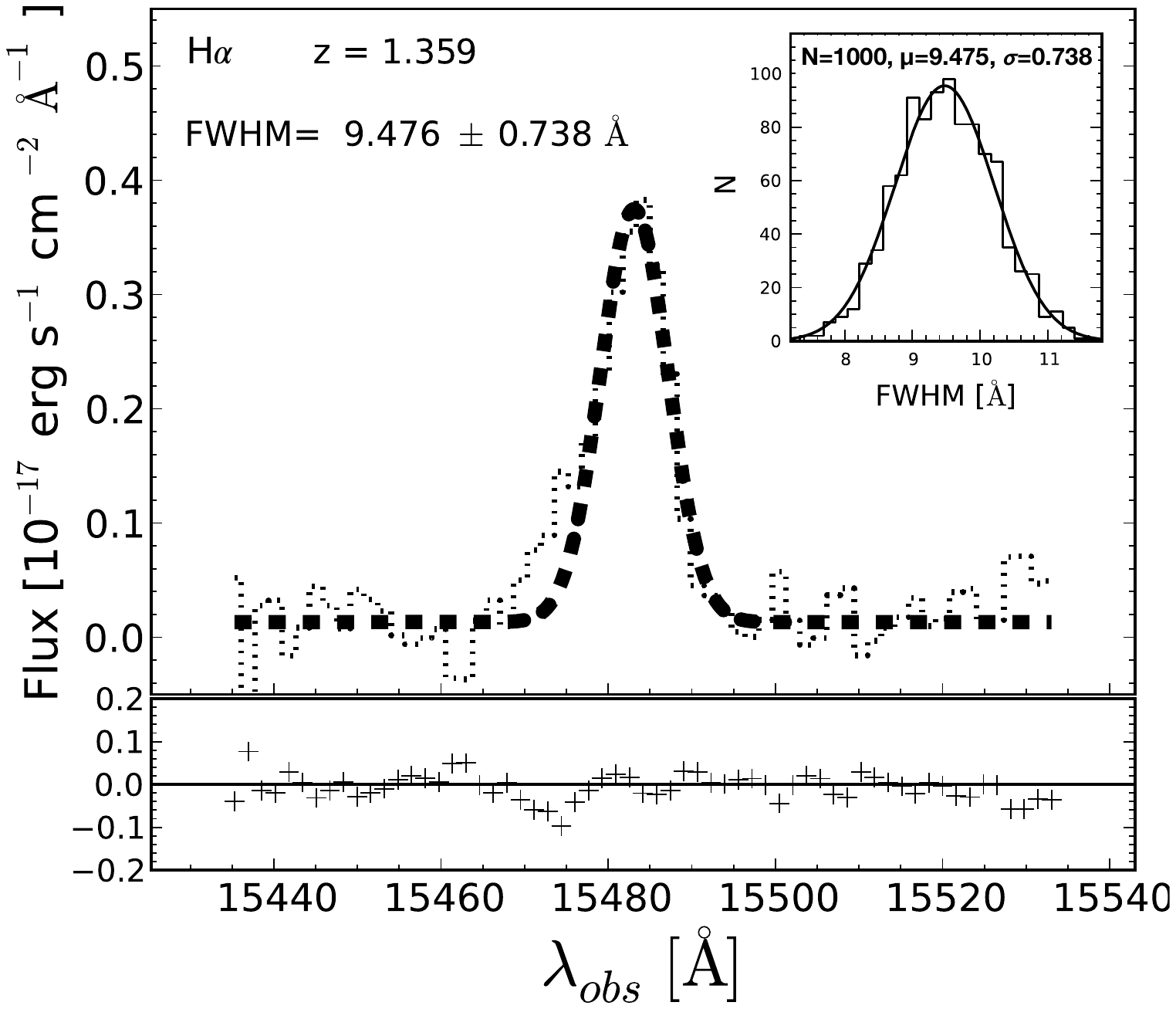}} &
   {\hspace*{0.15cm}\includegraphics[width=.39\textwidth]{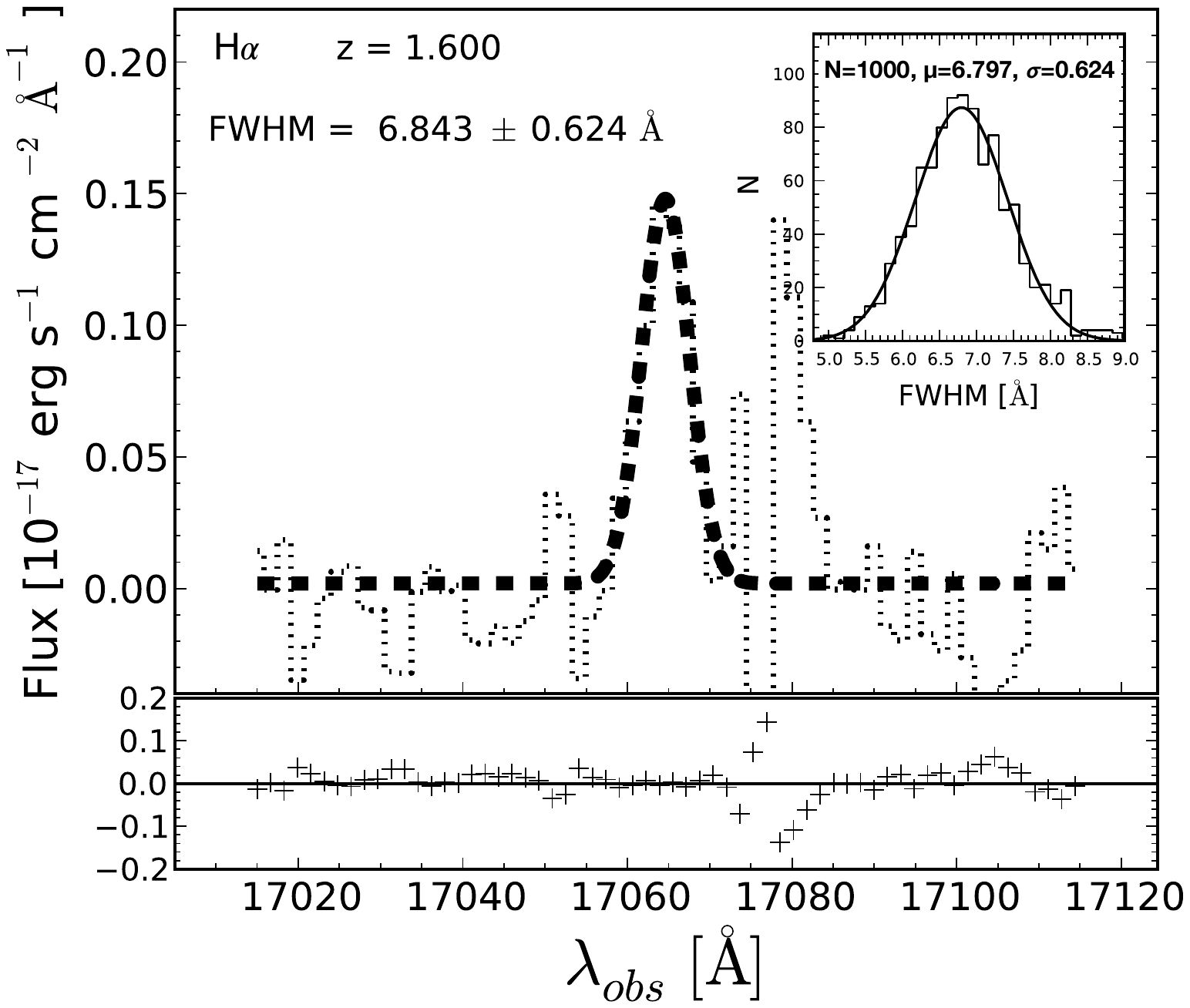}}
\end{array}$
\end{center}
\end{figure*}

\begin{figure*}
\vspace*{-0.7cm}
\begin{center}$
\begin{array}{cc}
\hspace*{0.25cm}
   {\includegraphics[width=.376\textwidth]{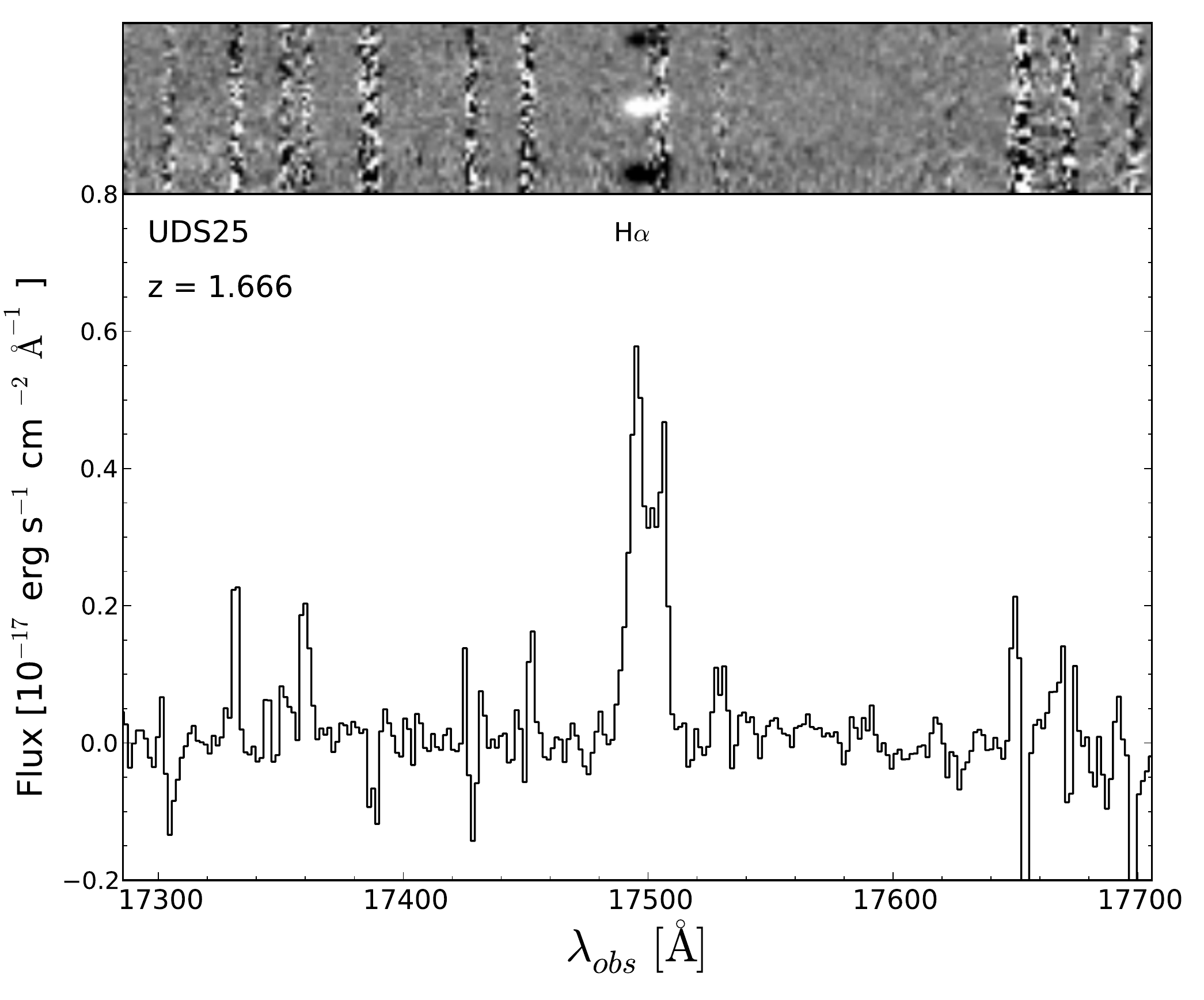}} &
   {\hspace*{0.31cm}\includegraphics[width=.383\textwidth]{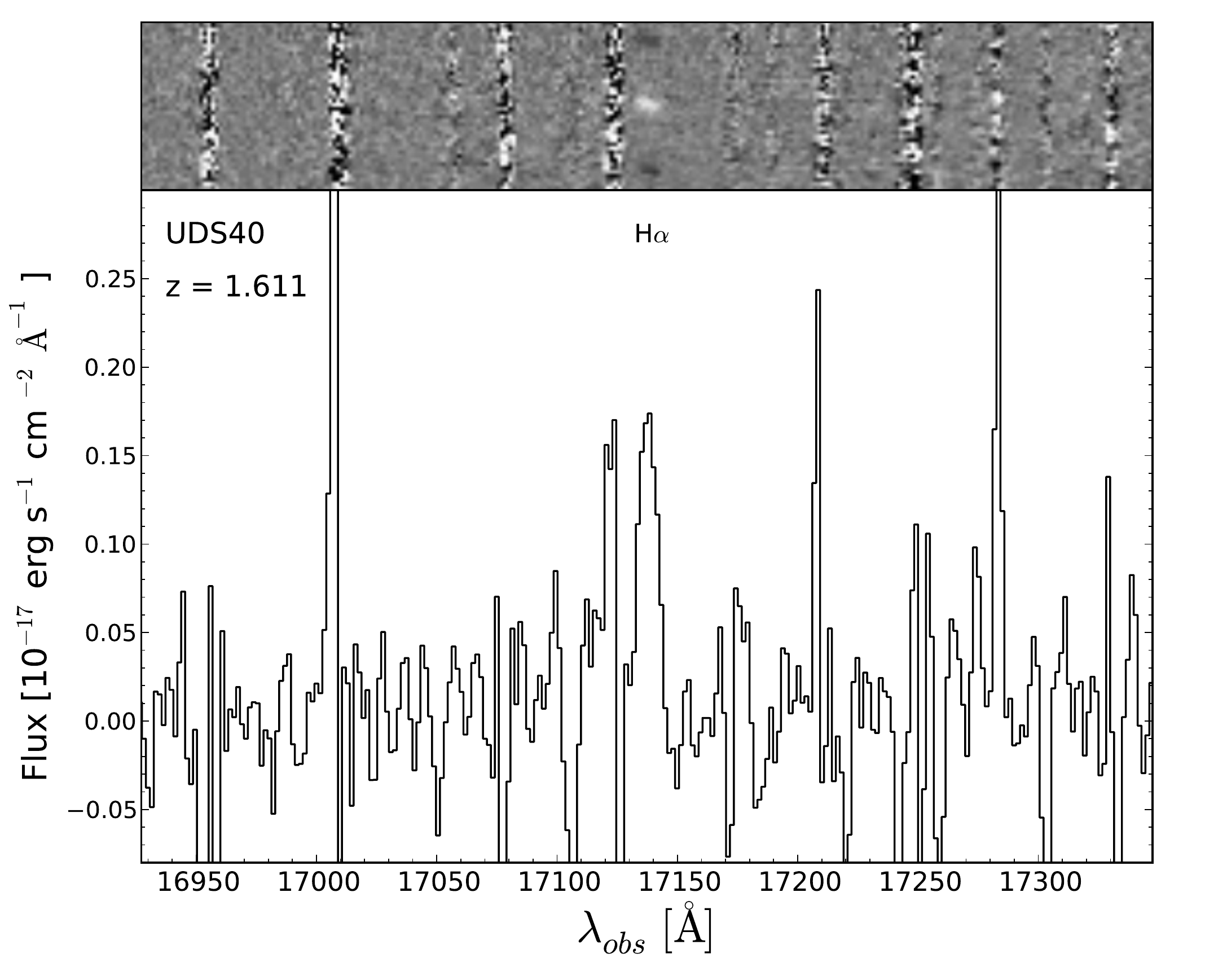}} 
\end{array}$
\end{center}

\vspace*{-0.7cm}
\begin{center}$
\begin{array}{cc}
   {\hspace*{-0.03cm}\includegraphics[width=.382\textwidth]{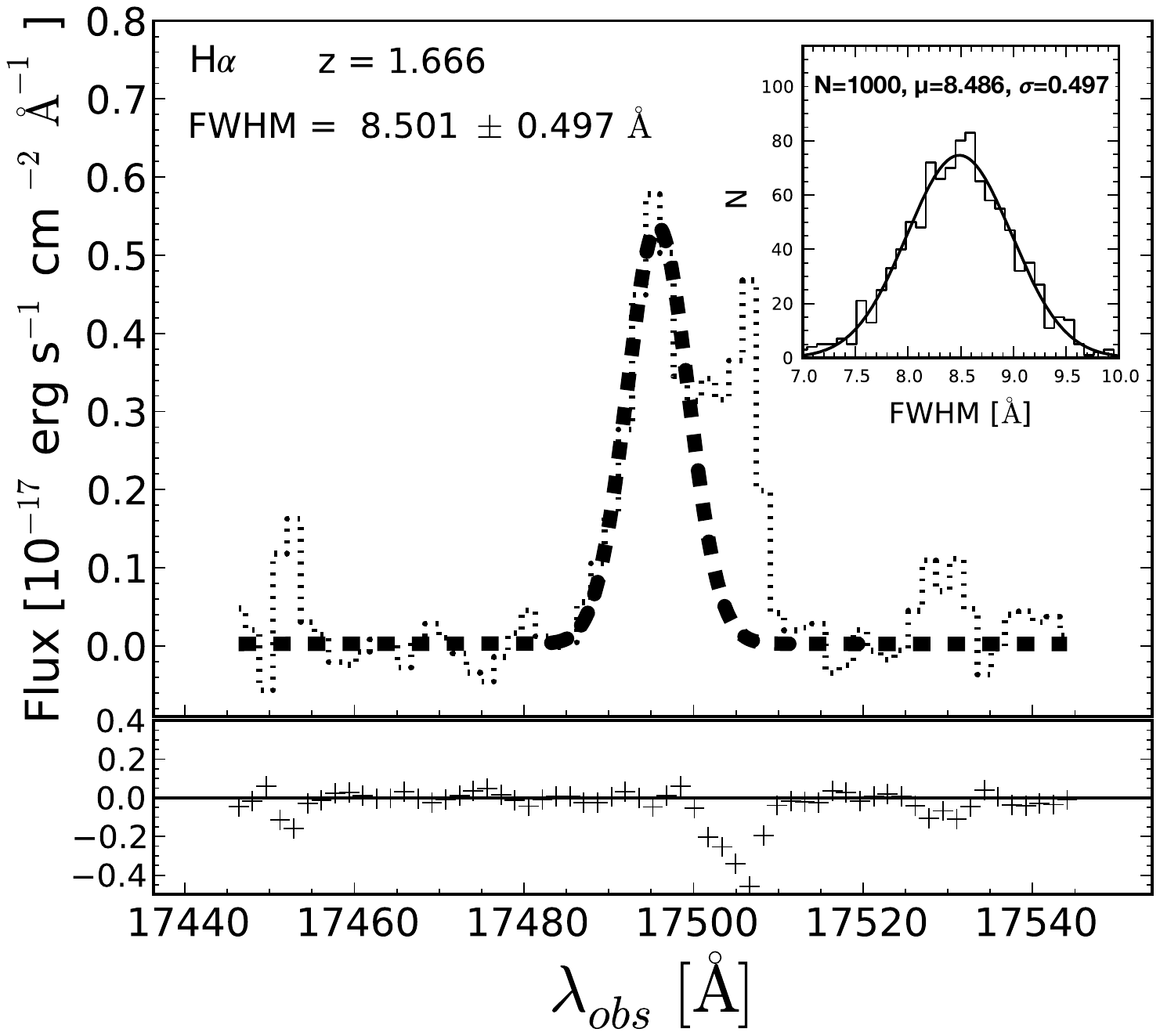}} &
   {\hspace*{0.21cm}\includegraphics[width=.39\textwidth]{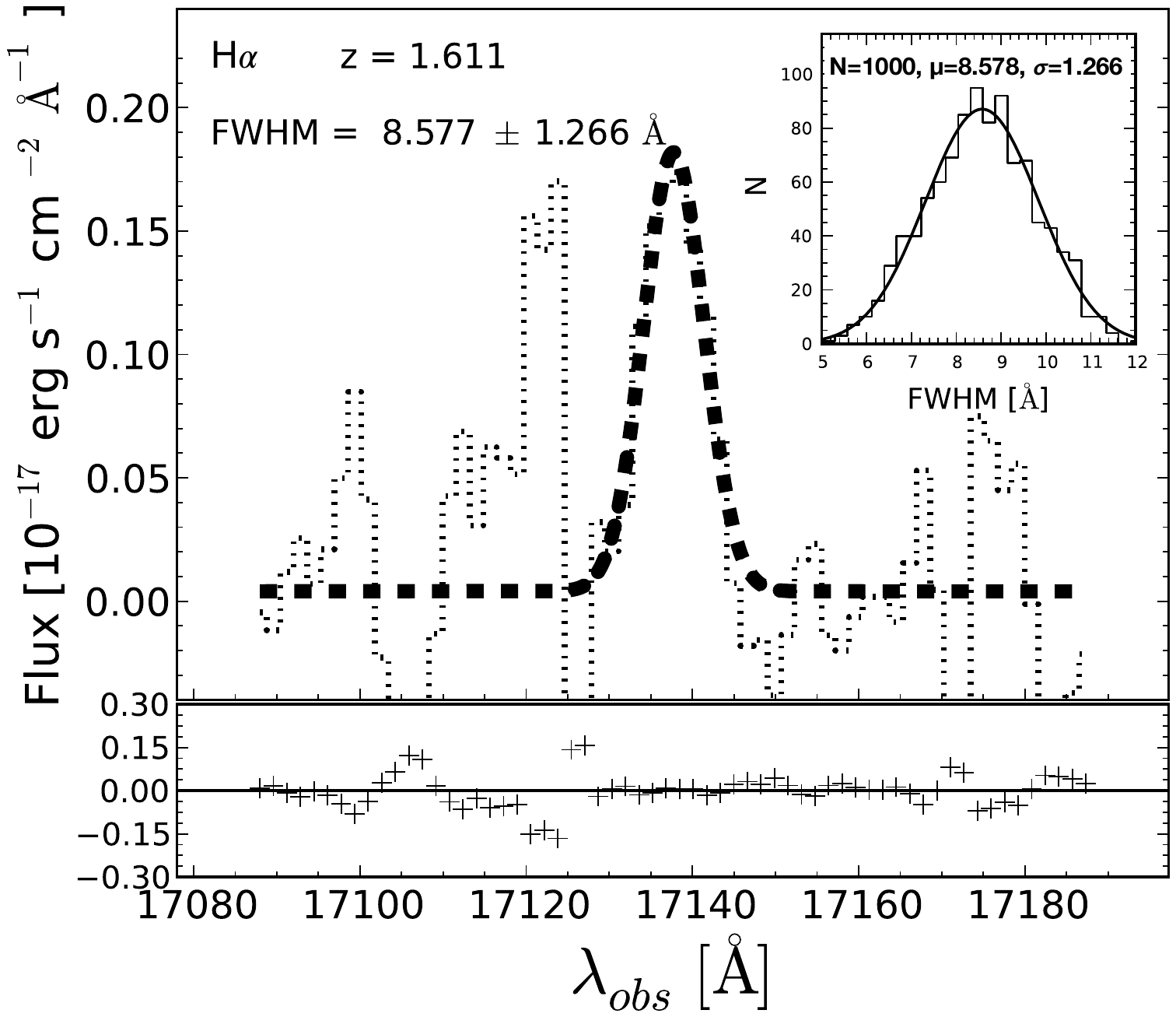}}
\end{array}$
\end{center}
\end{figure*}

\clearpage
\begin{figure*}
\begin{center}$
\hspace*{-1cm}
\begin{array}{cc}
   {\hspace*{-0.4cm}\includegraphics[width=.375\textwidth]{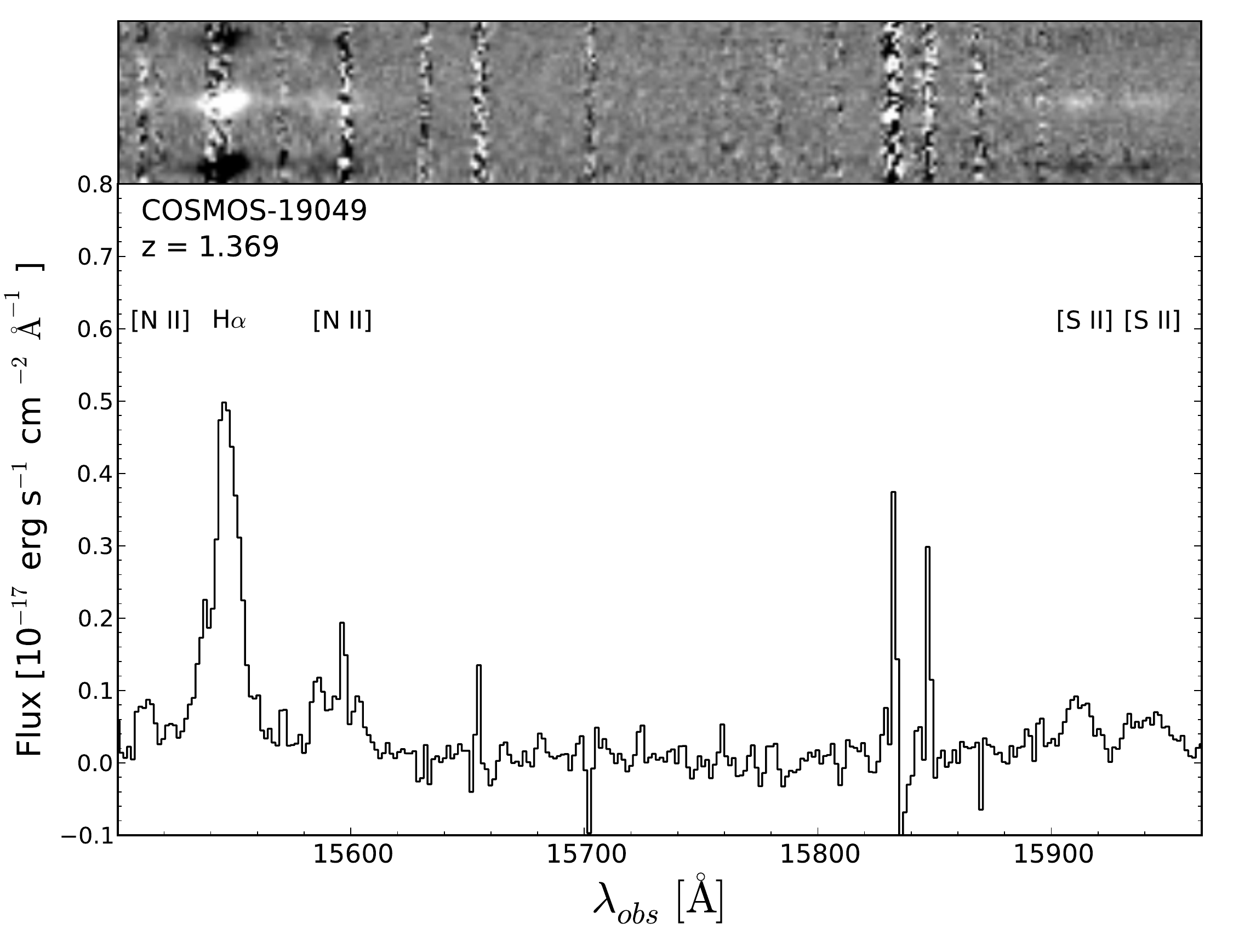}} &
   {\hspace*{0.45cm}\includegraphics[width=0.278\textwidth]{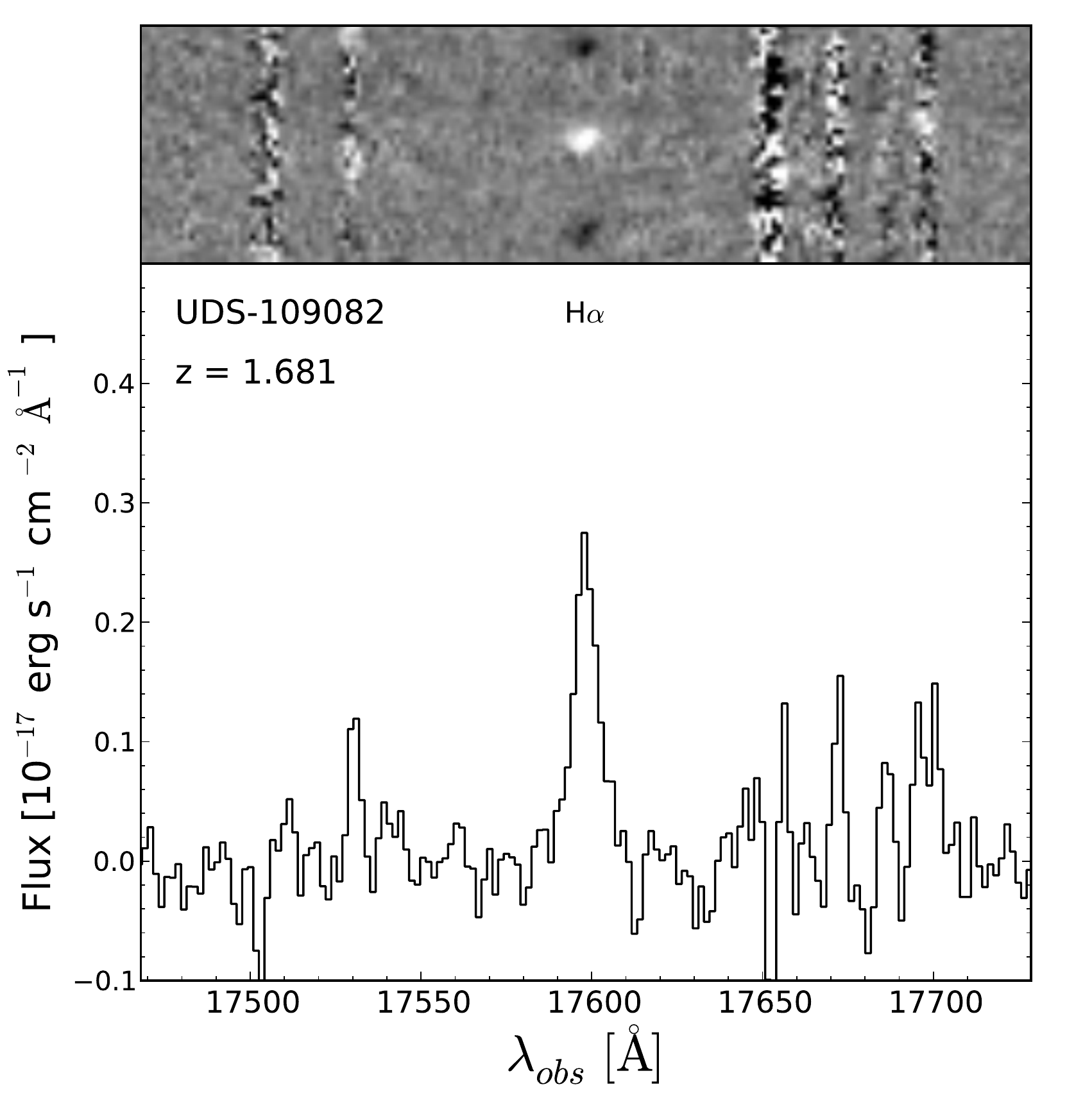}} 
\end{array}$
\end{center}

\vspace*{-0.7cm}
\begin{center}$
\begin{array}{cc}
   {\hspace*{-0.05cm}\includegraphics[width=.383\textwidth]{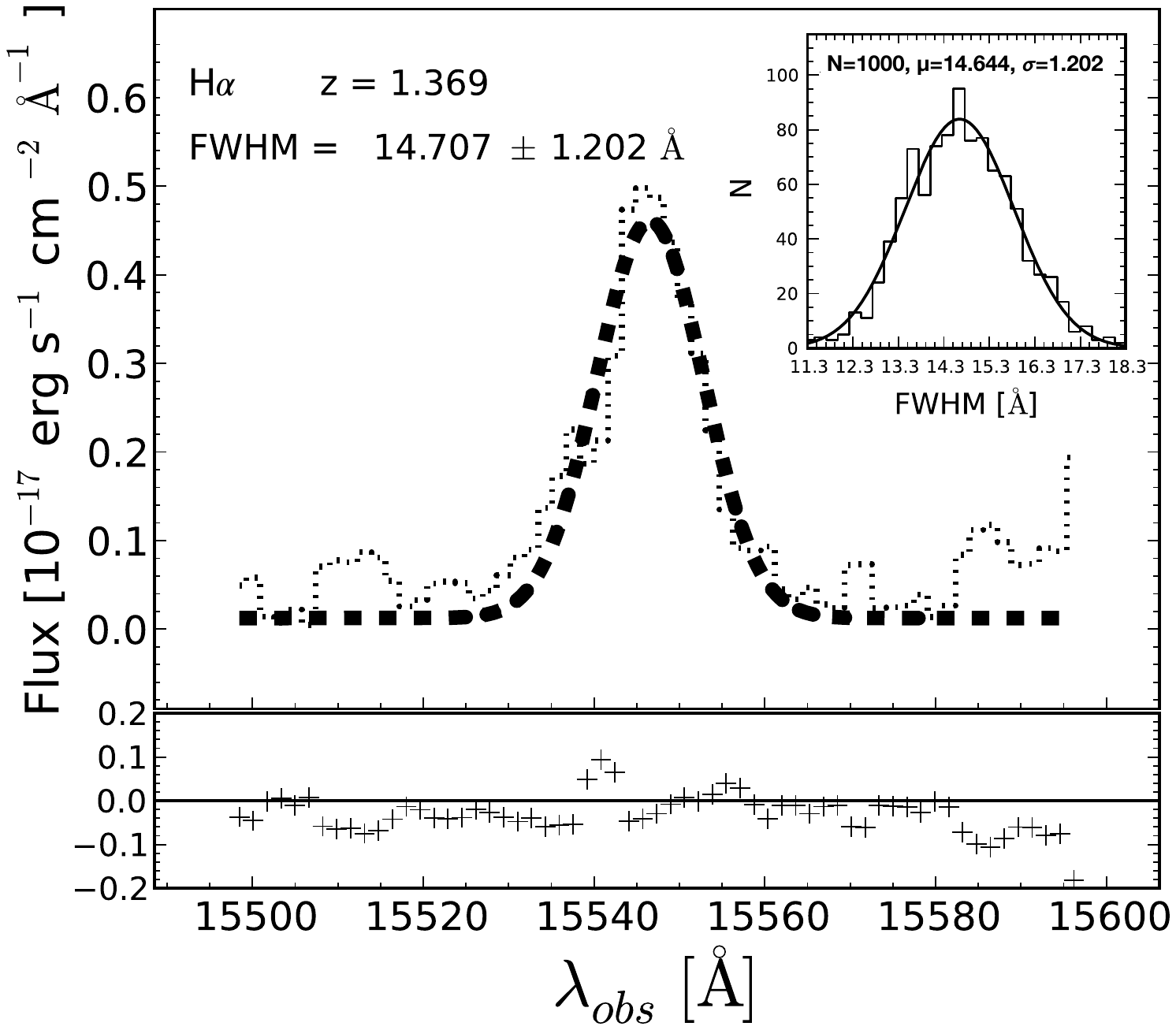}} &
  {\hspace*{0.2cm}\includegraphics[width=.389\textwidth]{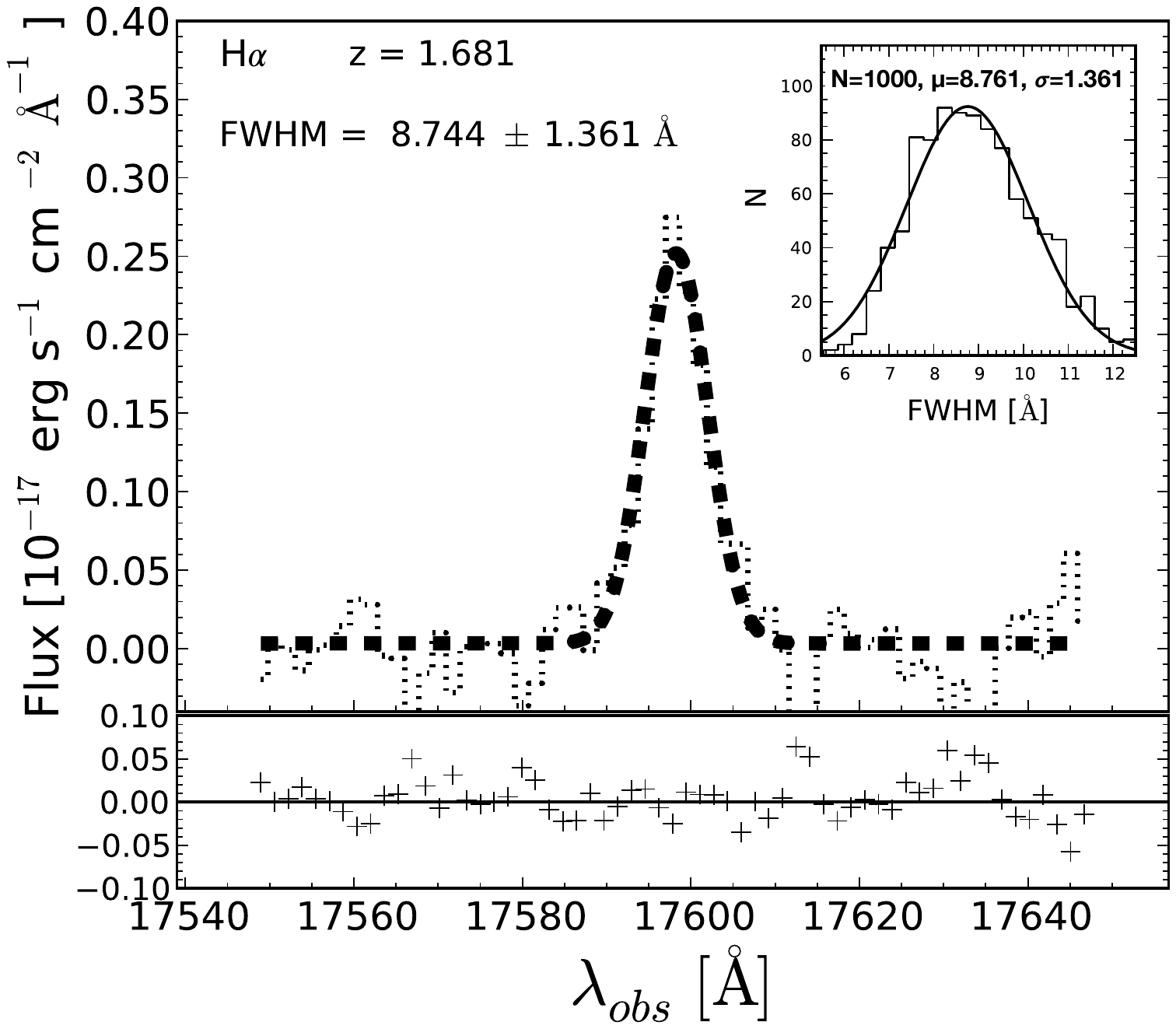}} 
\end{array}$
\end{center}
\caption{First and third rows show the H$\alpha$ 1D and 2D spectra. Second and fourth rows show the fits  to H$\alpha$,  the distribution of FWHM obtained from the Montecarlo simulations in the inset at the upper right corner and the residuals in the box underneath.}
\label{1D and 2D spectra of Halpha}
\end{figure*}

\clearpage
\begin{figure*}
\begin{center}$
\begin{array}{cc}
   {\hspace*{0.5cm}\includegraphics[width=.37\textwidth]{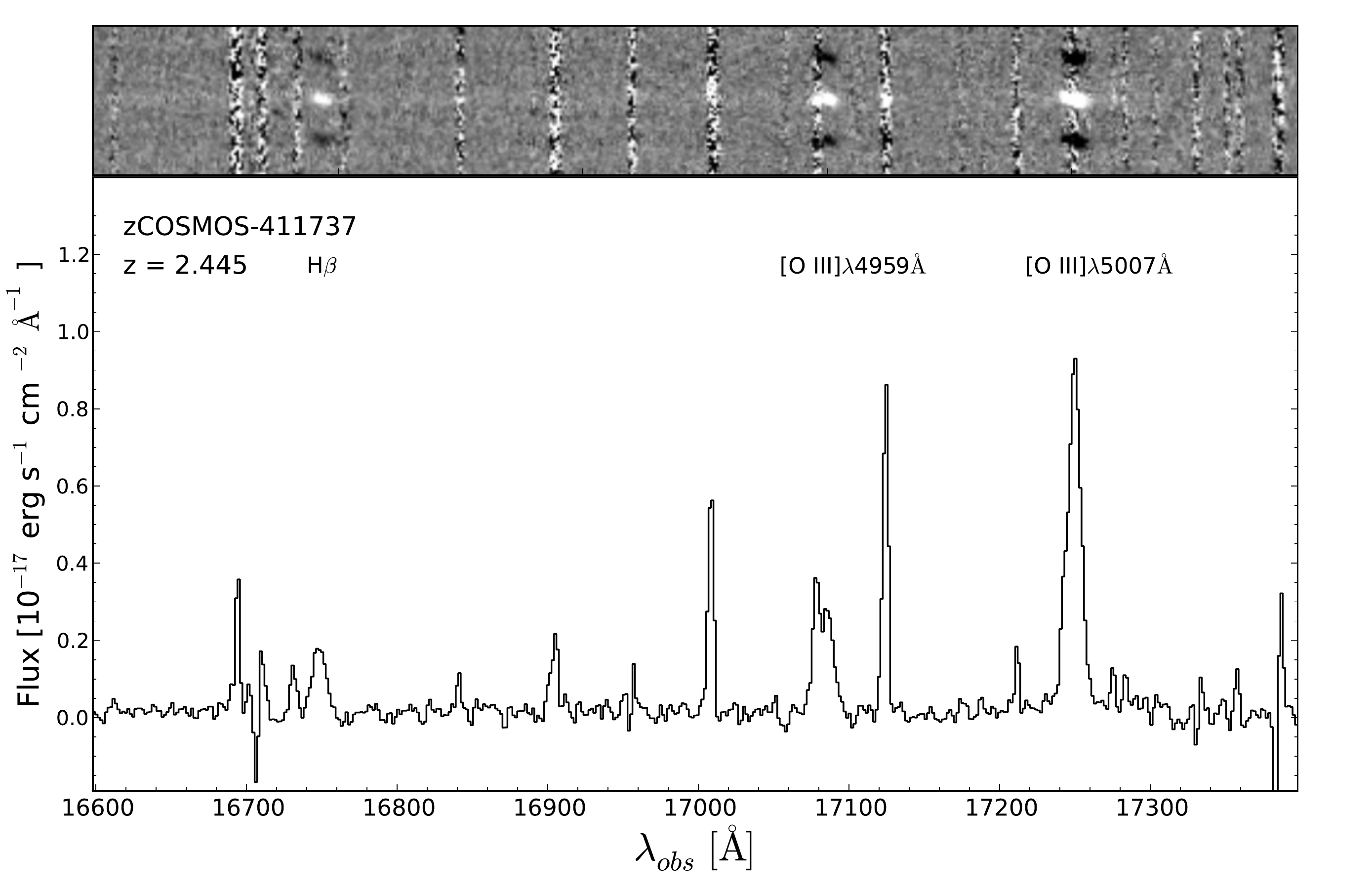}} &
   {\hspace*{0.5cm}\includegraphics[width=.37\textwidth]{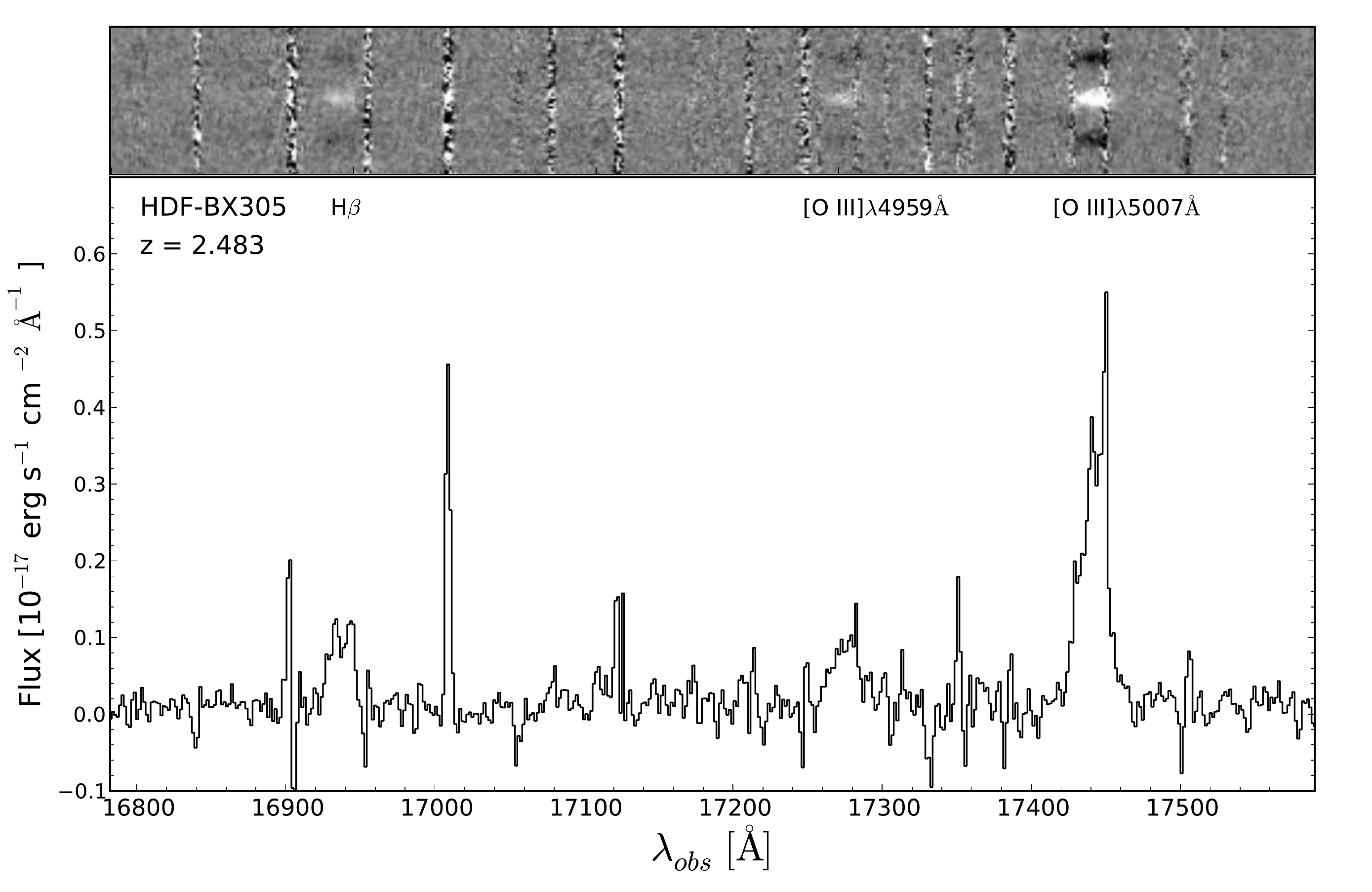}}
\end{array}$
\end{center}
\end{figure*}

\begin{figure*}
\vspace*{-0.67cm}
\begin{center}$
\begin{array}{cc}
   {\hspace*{-0.02cm}\includegraphics[width=.385\textwidth]{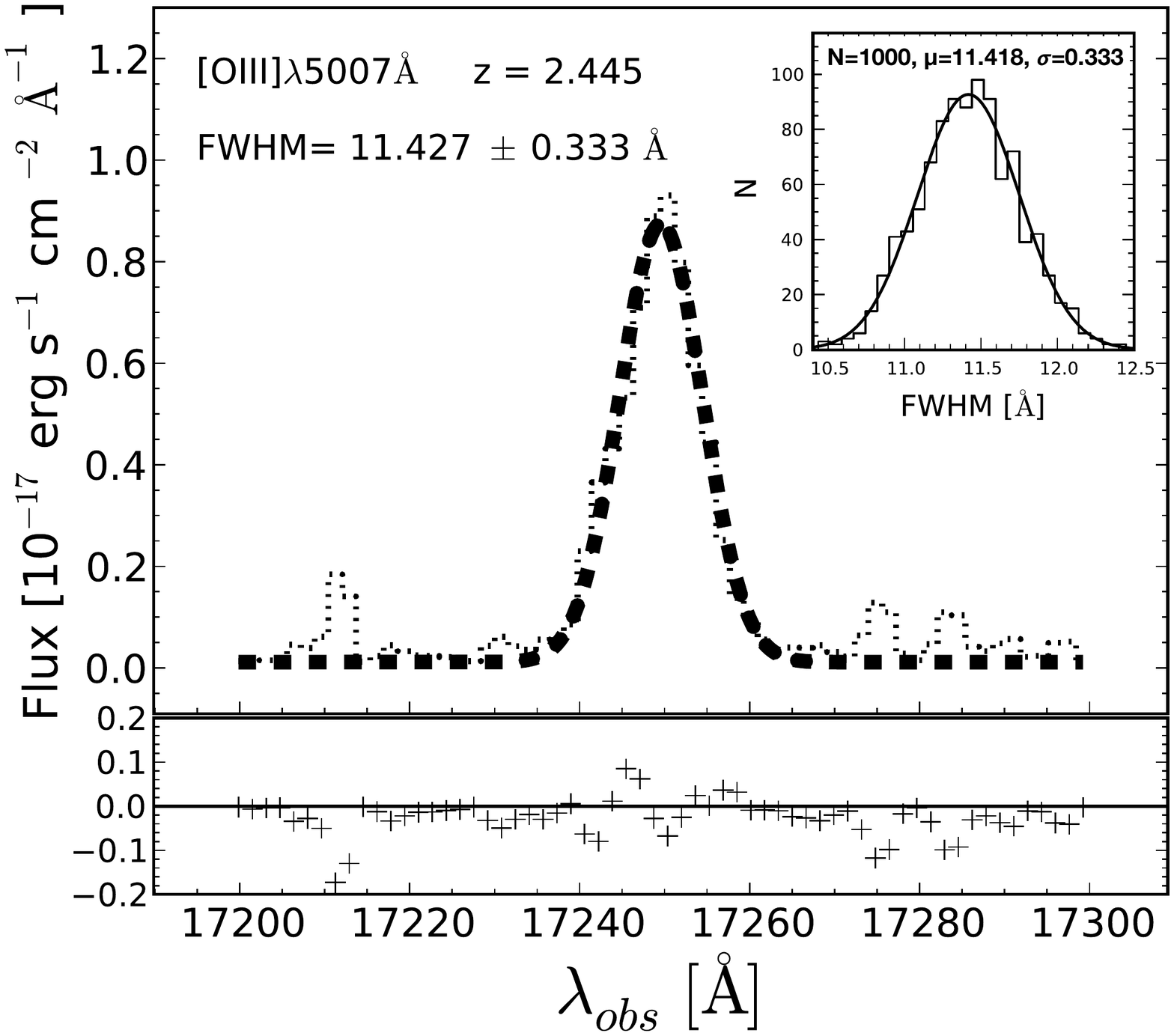}}  &
   {\hspace*{0.32cm}\includegraphics[width=.385\textwidth]{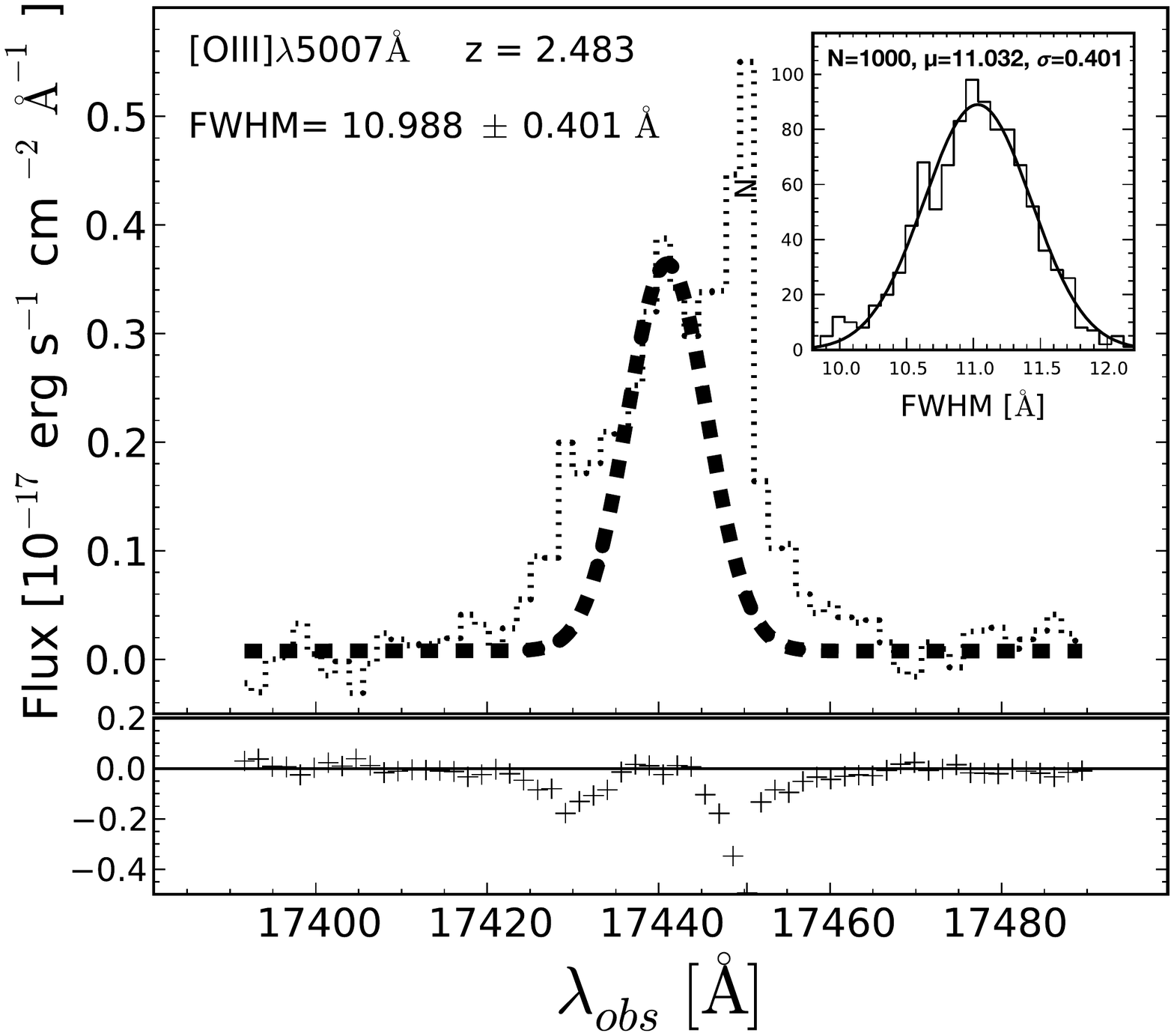}}
\end{array}$
\end{center}

\vspace*{-0.6cm}
\begin{center}$
\begin{array}{cc}
   {\hspace*{0.01cm}\includegraphics[width=.385\textwidth]{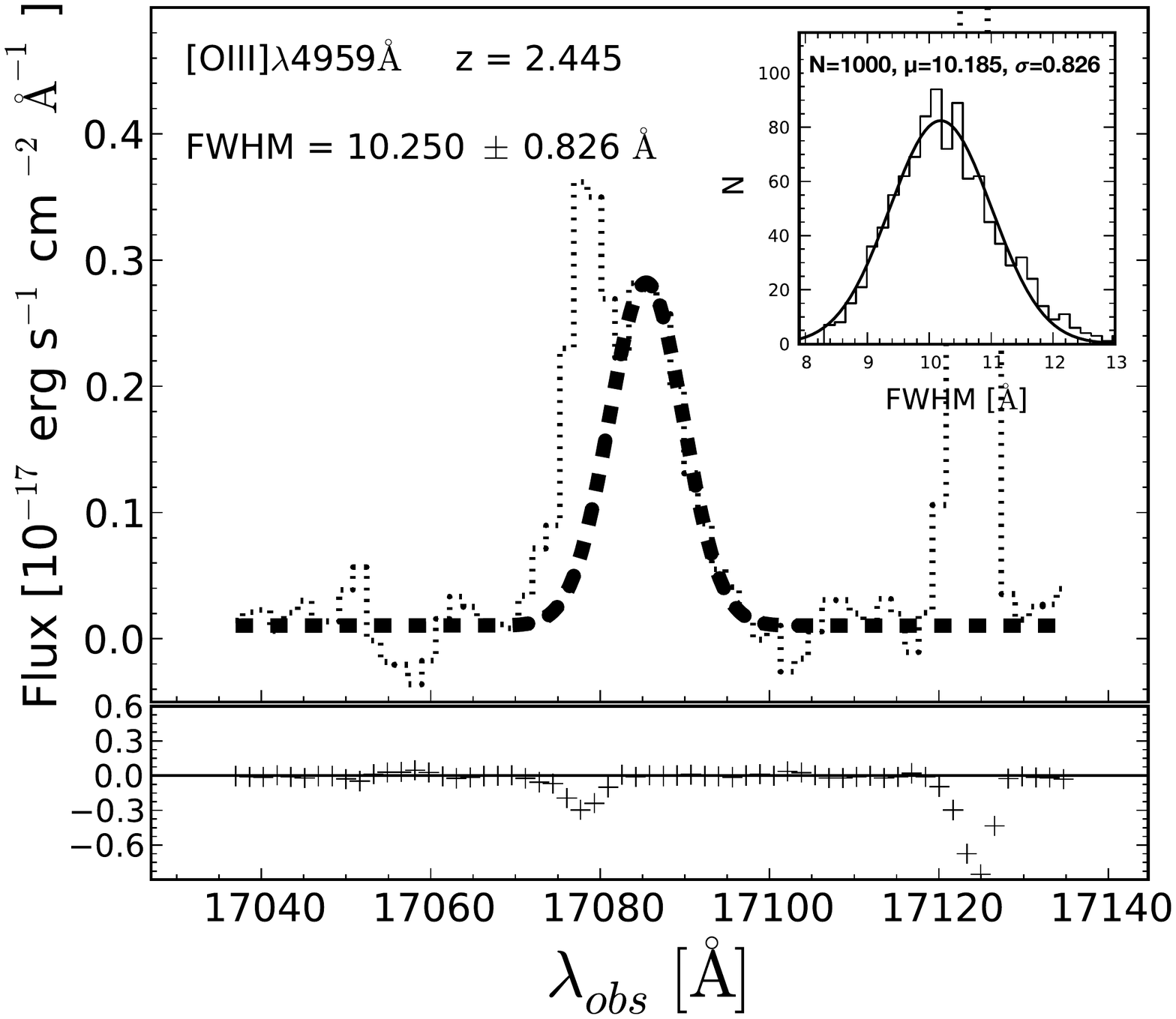}}  &
   {\hspace*{0.2cm}\includegraphics[width=.393\textwidth]{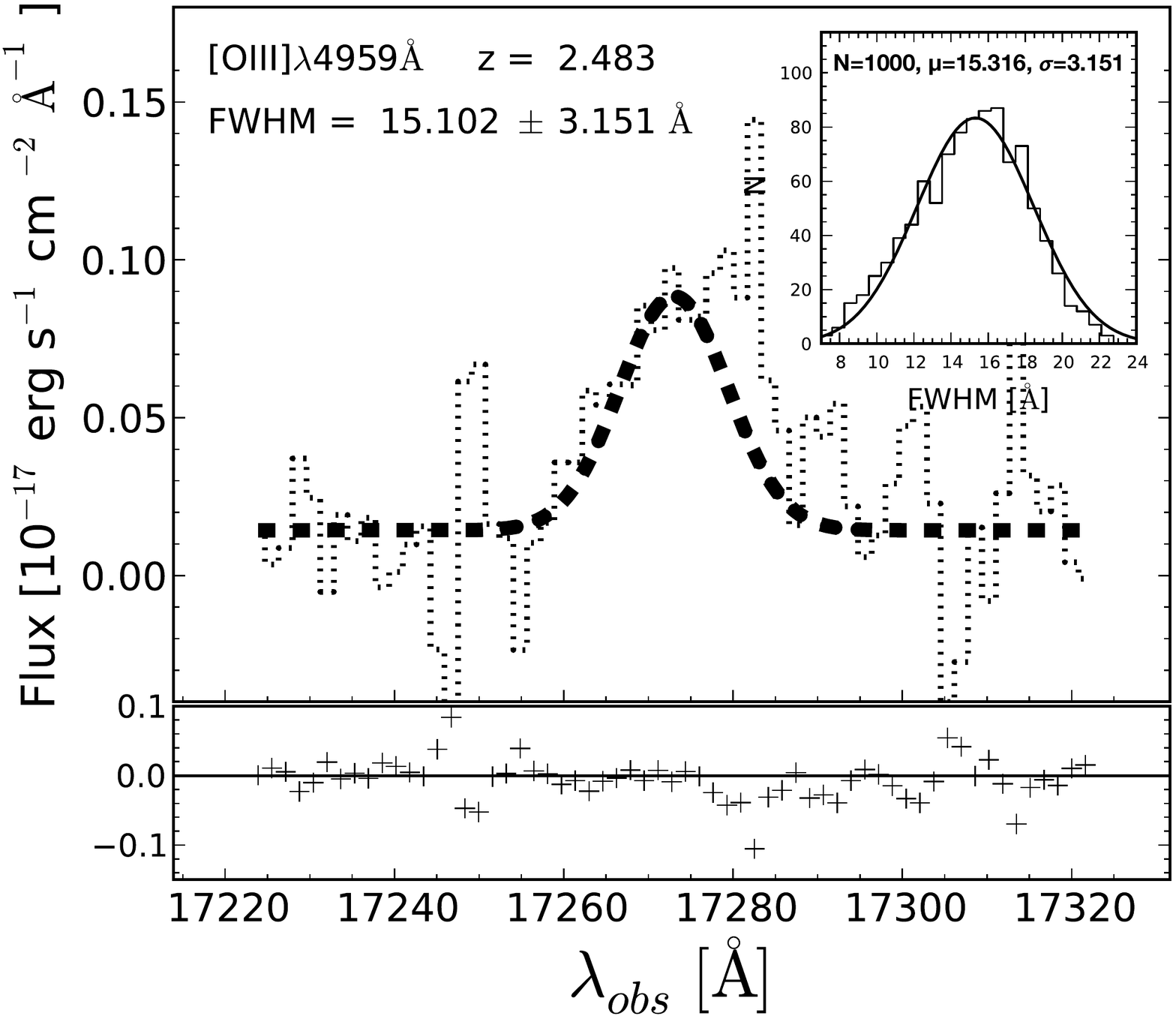}}
\end{array}$
\end{center}

\vspace*{-0.6cm}
\begin{center}$
\begin{array}{cc}
   {\hspace*{-0.1cm}\includegraphics[width=.394\textwidth]{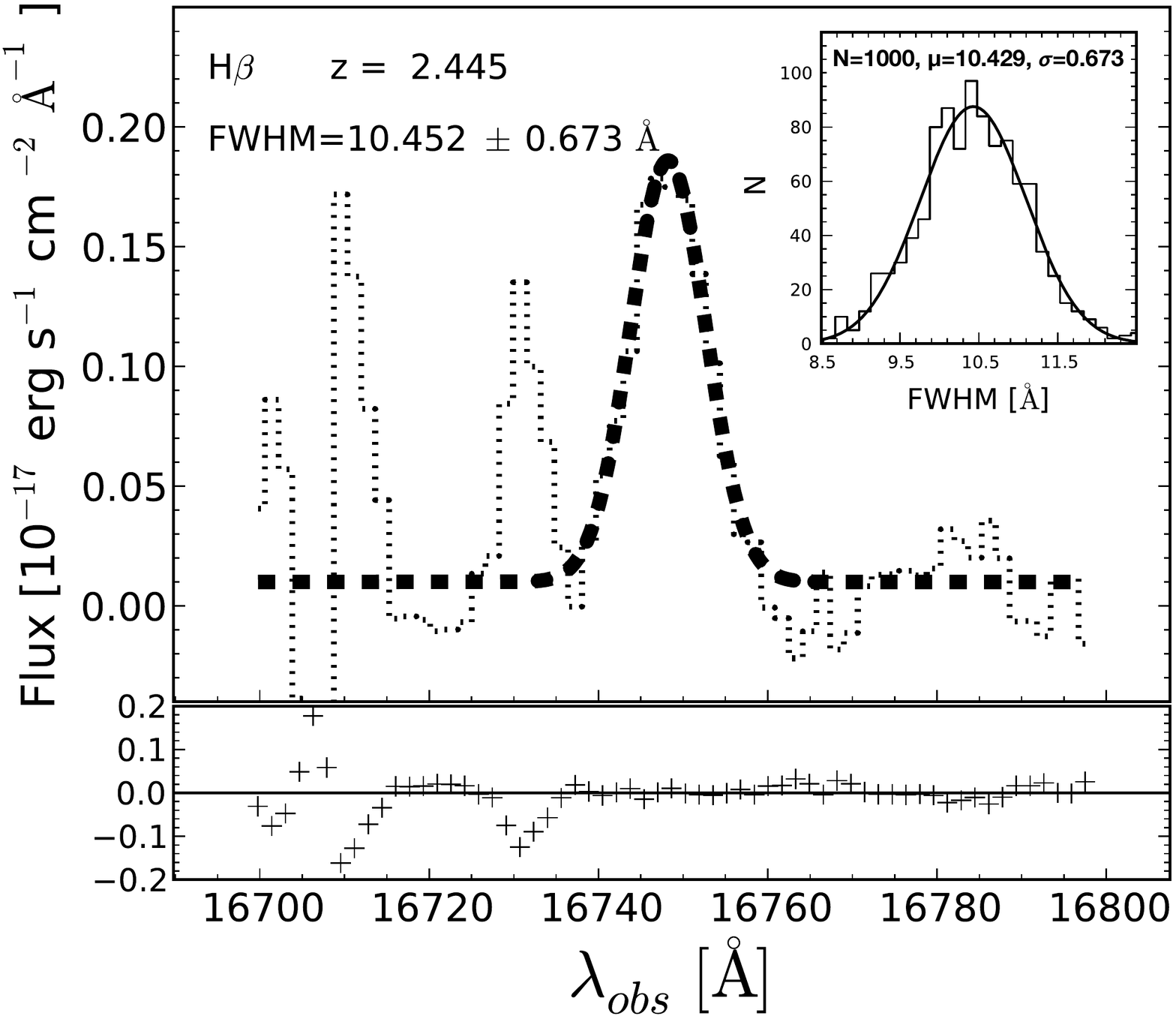}}  &
   {\hspace*{0.17cm}\includegraphics[width=.392\textwidth]{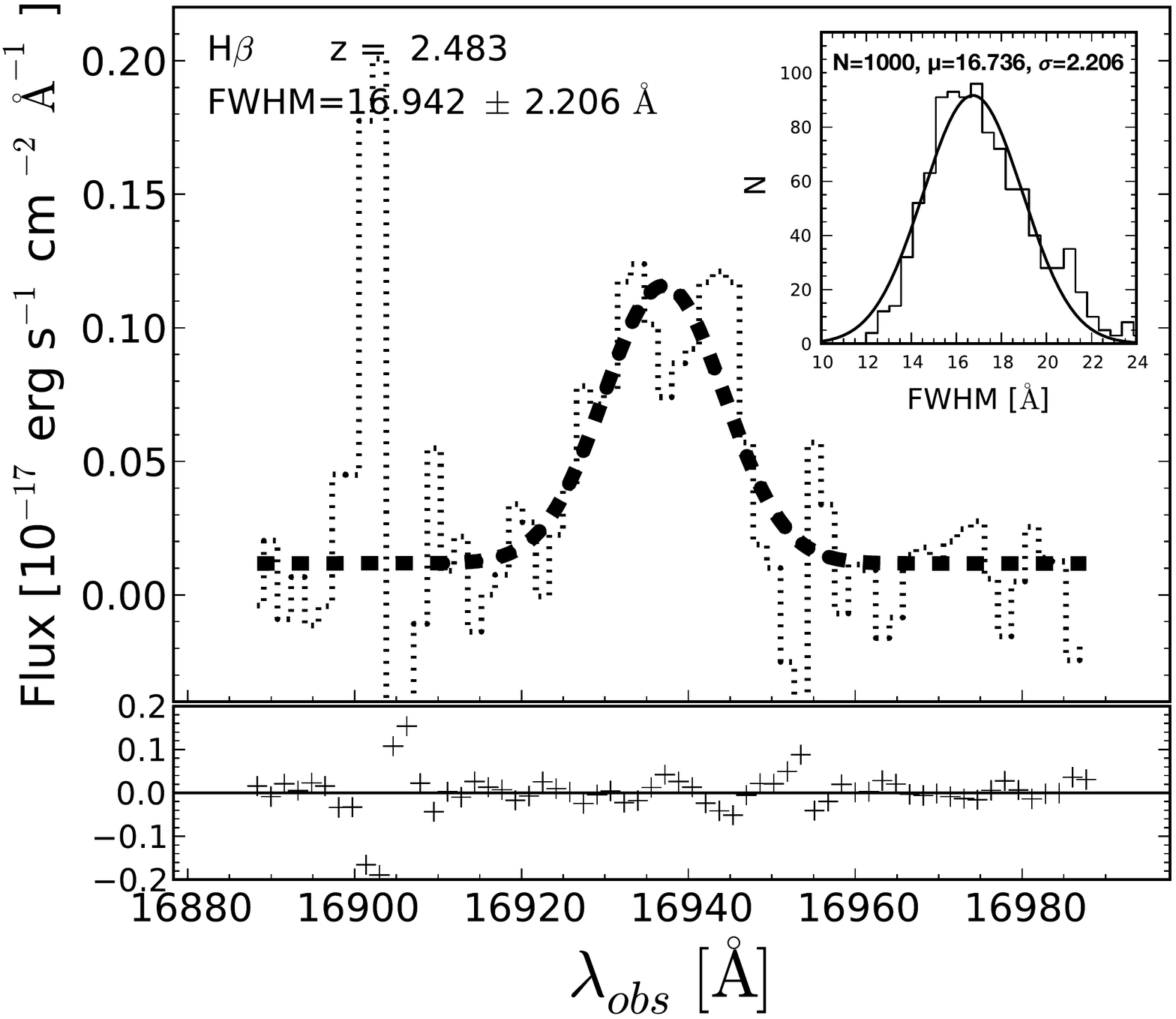}}
\end{array}$
\end{center}
\end{figure*}

\clearpage
\begin{figure*}
\begin{center}$
\begin{array}{cc}
   {\hspace*{0.5cm}\includegraphics[width=.365\textwidth]{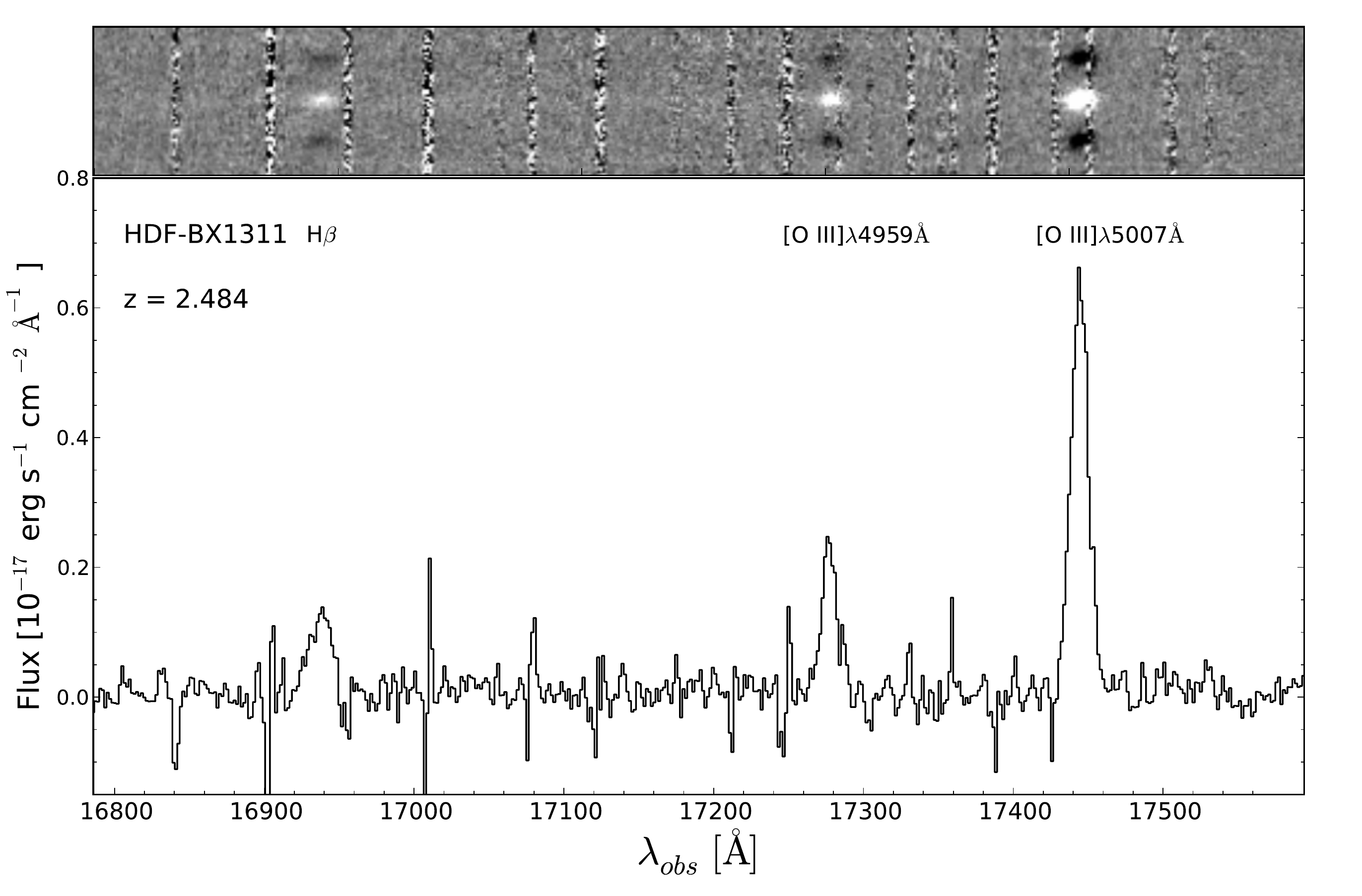}} &
   {\hspace*{0.6cm}\includegraphics[width=.37\textwidth]{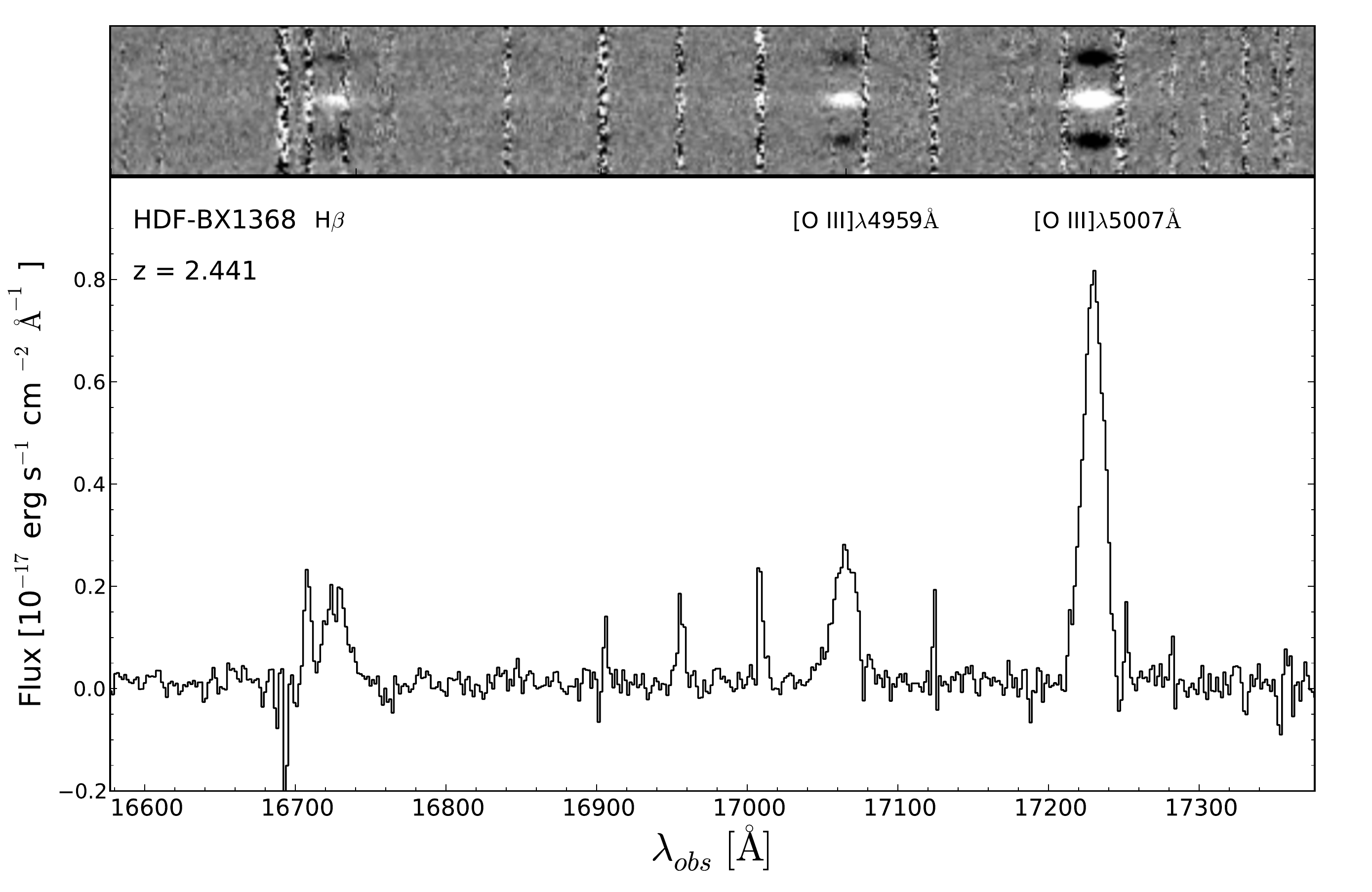}}
\end{array}$
\end{center}

\vspace*{-0.65cm}
\begin{center}$
\begin{array}{cc}
   {\hspace*{-0.02cm}\includegraphics[width=.385\textwidth]{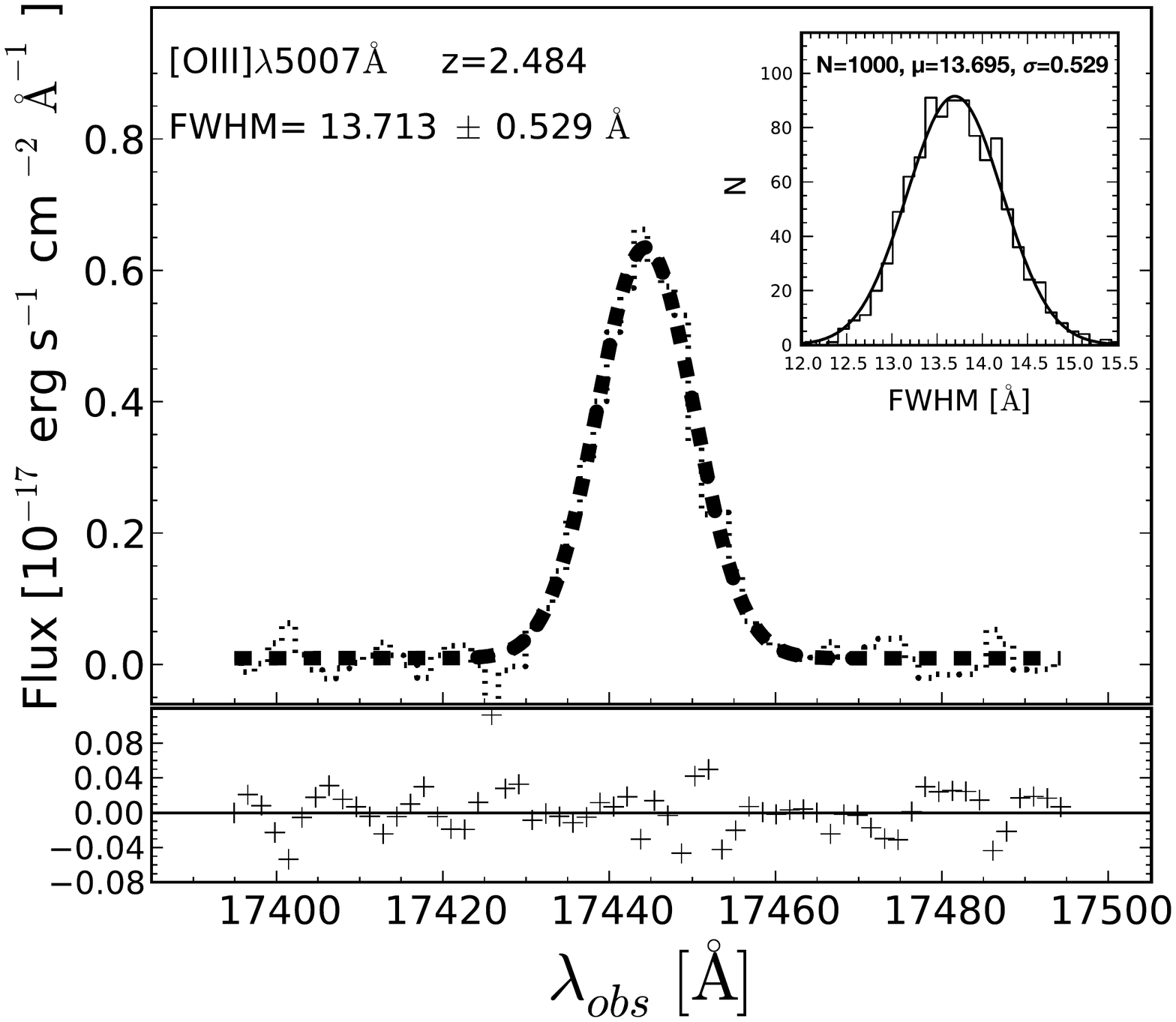}} &
   {\hspace*{0.32cm}\includegraphics[width=.385\textwidth]{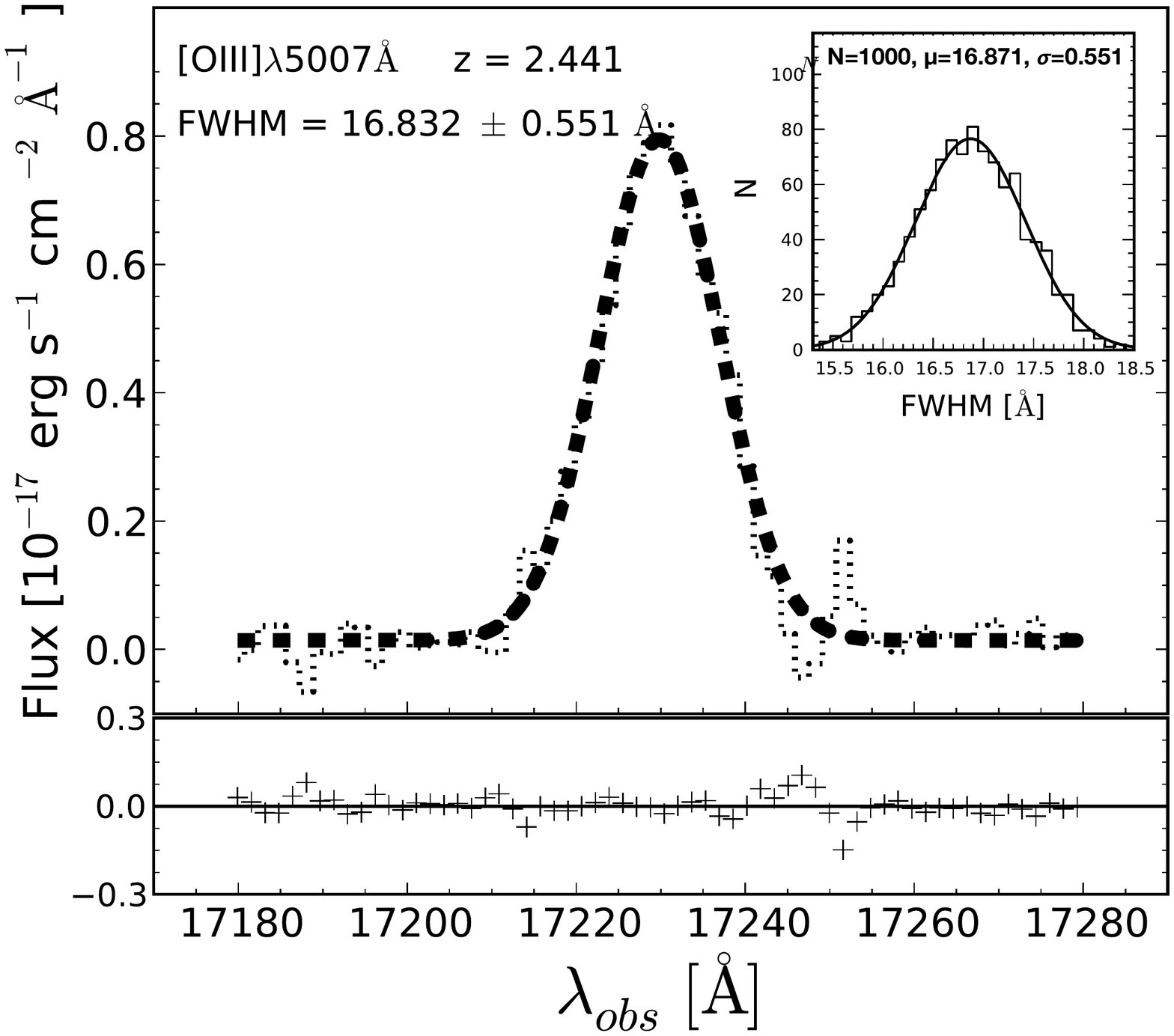}}
\end{array}$
\end{center}

\vspace*{-0.6cm}
\begin{center}$
\begin{array}{cc}
   {\hspace*{-0.1cm}\includegraphics[width=.392\textwidth]{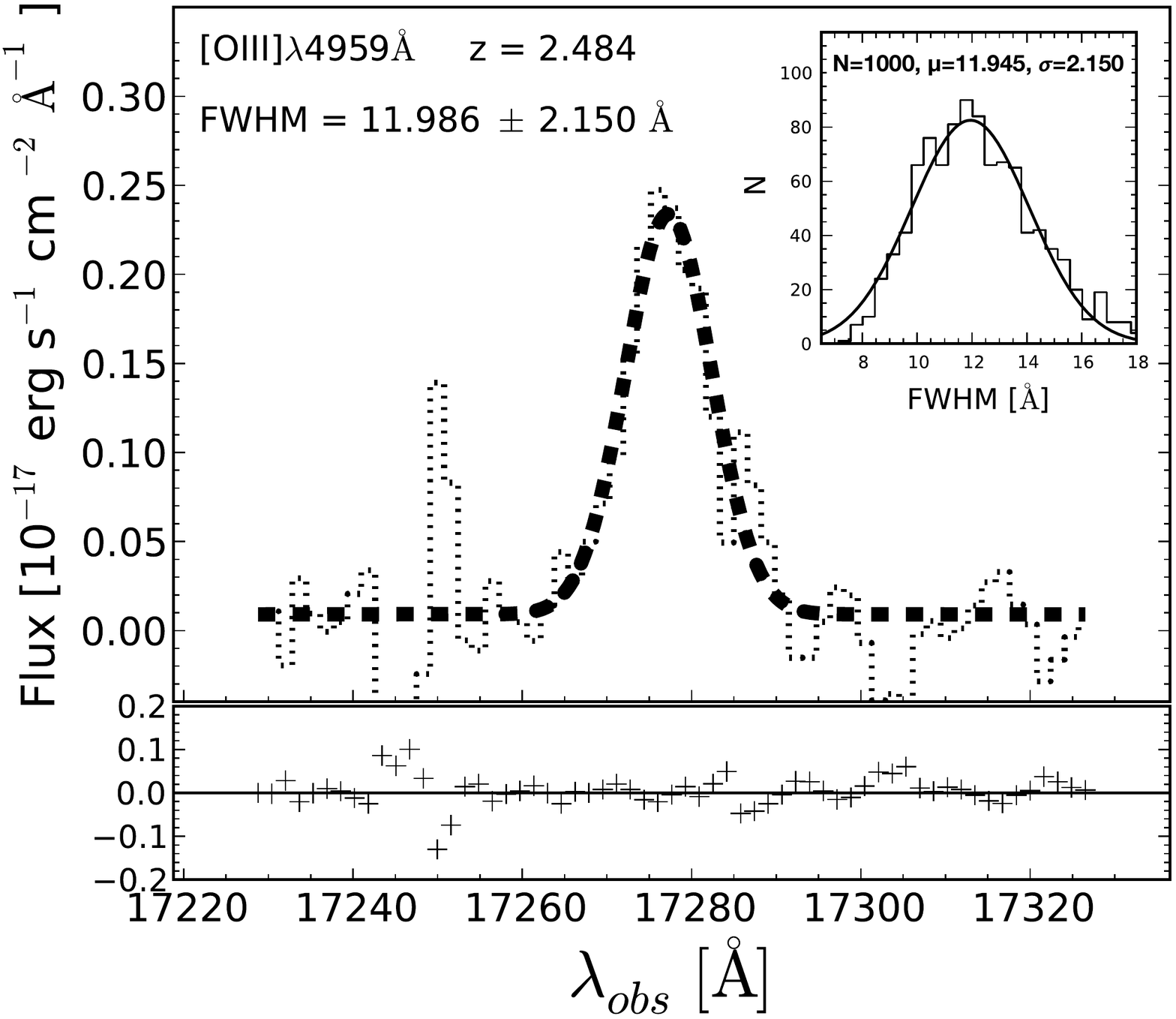}}
   {\hspace*{0.55cm}\includegraphics[width=.392\textwidth]{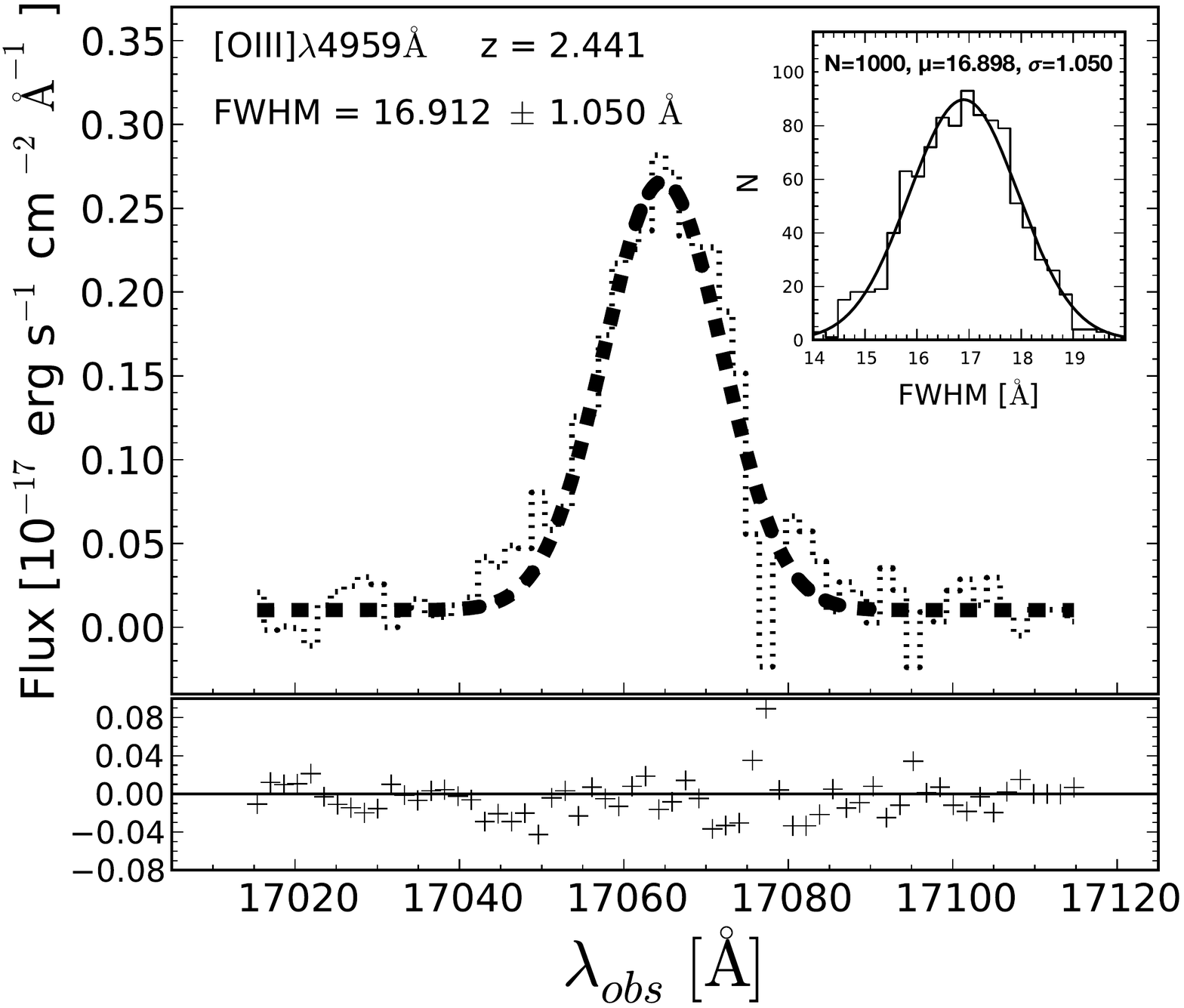}}
\end{array}$
\end{center}

\vspace*{-0.6cm}
\begin{center}$
\begin{array}{cc}
   {\hspace*{-0.1cm}\includegraphics[width=.392\textwidth]{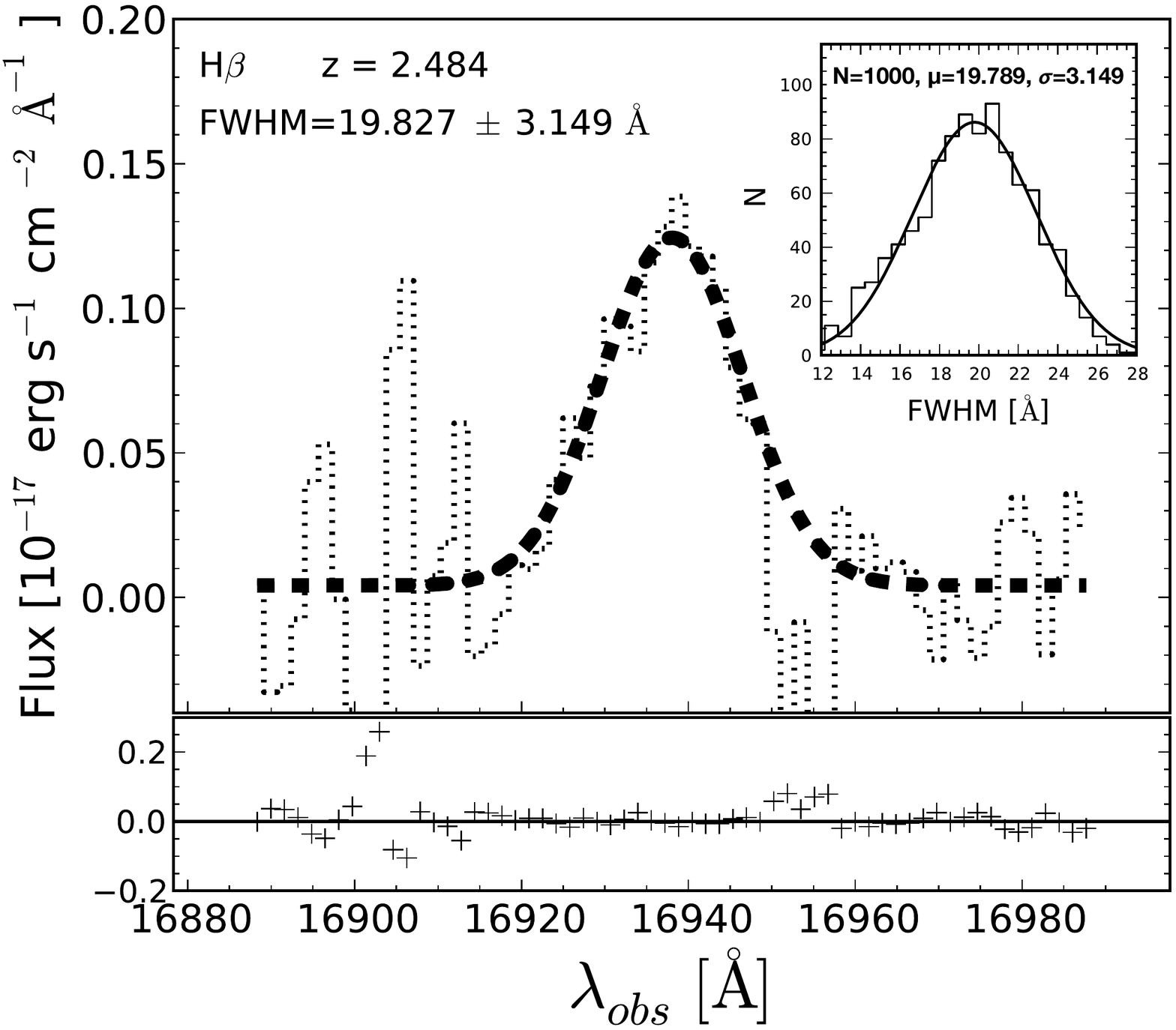}}
   {\hspace*{0.55cm}\includegraphics[width=.392\textwidth]{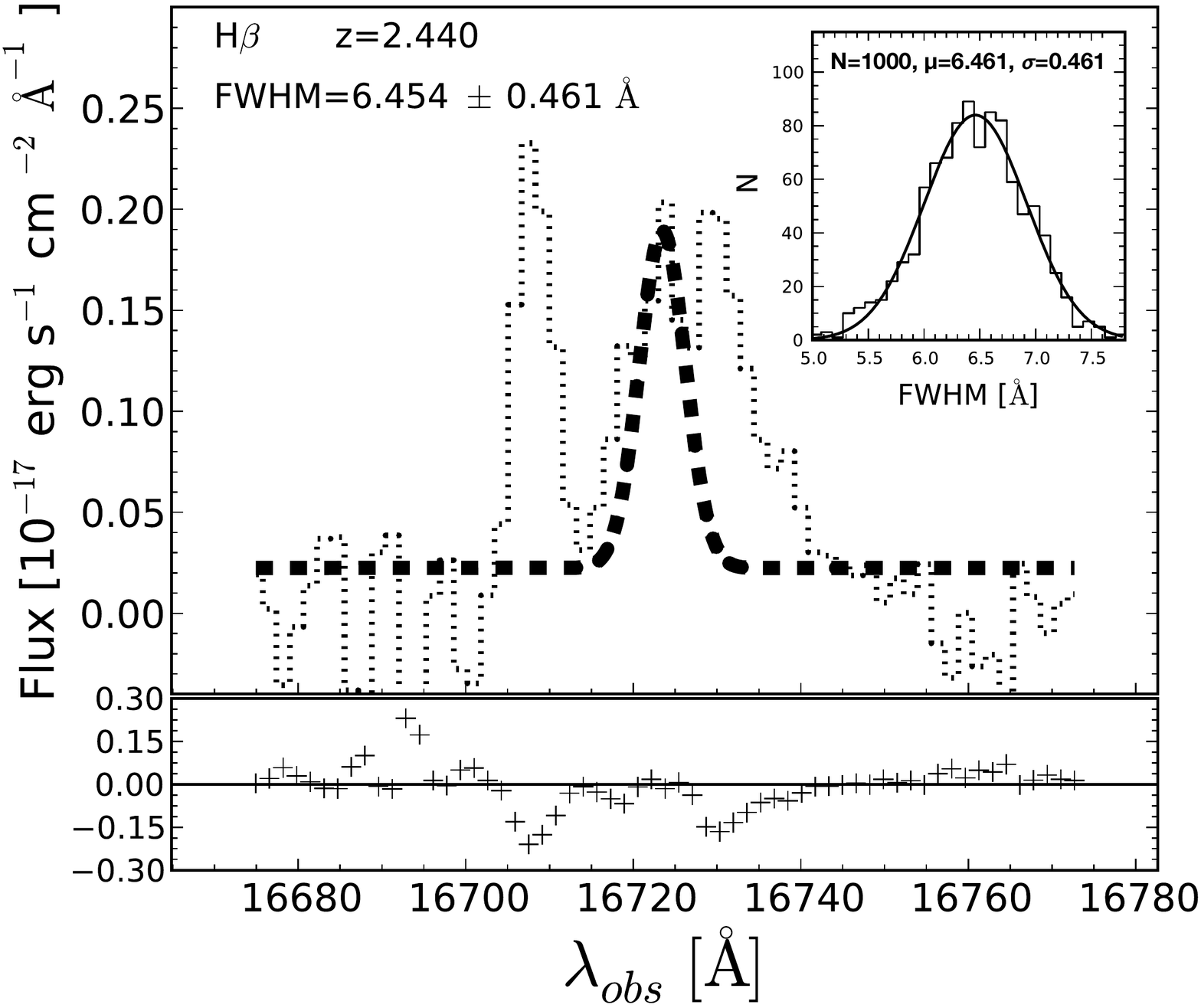}}  
\end{array}$
\end{center}
\end{figure*}

\clearpage
\begin{figure*}
\begin{center}$
\begin{array}{cc}
   {\hspace*{0.4cm}\includegraphics[width=.369\textwidth]{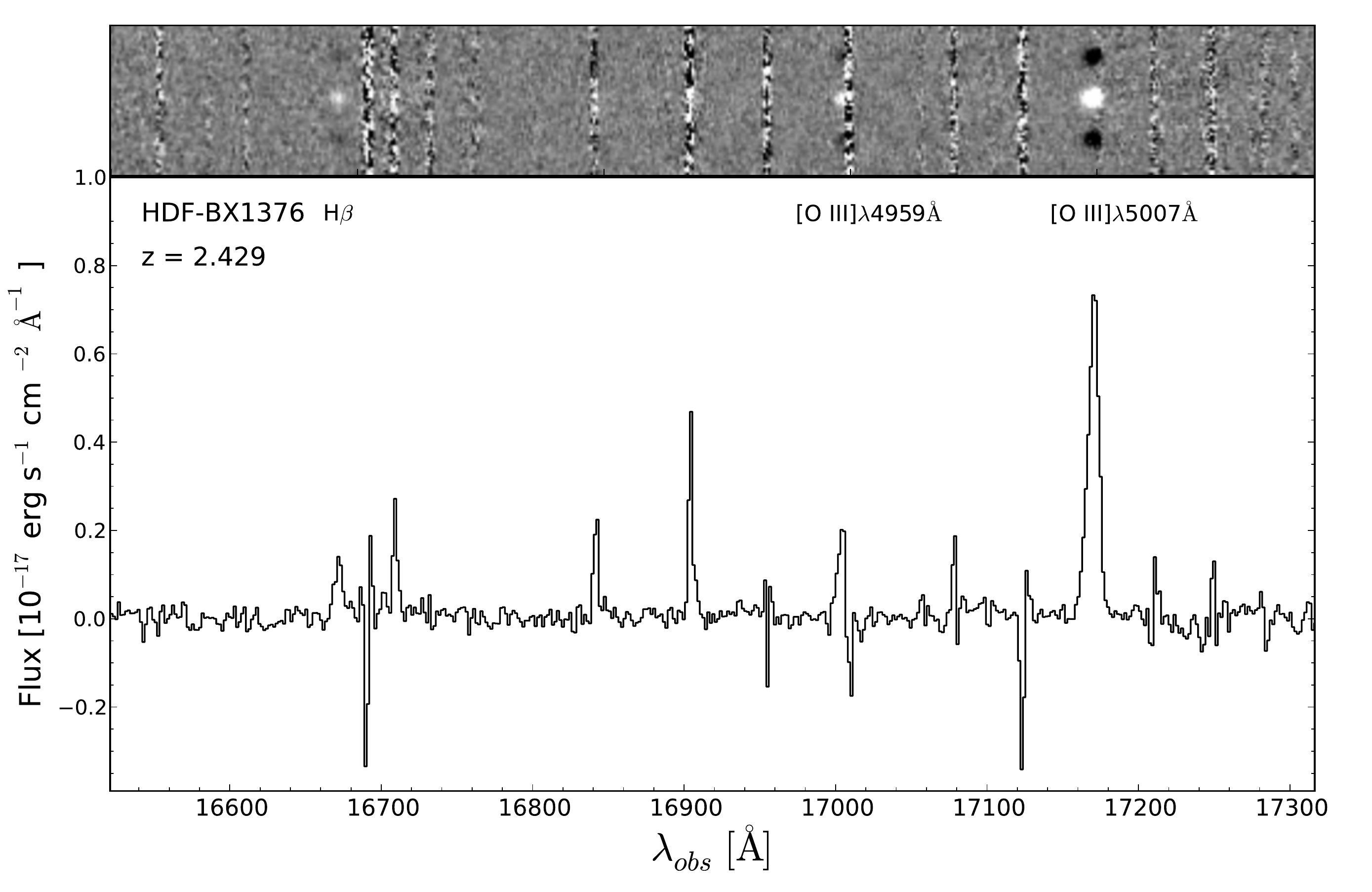}} &
   {\hspace*{0.55cm}\includegraphics[width=.371\textwidth]{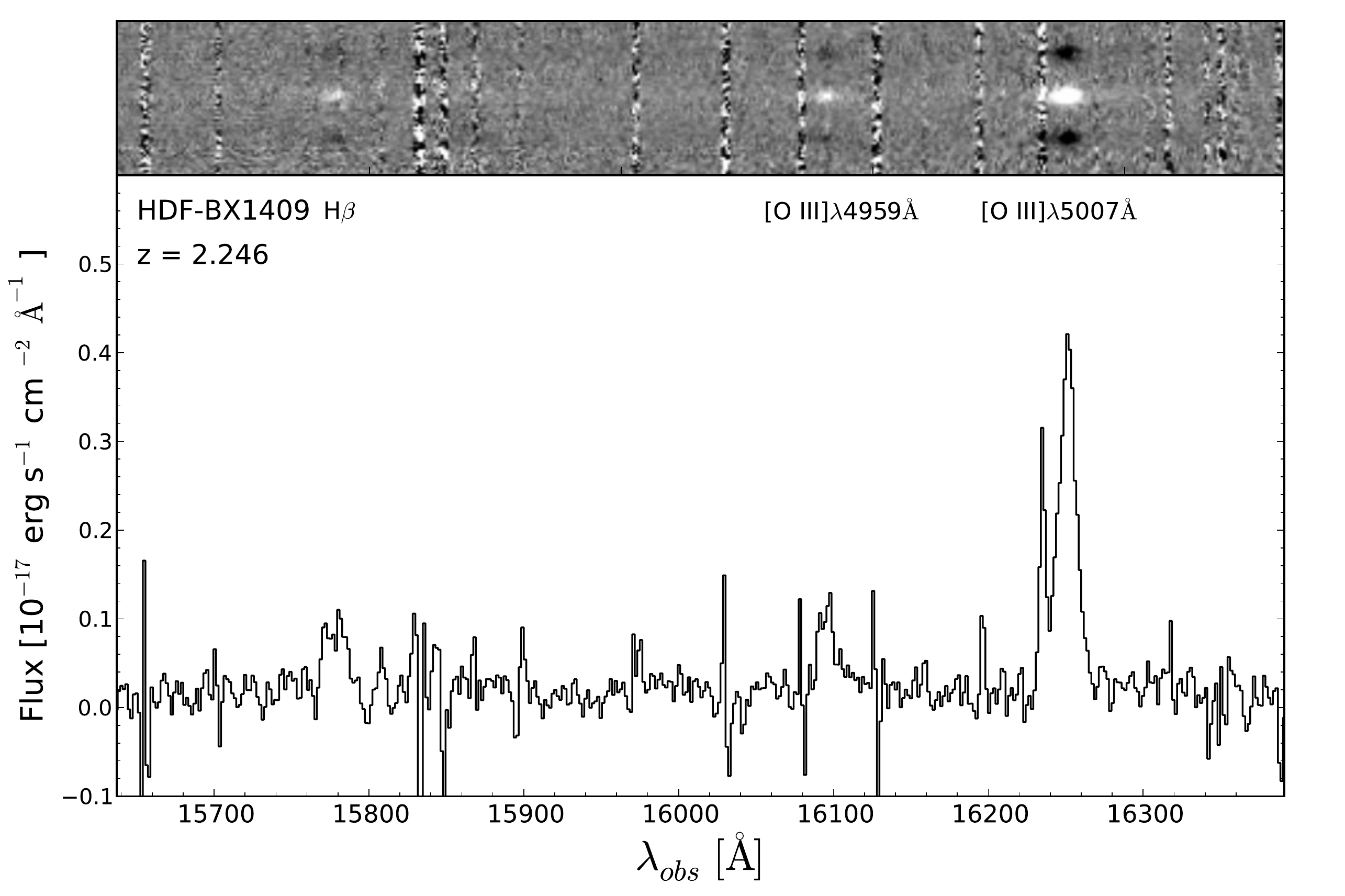}}
\end{array}$
\end{center}

\vspace*{-0.63cm}
\begin{center}$
\begin{array}{cc}
   {\hspace*{-0.02cm}\includegraphics[width=.385\textwidth]{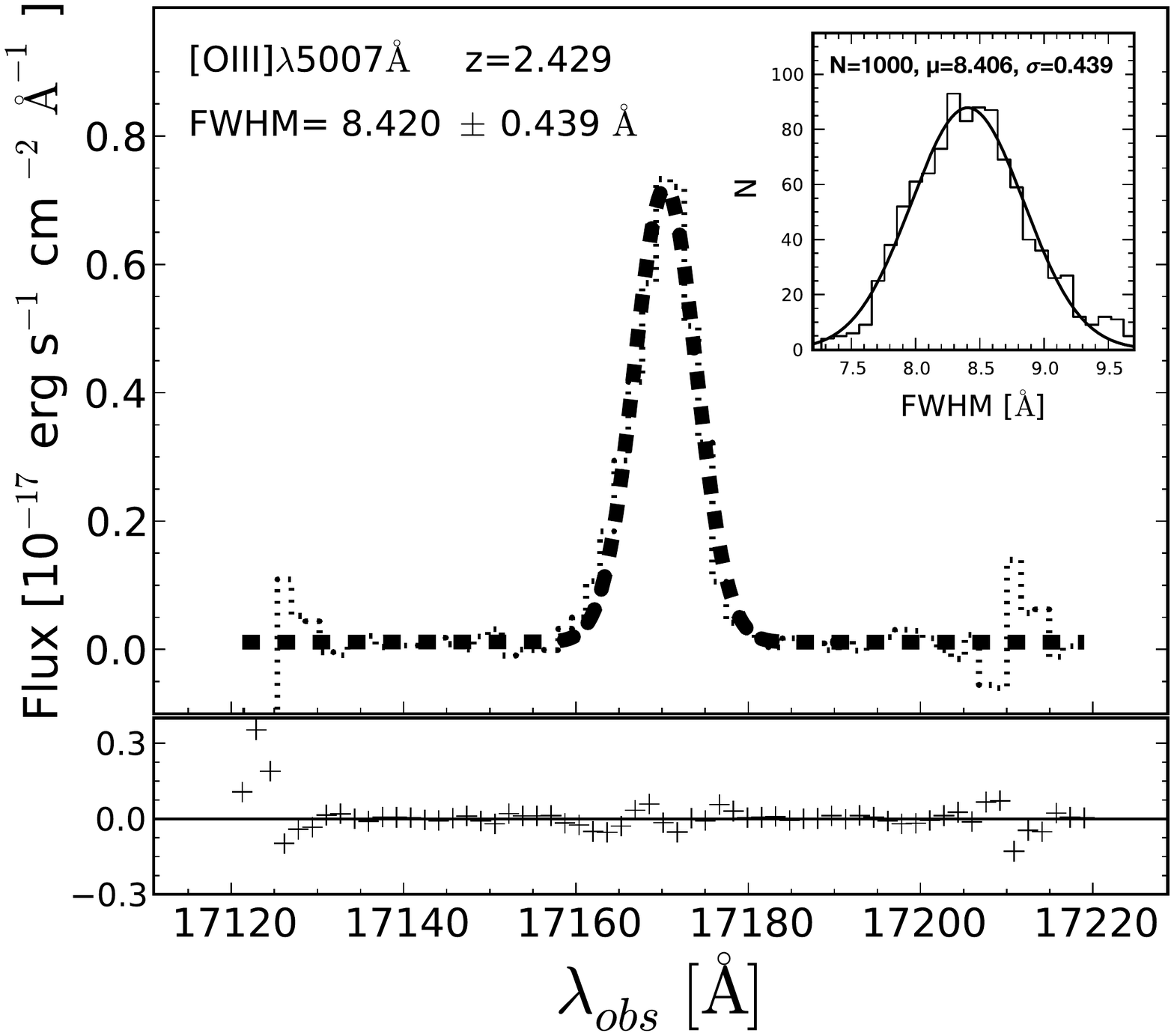}} &
   {\hspace*{0.32cm}\includegraphics[width=.385\textwidth]{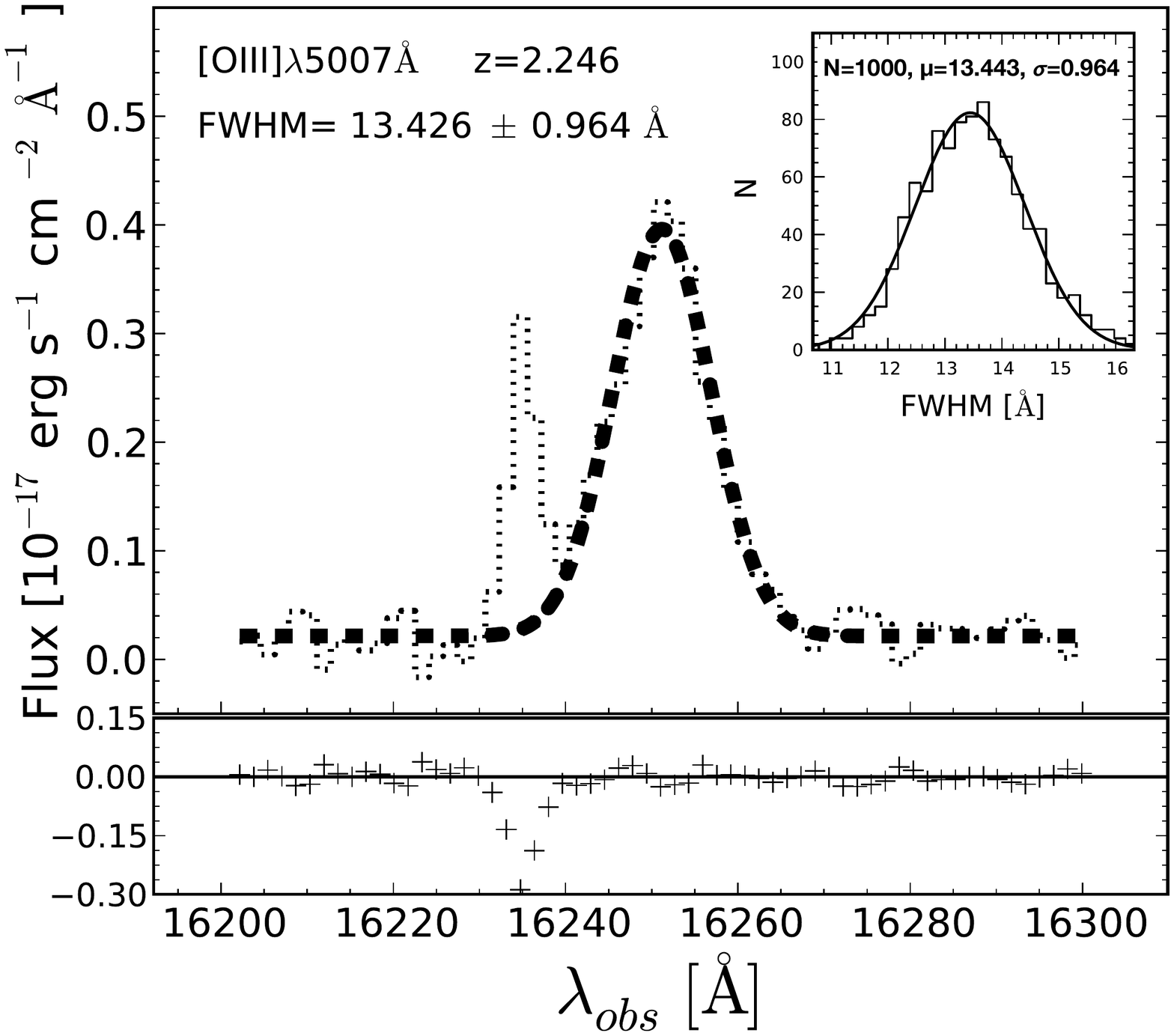}}
\end{array}$
\end{center}

\vspace*{-0.63cm}
\begin{center}$
\begin{array}{cc}
   {\hspace*{-0.1cm}\includegraphics[width=.396\textwidth]{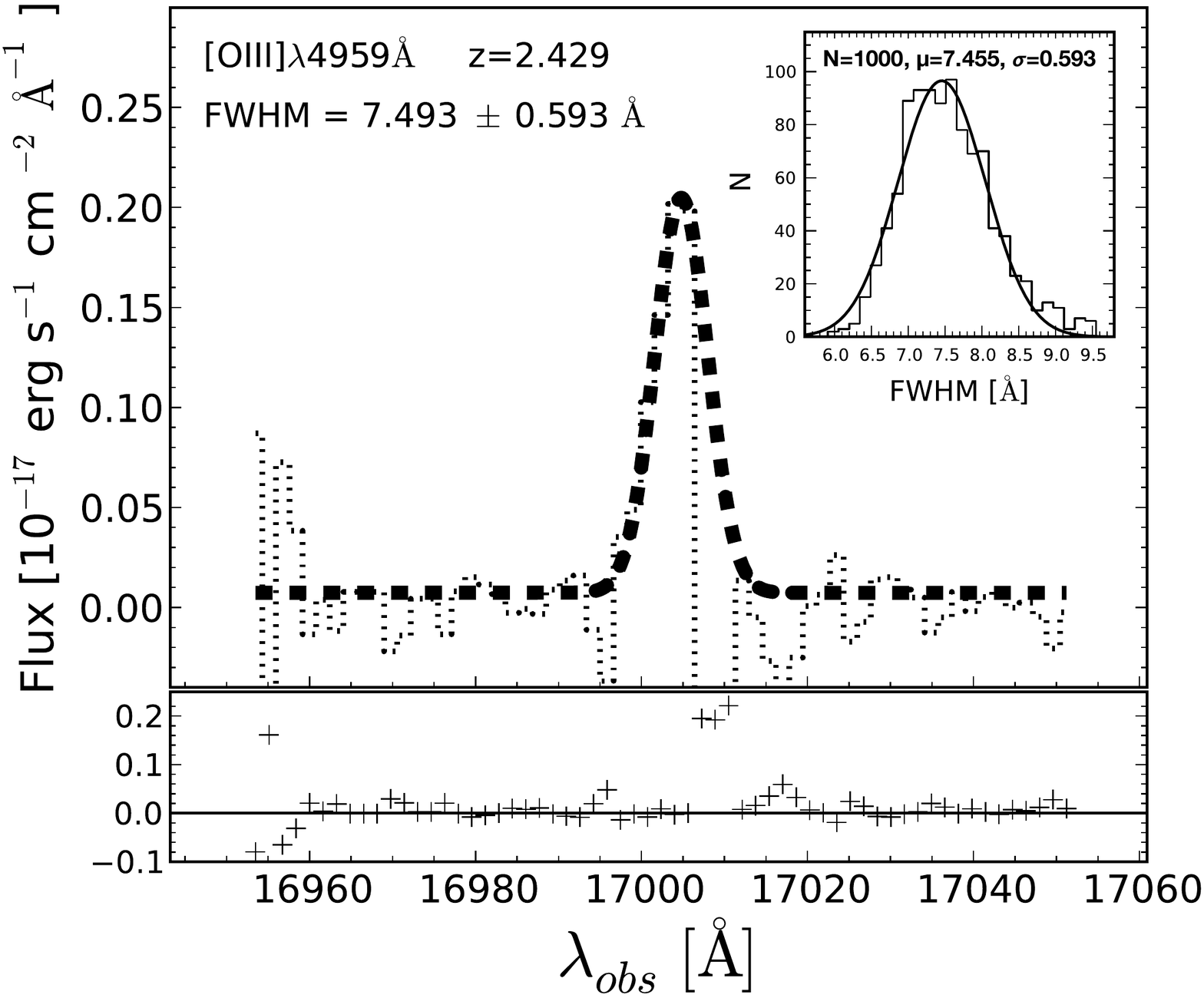}} &
   {\hspace*{0.1cm}\includegraphics[width=.392\textwidth]{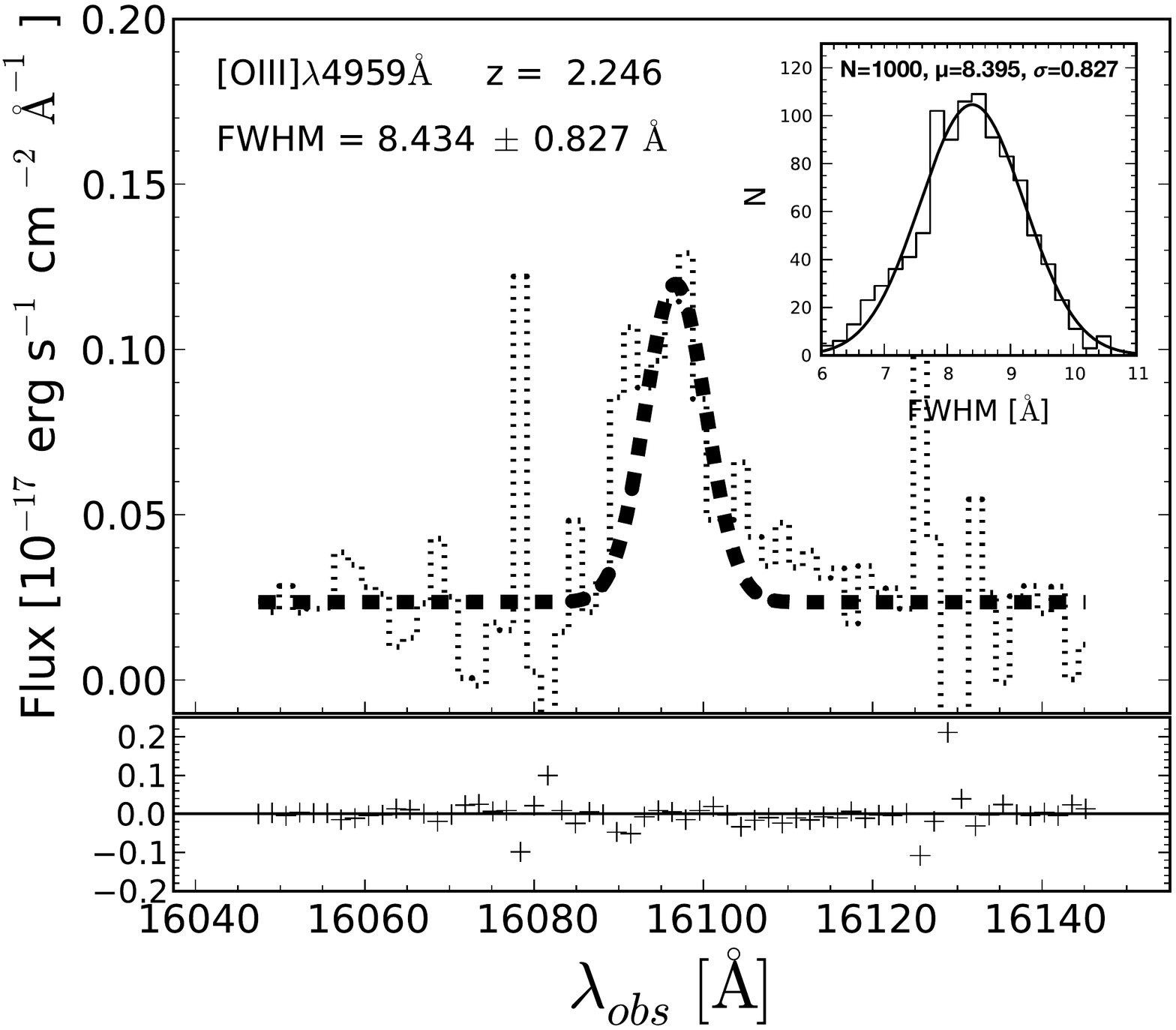}}
\end{array}$
\end{center}

\vspace*{-0.65cm}
\begin{center}$
\begin{array}{cc}
   {\hspace*{-0.1cm}\includegraphics[width=.392\textwidth]{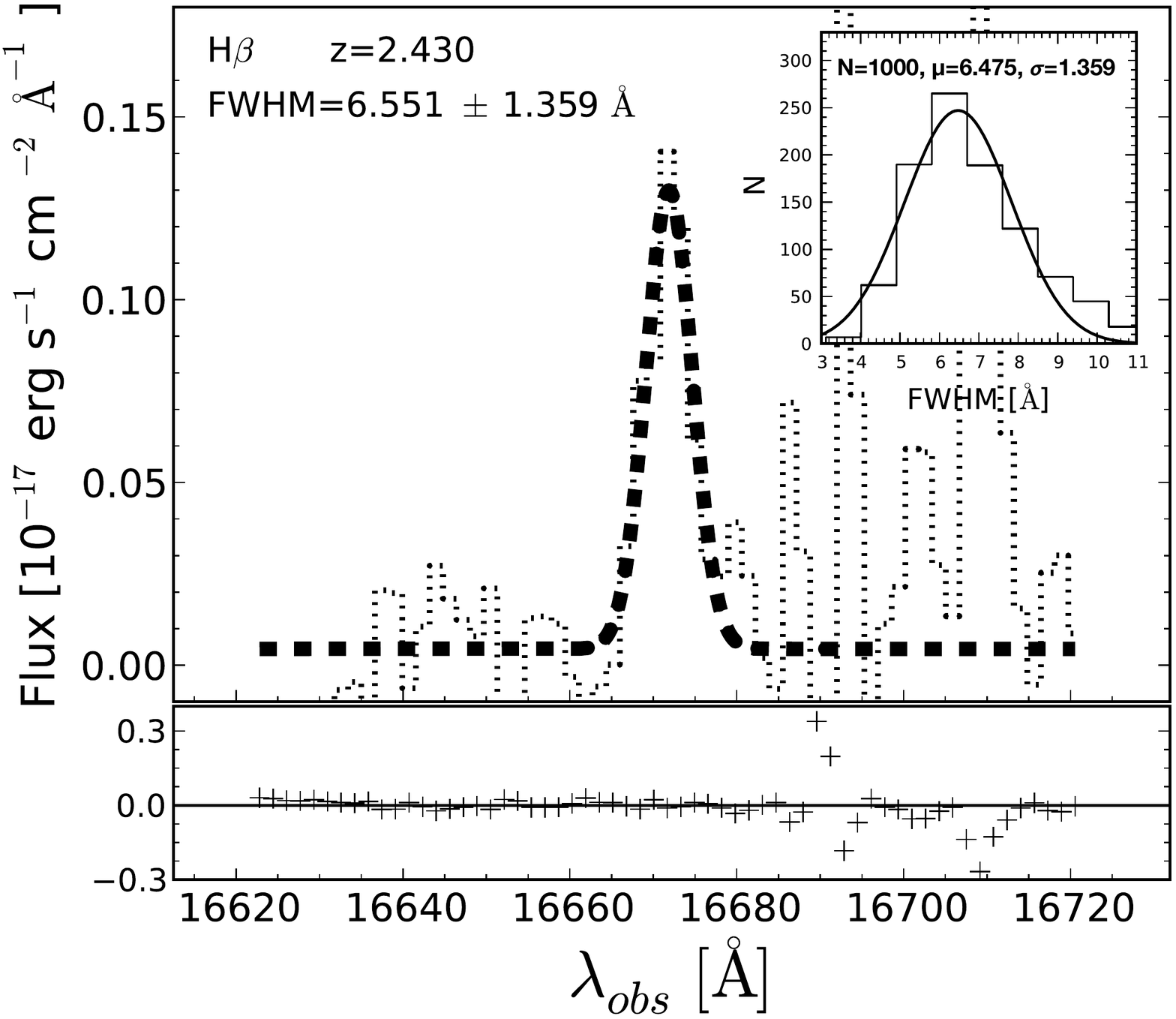}} &
   {\includegraphics[width=.402\textwidth]{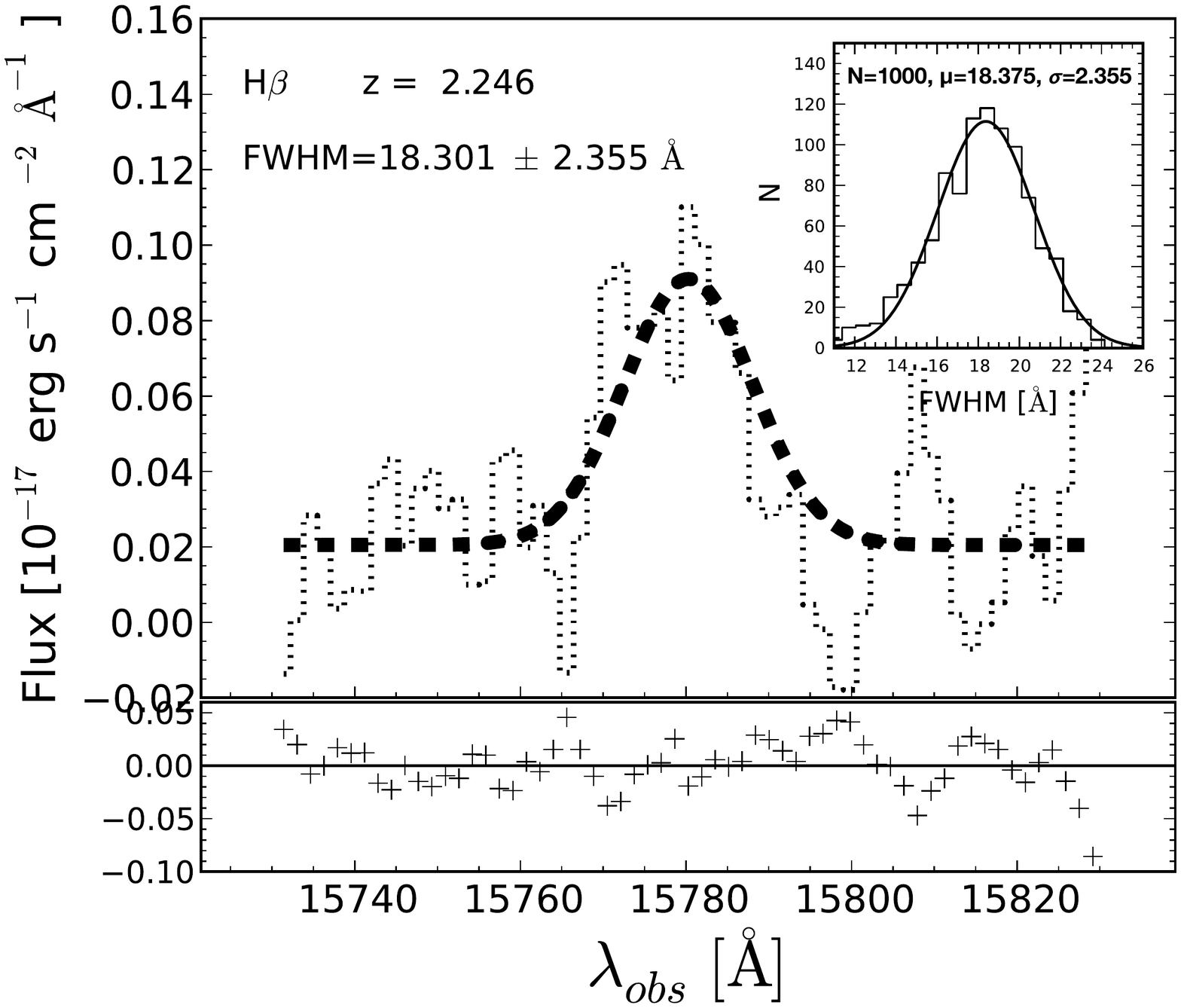}}
\end{array}$
\end{center}
\end{figure*}

\clearpage
\begin{figure*}
\begin{center}$
\begin{array}{cc}
   {\hspace*{0.4cm}\includegraphics[width=.369\textwidth]{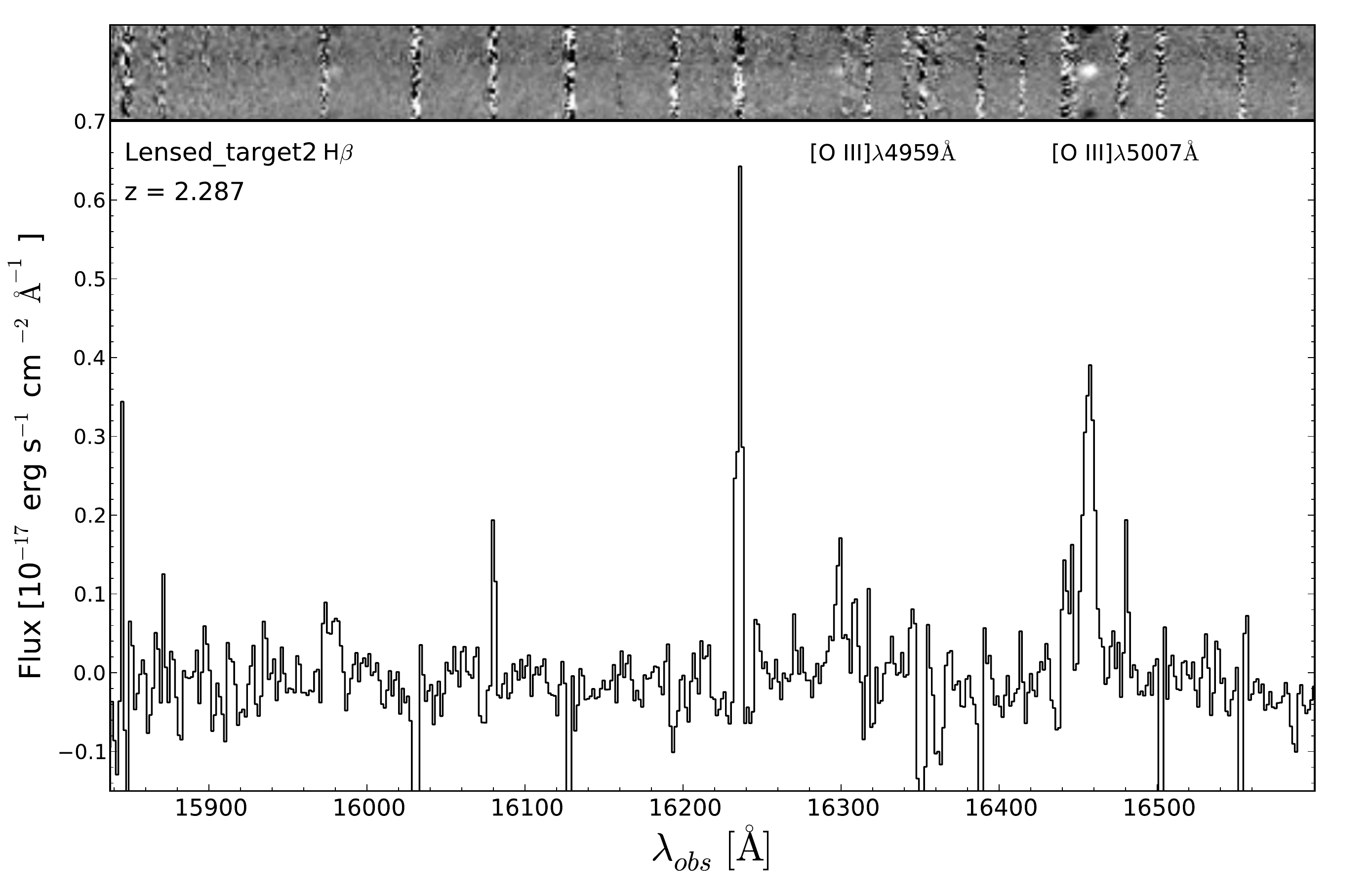}} &
   {\hspace*{0.65cm}\includegraphics[width=.365\textwidth]{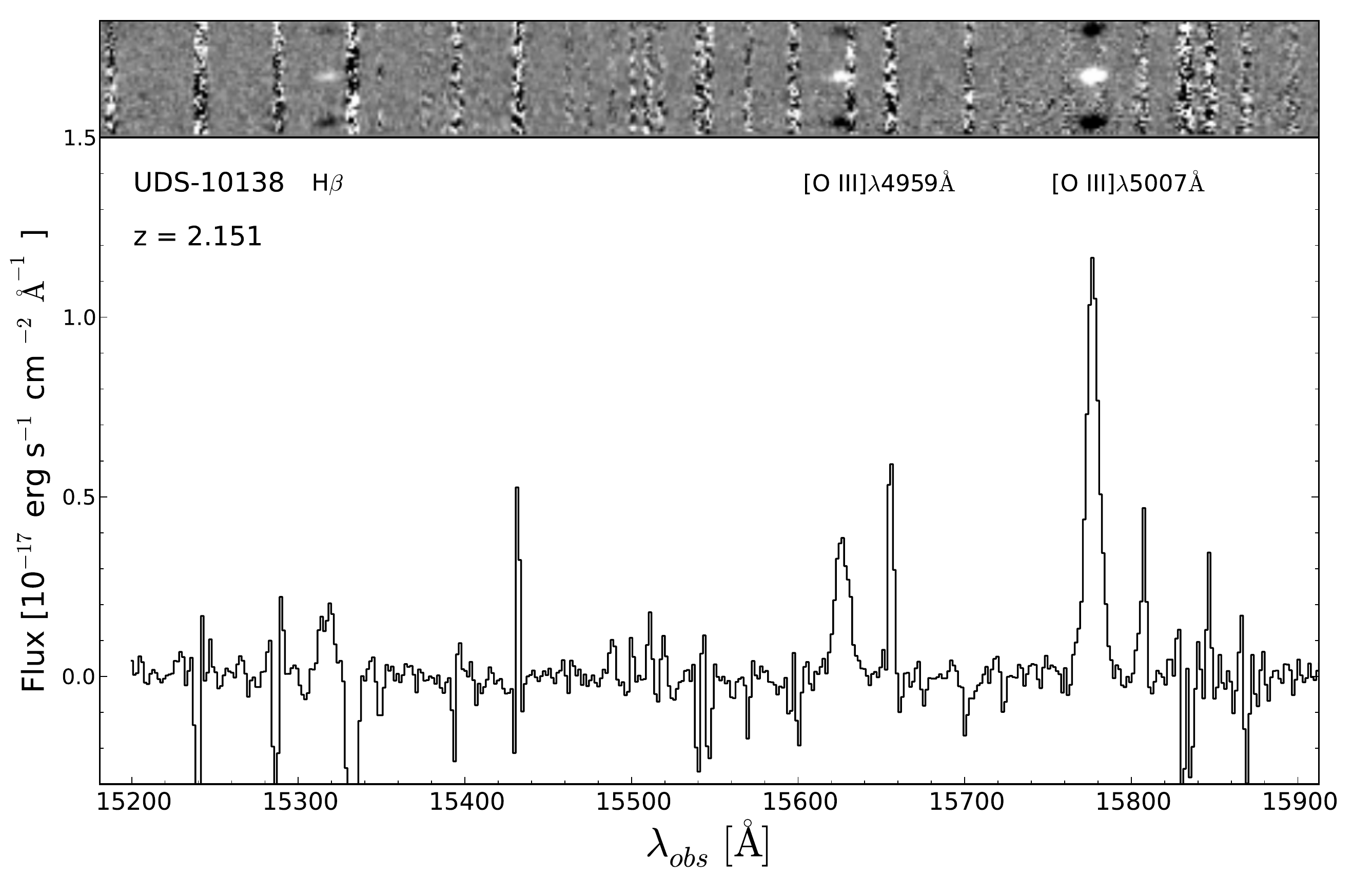}}
\end{array}$
\end{center}

\vspace*{-0.6cm}
\begin{center}$
\begin{array}{cc}
   {\hspace*{-0.02cm}\includegraphics[width=.385\textwidth]{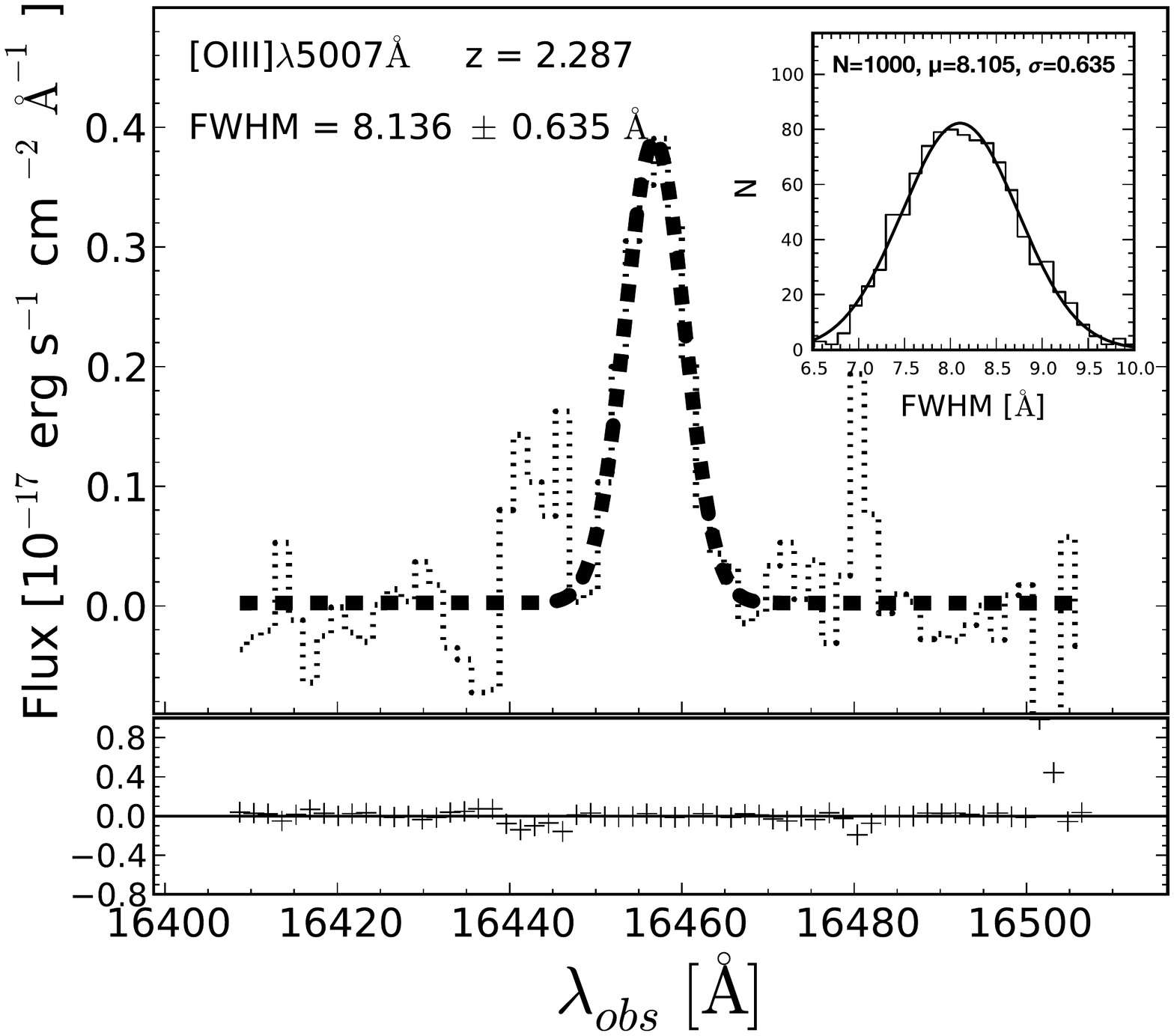}}  &
   {\hspace*{0.32cm}\includegraphics[width=.385\textwidth]{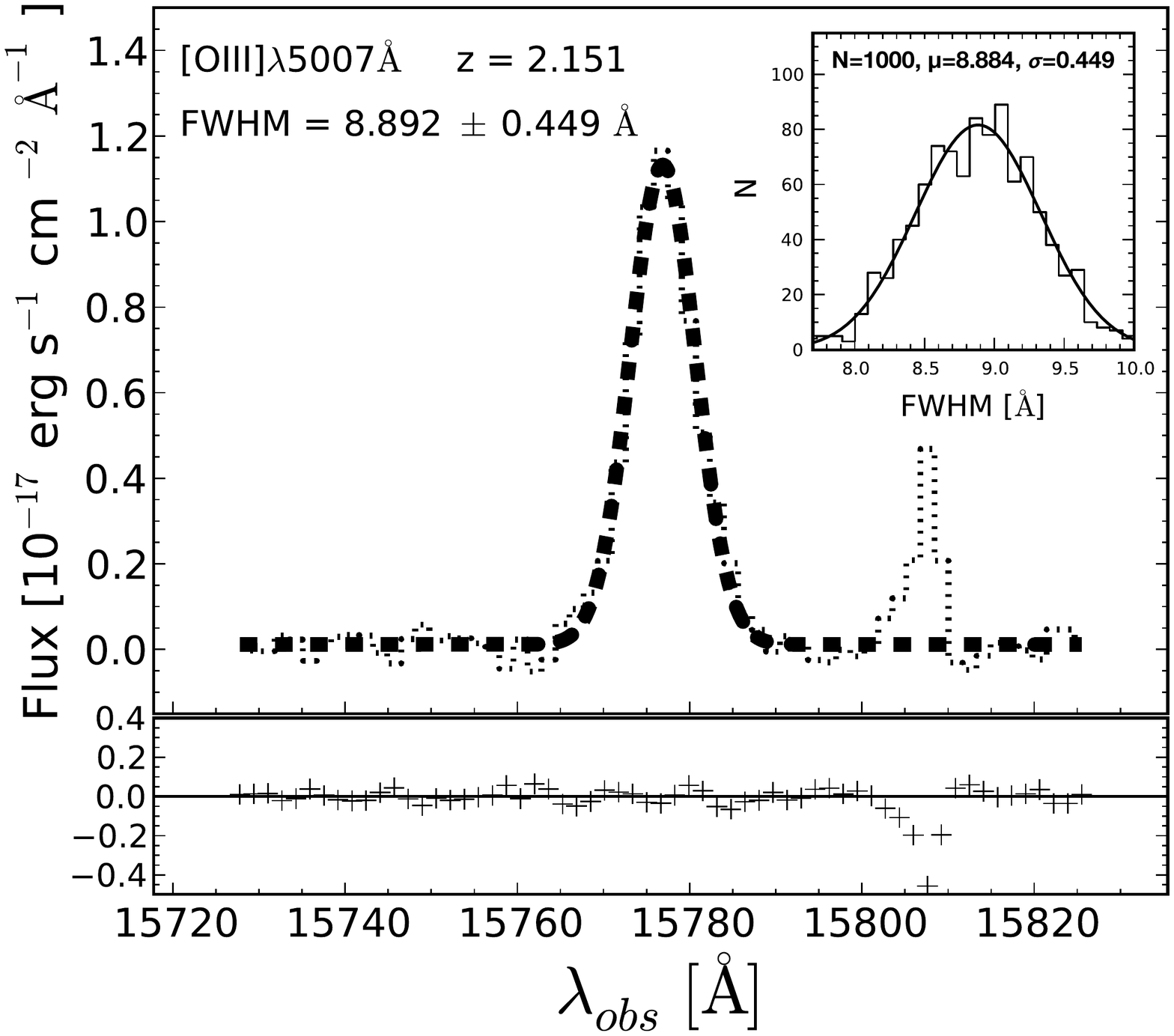}}
\end{array}$
\end{center}

\vspace*{-0.61cm}
\begin{center}$
\begin{array}{cc}
   {\hspace*{-0.12cm}\includegraphics[width=.393\textwidth]{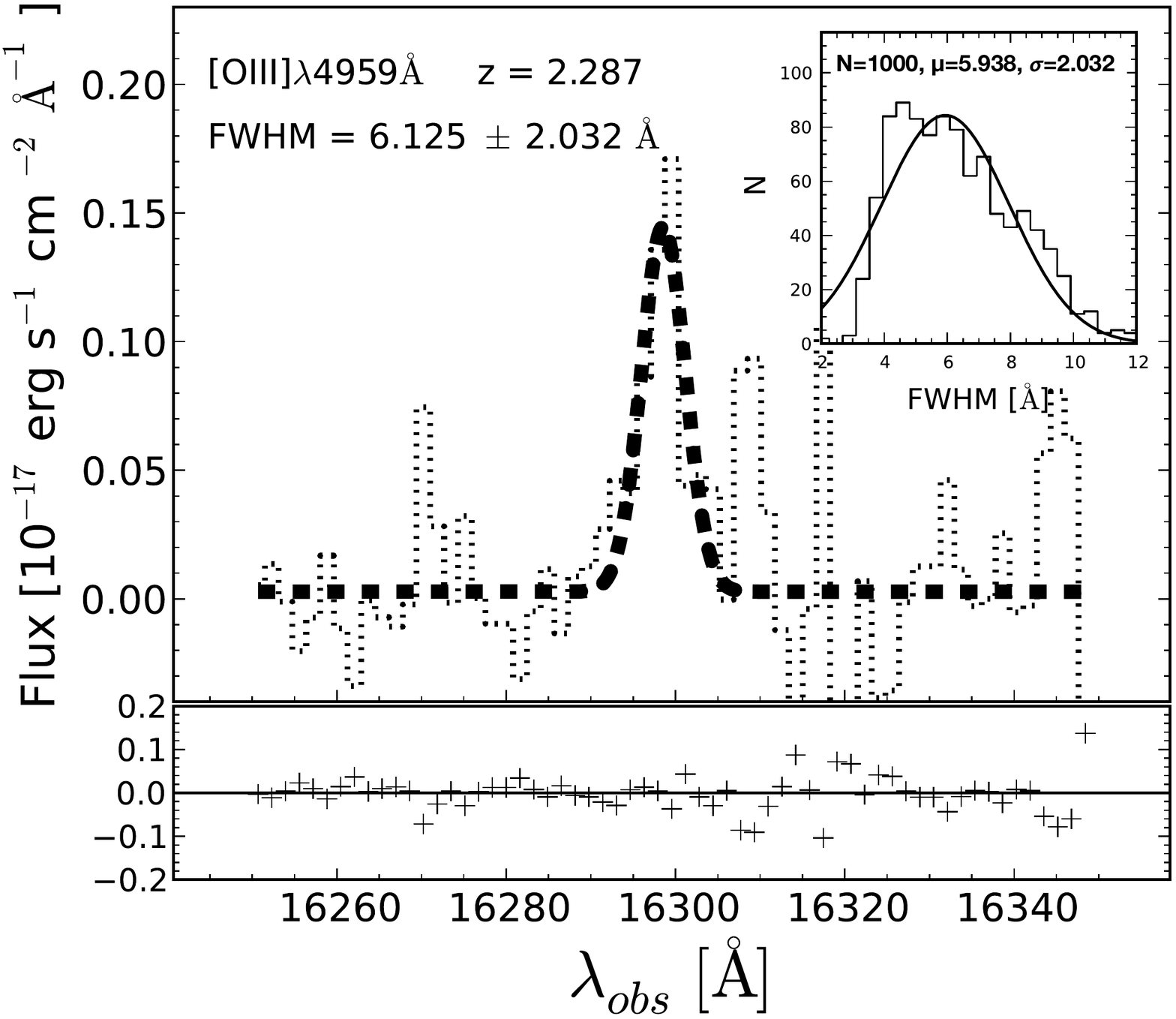}}  &
   {\hspace*{0.27cm}\includegraphics[width=.384\textwidth]{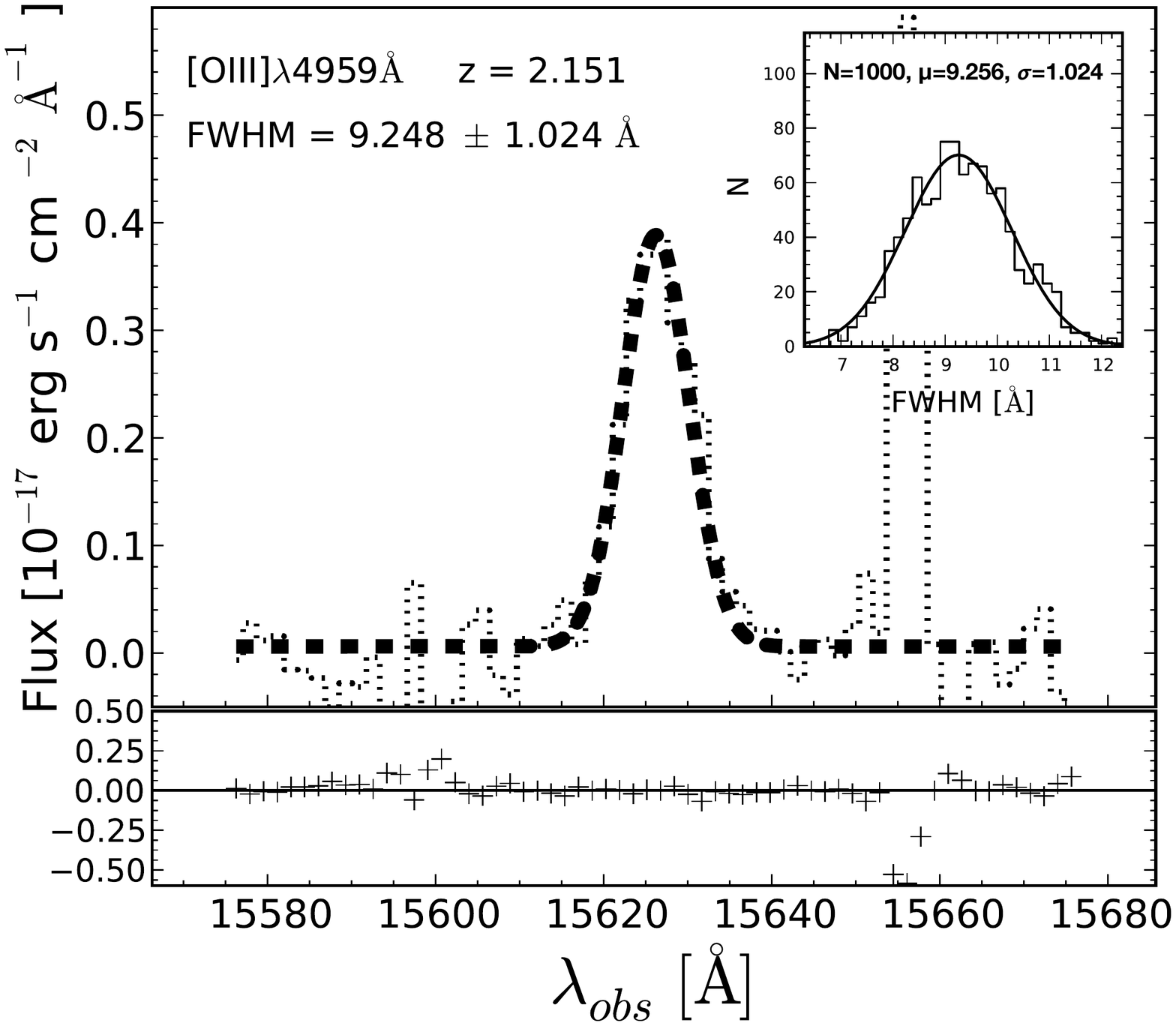}}
\end{array}$
\end{center}

\vspace*{-0.61cm}
\begin{center}$
\begin{array}{cc}
   {\hspace*{-0.25cm}\includegraphics[width=.402\textwidth]{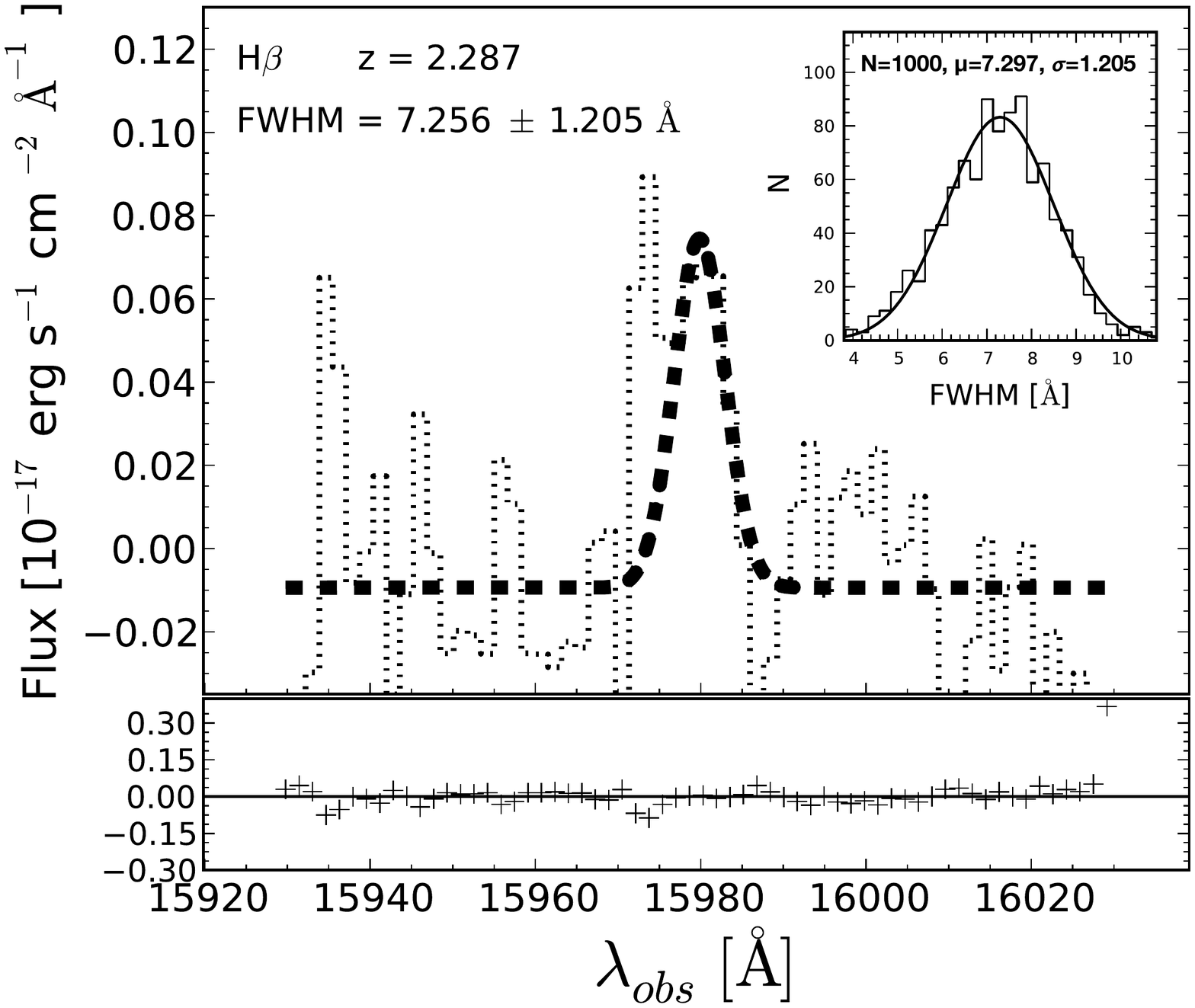}} &
   {\hspace*{0.2cm}\includegraphics[width=.392\textwidth]{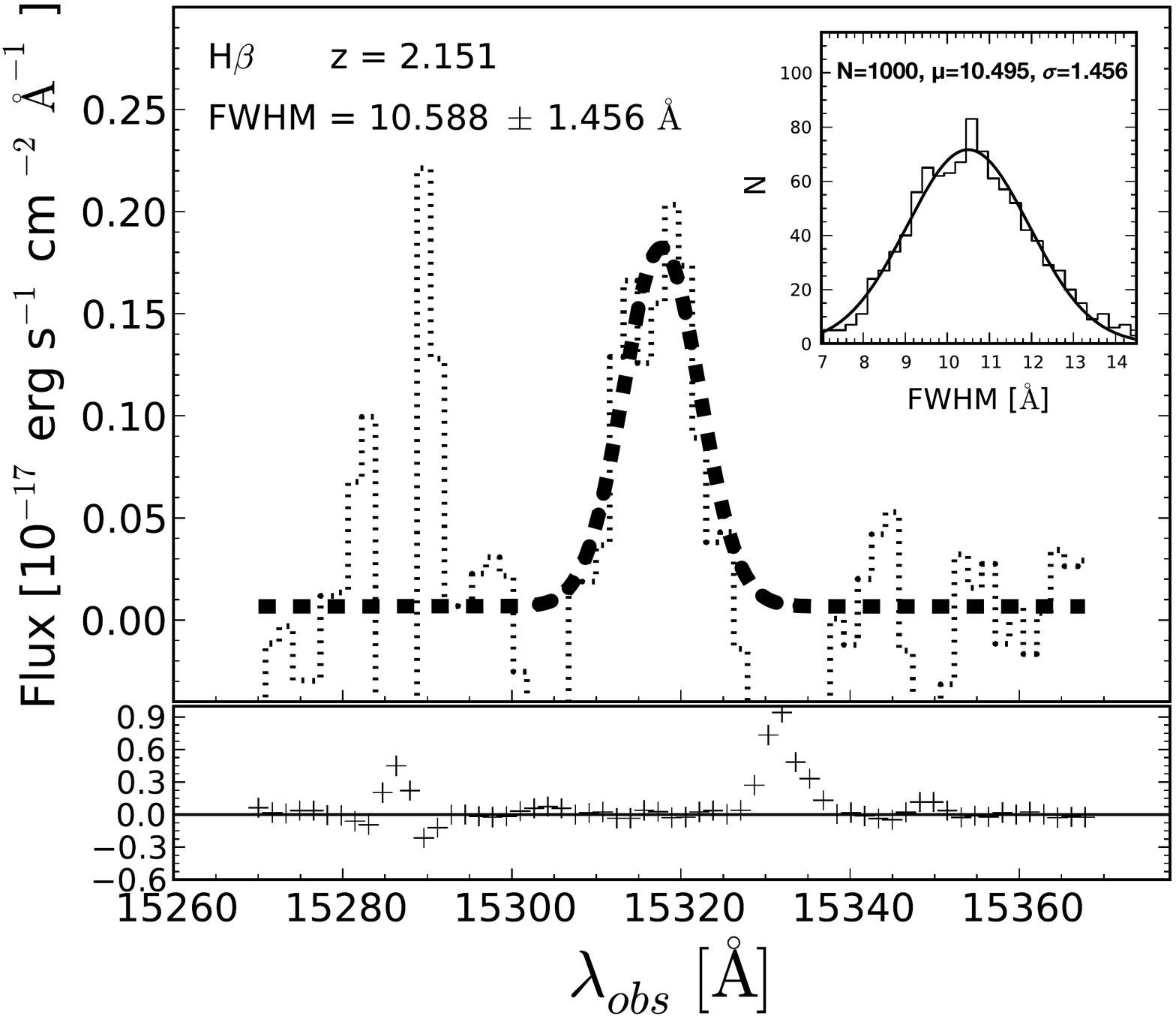}}
\end{array}$
\end{center}
\end{figure*}

\clearpage
\begin{figure*}
\vspace*{-0.3cm}
\begin{center}$
\begin{array}{cc}
   {\hspace*{0.5cm}\includegraphics[width=.368\textwidth]{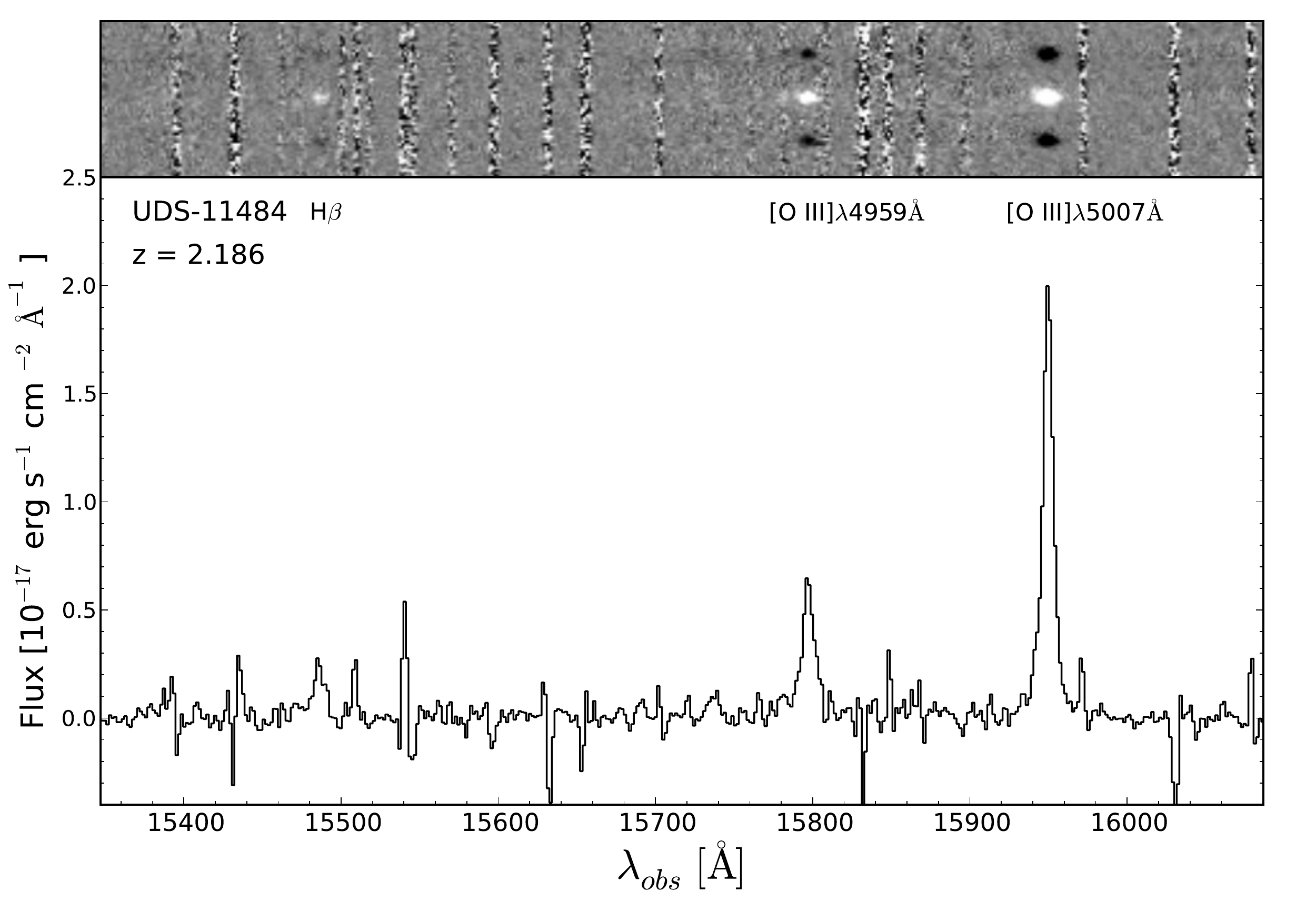}} &
   {\hspace*{0.65cm}\includegraphics[width=.37\textwidth]{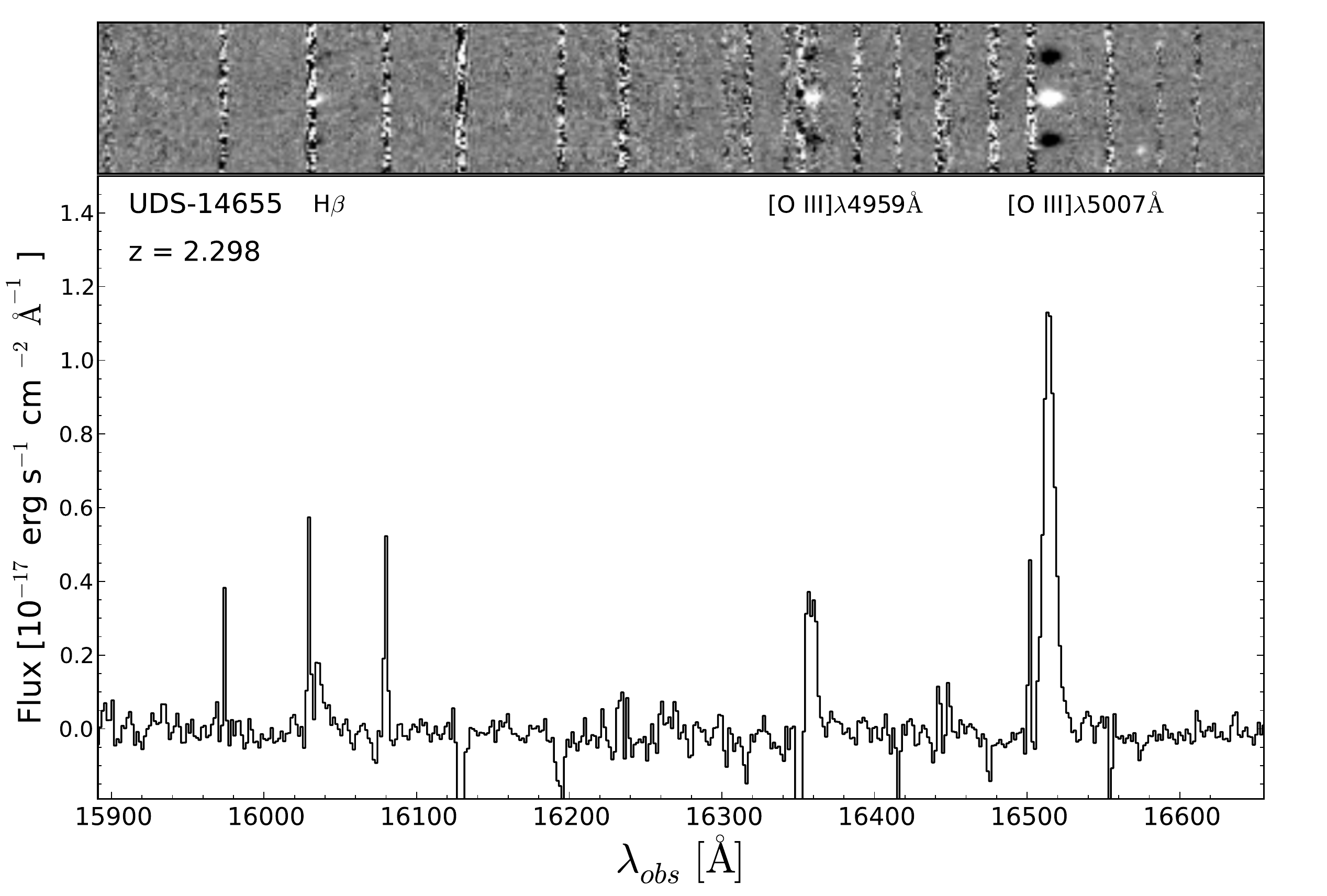}}
\end{array}$
\end{center}

\vspace*{-0.6cm}
\begin{center}$
\begin{array}{cc}
   {\hspace*{-0.02cm}\includegraphics[width=.385\textwidth]{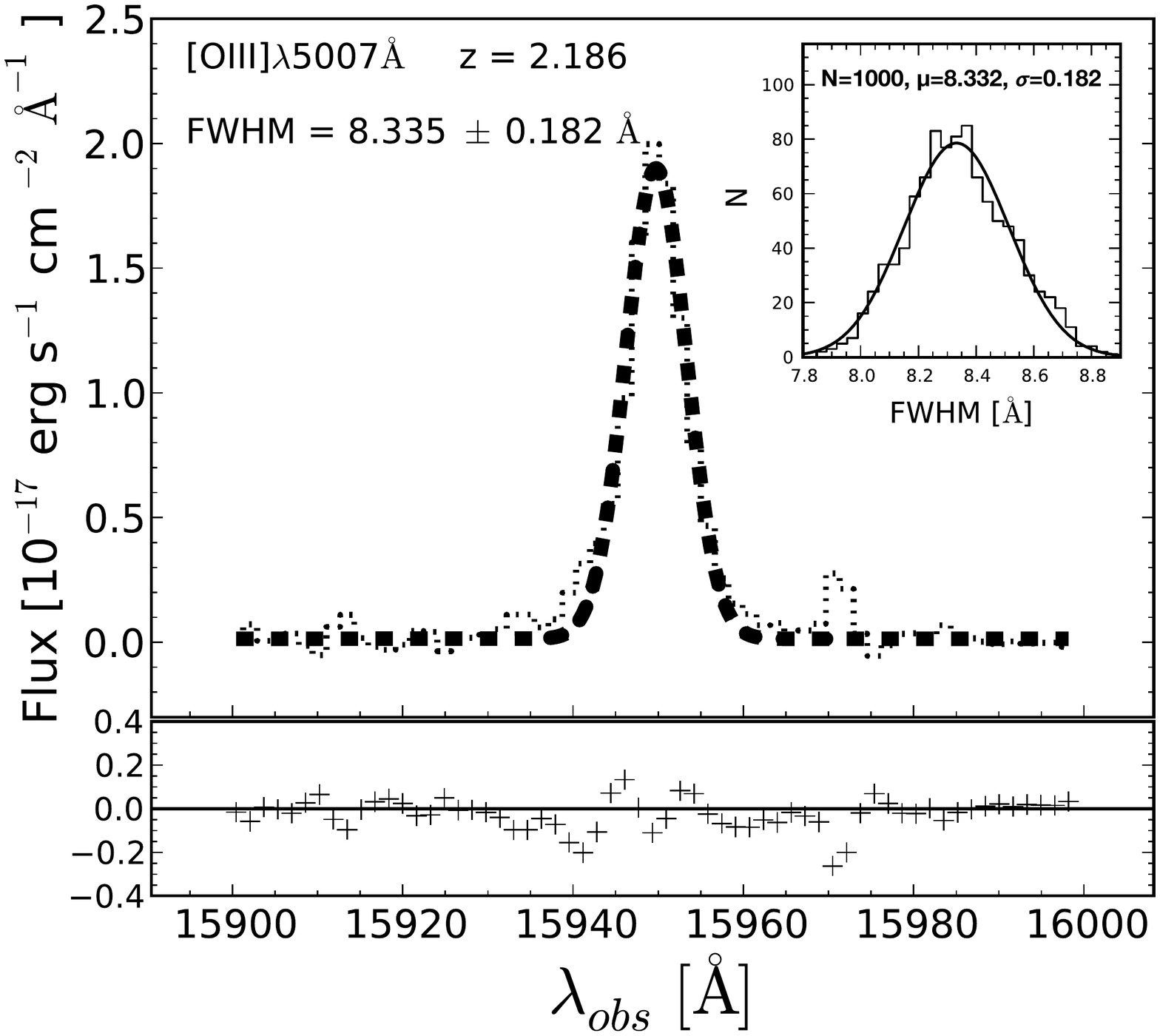}}  &
   {\hspace*{0.32cm}\includegraphics[width=.385\textwidth]{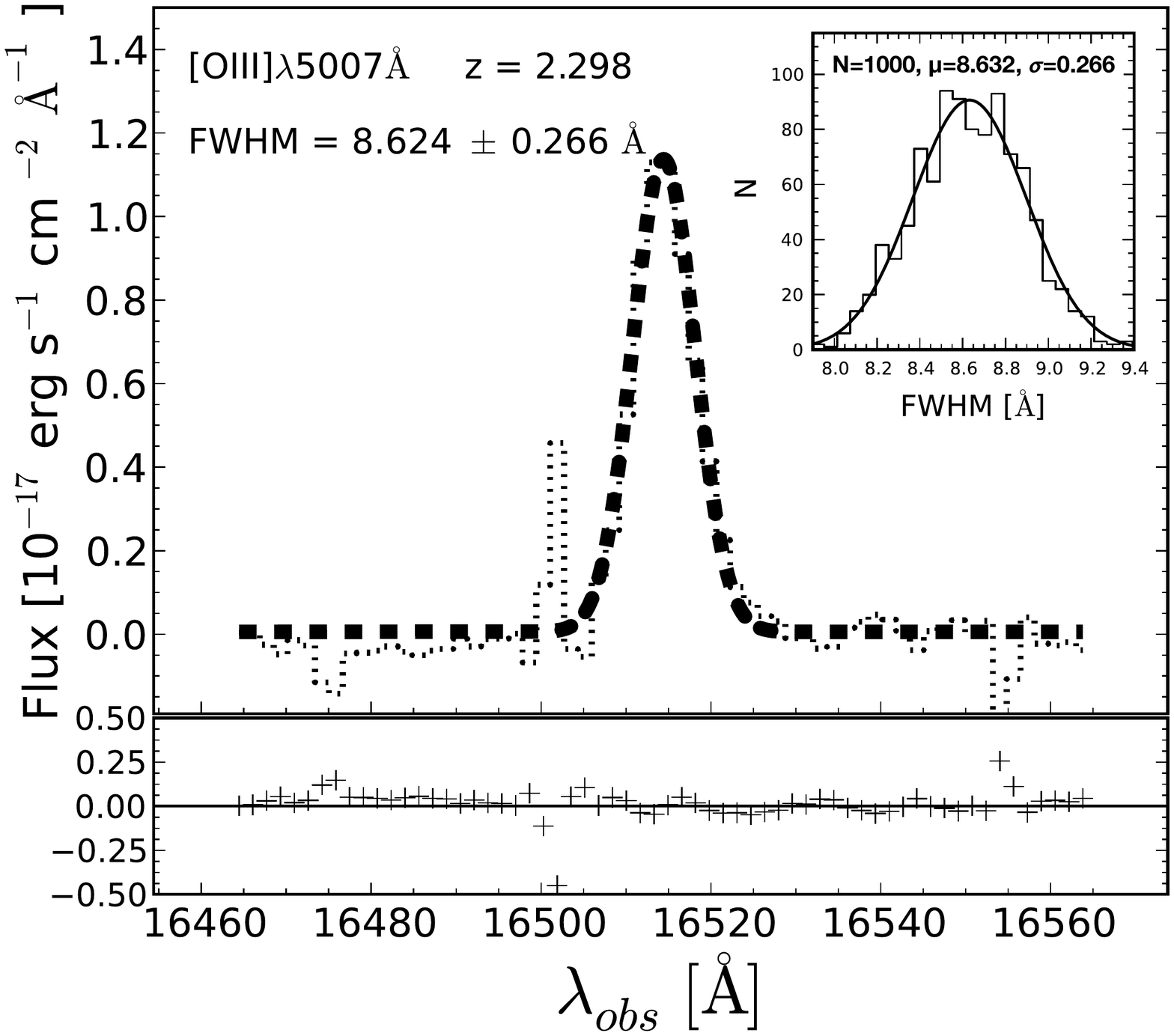}}
\end{array}$
\end{center}

\vspace*{-0.63cm}
\begin{center}$
\begin{array}{cc}
   {\hspace*{0.15cm}\includegraphics[width=.383\textwidth]{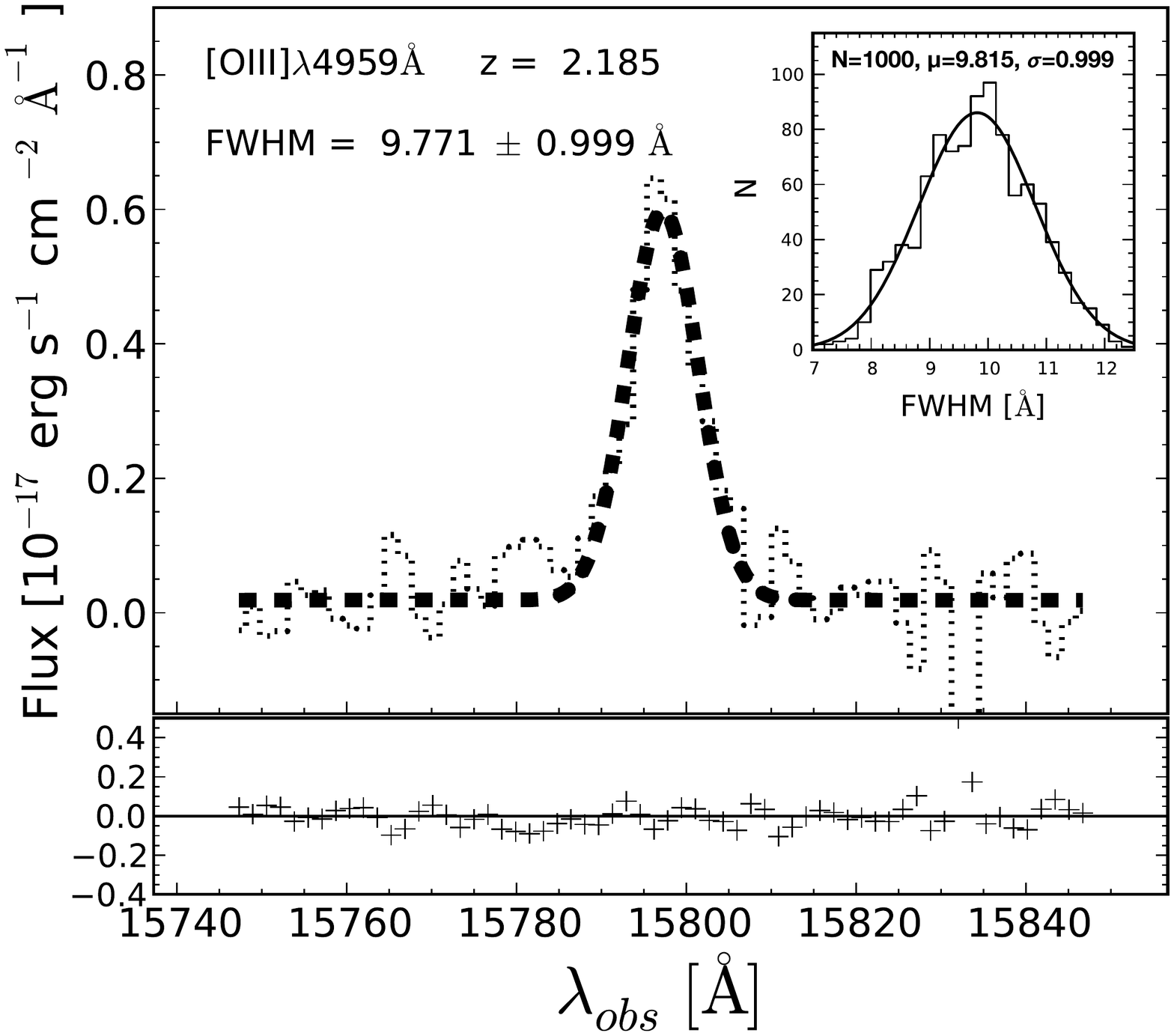}}  &
   {\hspace*{0.35cm}\includegraphics[width=.394\textwidth]{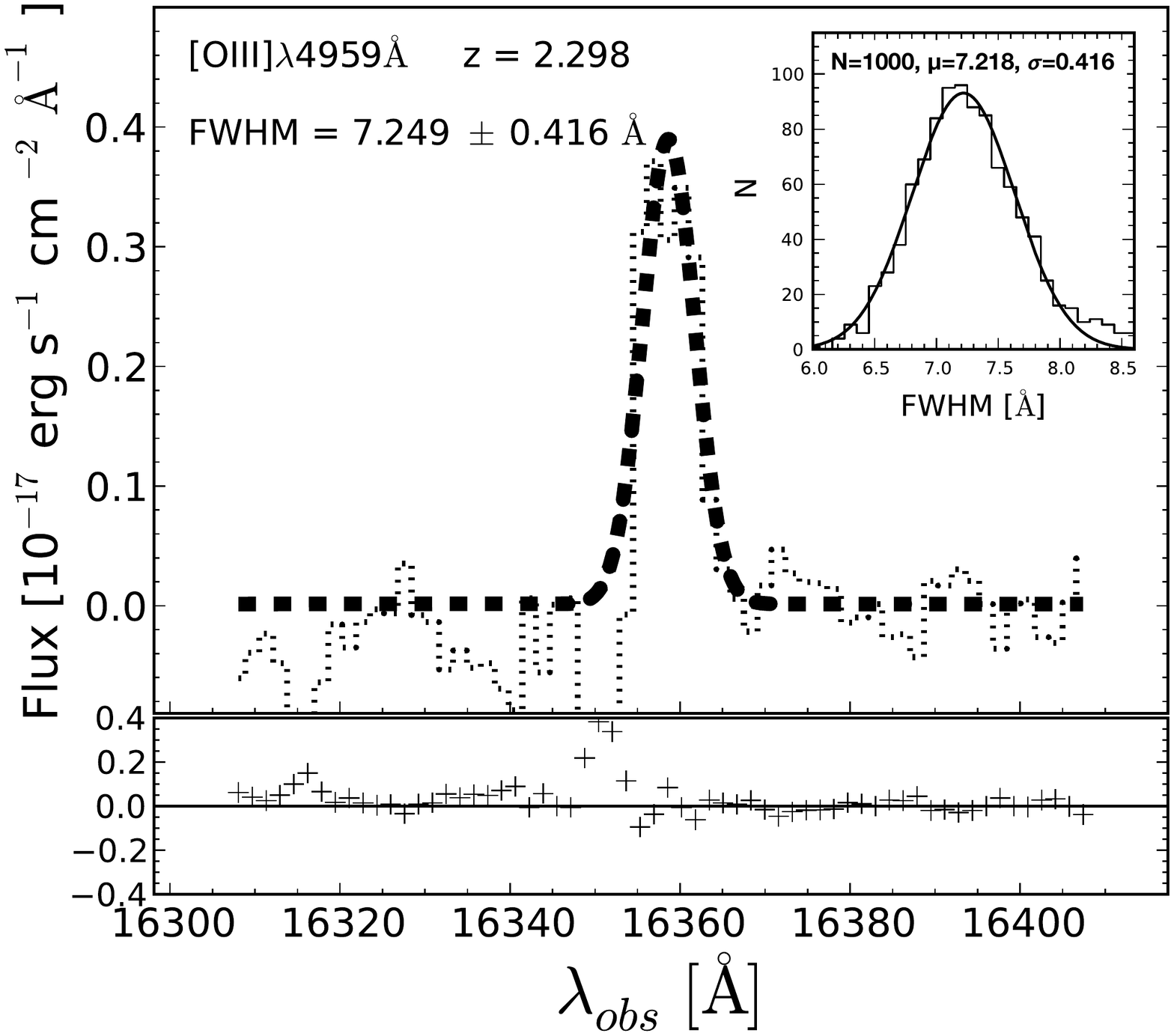}}
\end{array}$
\end{center}

\vspace*{-0.61cm}
\begin{center}$
\begin{array}{cc}
   {\hspace*{-0.13cm}\includegraphics[width=.392\textwidth]{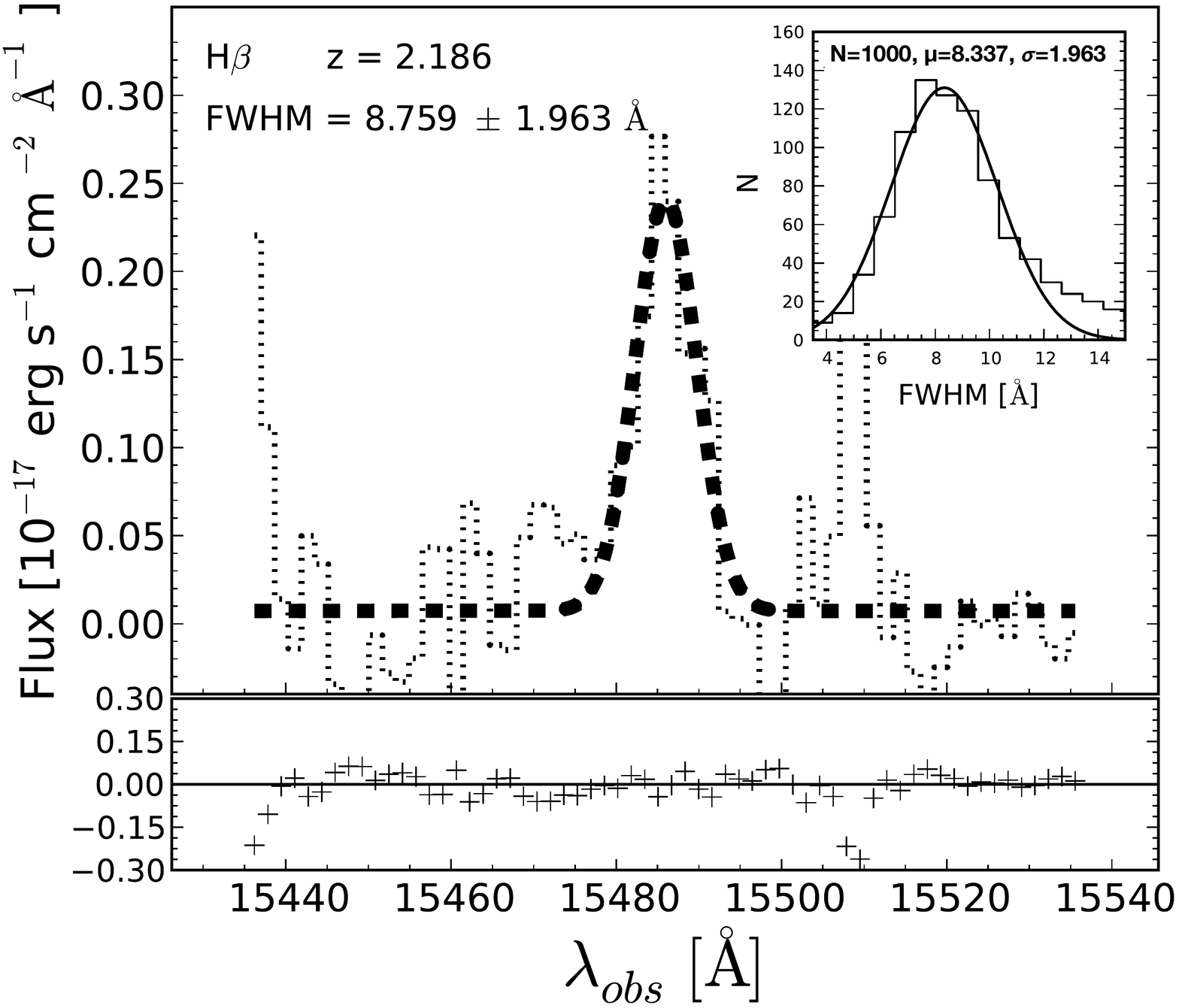}}  &
   {\hspace*{0.2cm}\includegraphics[width=.392\textwidth]{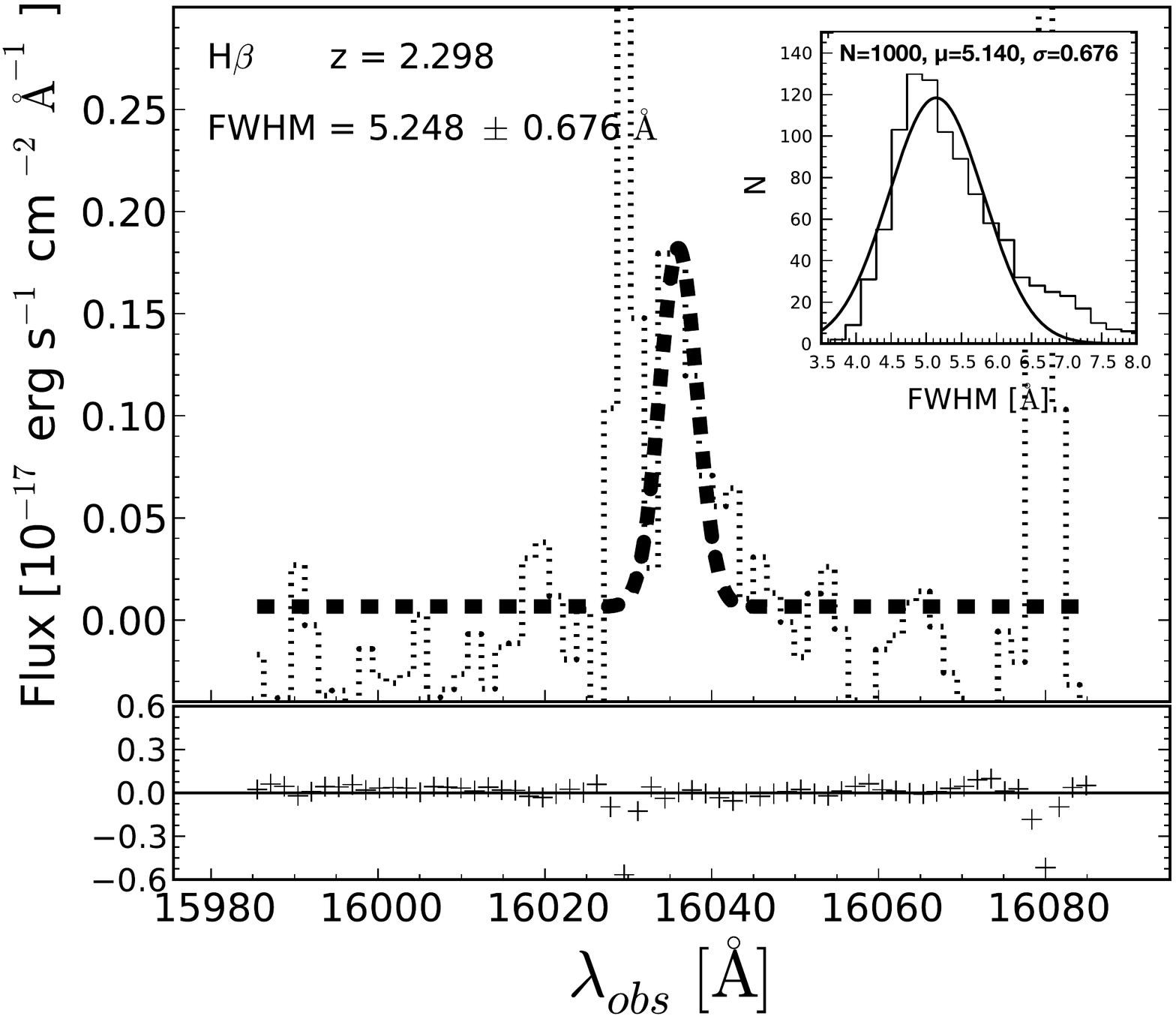}}
\end{array}$
\end{center}
\end{figure*}

\clearpage
\begin{figure*}
\vspace*{-0.3cm}
\begin{center}$
\begin{array}{c}
   {\hspace*{0.5cm}\includegraphics[width=.37\textwidth]{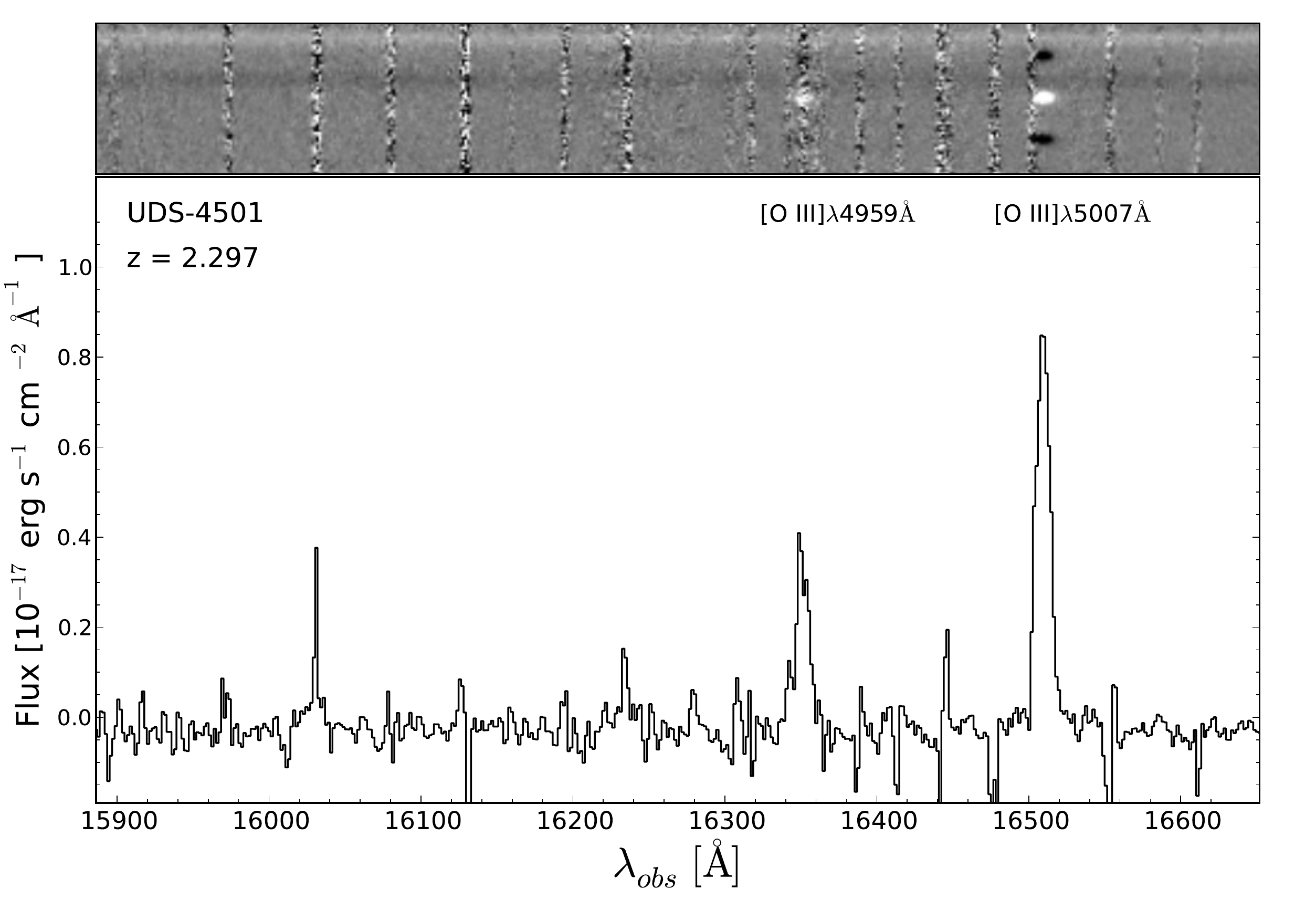}}
\end{array}$
\end{center}

\vspace*{-0.61cm}
\begin{center}$
\begin{array}{c}
   {\includegraphics[width=.39\textwidth]{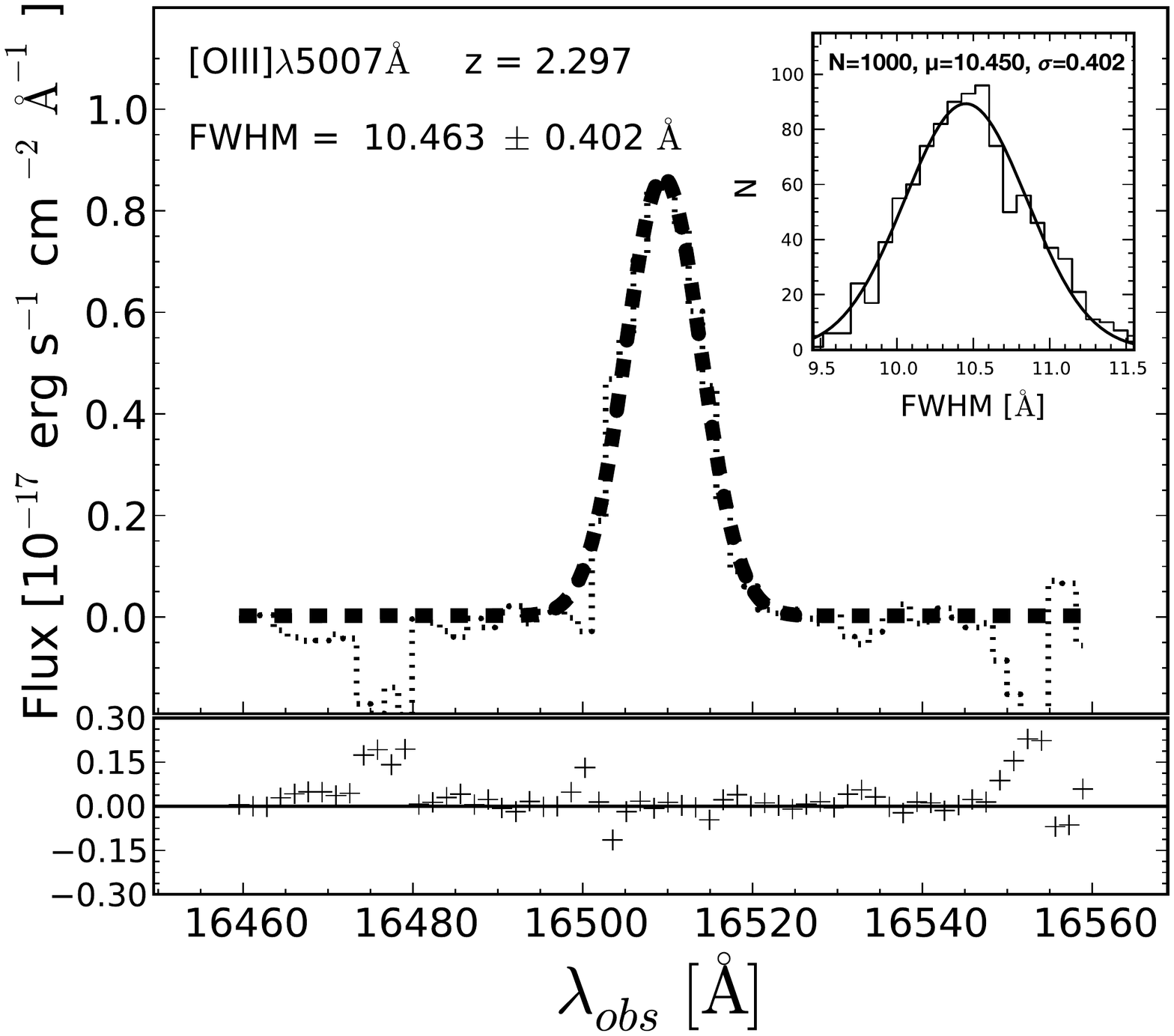}} 
\end{array}$
\end{center}
\end{figure*}

\begin{figure*}
\vspace*{-0.67cm}
\begin{center}$
\begin{array}{c}
   {\includegraphics[width=.39\textwidth]{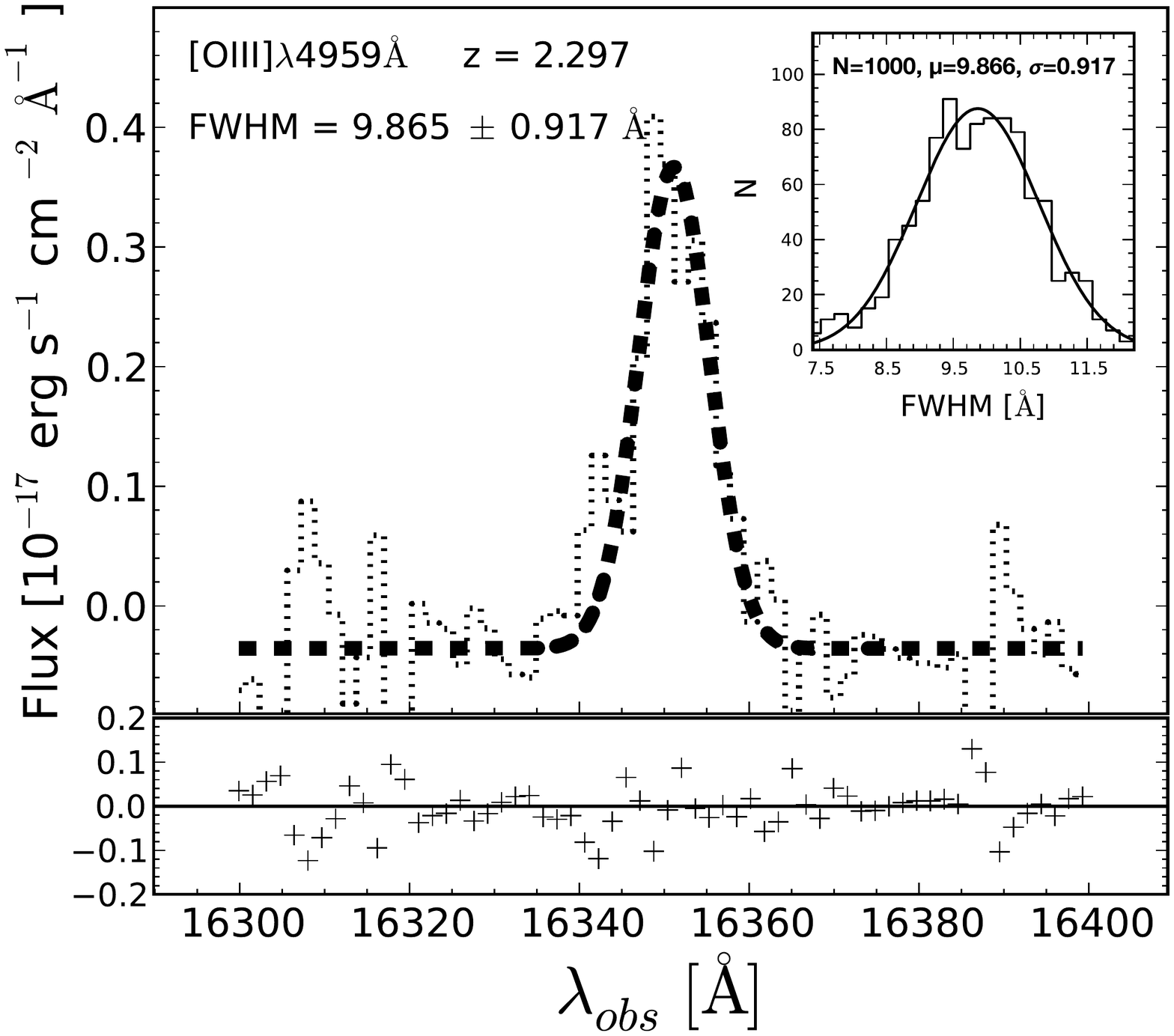}}
\end{array}$
\end{center}
\caption{First row shows the 1D and 2D spectra of [$\Oiii$]$\lambda\lambda$4959,5007 \AA\ and H$\beta$. Second, third and fourth rows show the fits to the  [$\Oiii$]$\lambda\lambda$4959,5007 \AA\ and H$\beta$ lines, if  detected, the distribution of FWHM obtained from the Montecarlo simulations in the inset at the upper right corner and the residuals in the box underneath.}
\label{1D and 2D spectra of O III}
\end{figure*}

\clearpage

\end{document}